\documentclass[%
 reprint,
 amsmath,amssymb,
 aps,
nofootinbib
]{revtex4-2}

\usepackage{graphicx}
\usepackage{dcolumn}
\usepackage{bm}
\usepackage{tikz}
\usetikzlibrary{arrows.meta,
                calc, chains,
                quotes,
                positioning,
                shapes.geometric}

\usepackage{amsmath} 
\usepackage{listofitems} 
\usepackage[outline]{contour} 
\contourlength{1.4pt}
\usepackage[export]{adjustbox}

\tikzset{>=latex} 
\usepackage{xcolor}
\usetikzlibrary{positioning}
\colorlet{myred}{red!80!black}
\colorlet{myblue}{blue!80!black}
\colorlet{mygreen}{green!60!black}
\colorlet{myorange}{orange!70!red!60!black}
\colorlet{mydarkred}{red!30!black}
\colorlet{mydarkblue}{blue!40!black}
\colorlet{mydarkgreen}{green!30!black}
\tikzstyle{node}=[thick,circle,draw=myblue,minimum size=22,inner sep=0.5,outer sep=0.6]
\tikzstyle{node in}=[node, black,draw= black,fill= white]
\tikzstyle{node hidden}=[node,blue!20!black,draw=myblue!30!black,fill=myblue!20]
\tikzstyle{node convol}=[node,orange!20!black,draw=myorange!30!black,fill=myorange!20]
\tikzstyle{node out}=[node,black,draw=black,fill=green!20]
\tikzstyle{connect}=[thick,mydarkblue] 
\tikzstyle{connect arrow}=[-{Latex[length=4,width=3.5]},thick,mydarkblue,shorten <=0.5,shorten >=1]
\tikzset{ 
  node 1/.style={node in},
  node 2/.style={node hidden},
  node 3/.style={node out},
}

\begin{document}

\preprint{APS/123-QED}

\title{Extraction of the Sivers function with deep neural networks}

\author{I. P. Fernando}
\email{ishara@virginia.edu}
\author{D. Keller}%
 \email{dustin@virginia.edu}
\affiliation{%
 Department of Physics, University of Virginia\\
 Charlottesville, Virginia 22904, USA}%


\date{\today}

\begin{abstract}
Deep Neural Networks (DNNs) are a powerful and flexible tool for information extraction and modeling. In this study, we use DNNs to extract the Sivers functions by globally fitting Semi-Inclusive Deep Inelastic Scattering (SIDIS) data. To make predictions of this Transverse Momentum-dependent Distribution (TMD), we construct a minimally biased model using data from COMPASS and HERMES. The resulting Sivers function model, constructed using SIDIS data, is also used to make predictions for Drell-Yan kinematics specific to the valence and sea quarks, with careful consideration given to experimental errors, data sparsity, and complexity of phase space.
\end{abstract}

\maketitle

\tableofcontents

\section{\label{sec1}Introduction}


The development of modern information extraction techniques has far outpaced experimental progress. Even with decades of experimental results, there is limited data on Transverse Momentum Dependent Parton Distribution Functions (TMDs) for global analyses that can be applied to a 3D phenomenological interpretation. Theoretical efforts are providing an ever-evolving framework and toolbox to interpret the data, leading to new areas of research that optimize information extraction unique to spin physics and the internal structure of hadrons. Artificial intelligence (AI) is accelerating data-driven research with the superior capacity of deep neural networks (DNNs) to be used as function approximators. DNNs can learn to approximate any relationships contained within data, provided there are no limits to the number of neurons, layers, and computing power. With enough data and training, such approximations can be made to nearly any desired degree of accuracy.

The Sivers function is the most studied of the eight leading-twist TMD distributions that pertain to polarized nucleons. The Sivers distribution is naively time-reversal odd and is expected to be process-dependent, which leads to a distribution equal in magnitude but opposite in sign in Semi-Inclusive Deep Inelastic Scattering (SIDIS) compared to the Drell-Yan (DY) process \cite{Sivers_1990, Sivers_1991}.
The Sivers distribution, as measured from the transverse single-spin asymmetries, provides information on the correlation between the nucleon's spin and the angular distribution of outgoing hadrons in SIDIS or the di-muons in DY. The quark Sivers function, $\Delta^N f_{q/{N}^\uparrow}$, describes the number density in momentum space of unpolarized quarks inside the transversely polarized target nucleon with nuclear spin ($N^{\uparrow}$). A non-zero Sivers function indicates a contribution of quark orbital angular momentum to the target's spin.

TMD extraction and modeling with sensitivity to TMD evolution and factorization is critical and provides predictive power in the collinear limit characterizing the nonperturbative effects. Representing evolution also accounts for the momentum-dependent QCD interactions between partons inside the hadron, which affect their distribution and can significantly impact observables. Some theoretical approaches use the parton model approximation without TMD evolution and demonstrate good agreement with experimental results. At the time of their origin, these calculations assumed that evolution effects in asymmetries are suppressed, as asymmetries are ratios of cross-sections where evolution and higher-order effects should cancel out \cite{Anselmino_DY_2009, Anselmino2017, Cammarota_2020}. The more modern picture \cite{Aybat_2012} implies this is likely an oversimplification.  There is now strong evidence that the main effect of evolution takes place at lower energies ($<10$~GeV) \cite{Aybat_2012,Bury2021_JHEP}, so it is precisely this domain that offers the best opportunity to capture these important QCD relations in TMD distributions. On the other hand, studies that incorporated TMD evolution, as seen in \cite{Echevarria_2014, Echevarria_2021}, encountered difficulties and did not achieve better agreement with the Drell-Yan data compared to earlier analyses \cite{Kang2009, Anselmino2017, Cammarota_2020}. This situation poses a challenge in confirming the validity of any theoretical implementation with respect to the TMD factorization theorem or modern utilities required to preserve it, particularly the scale-dependent universal nonperturbative evolution kernel, the so-called Collins-Soper (CS) kernel. However, the work in \cite{Bury2021_JHEP, Bury2021_PRL} demonstrated the conditional universality of the Sivers function using a simultaneous fit to SIDIS, Drell-Yan, and W/Z boson production data, including TMD evolution and the universal nonperturbative CS kernel extracted in \cite{Scimemi_2020} from unpolarized measurements. In contrast, the working principle of DNNs as a means of extraction can facilitate the necessary flexibility to capture not only the TMD and DGLAP evolution features but also the full complexity of QCD. In this regard, a carefully designed DNN model accompanied by a systematic method of extraction can incorporate not only the evolution features but other components not yet quantifiable in the formalism. The inherent capacity for DNNs to reduce dimensionality and organize intricate correlations in the data can be used to help determine the correct theoretical implementation. The challenge, of course, is that once captured in a DNN model, disentangling these features into formal expressions consistent with factorization is nontrivial, and we expect it to be the focus of much future work.

The approach outlined in this article highlights the inherent flexibility of employing specialized DNNs in combination with candidate multiplicative expressions holding some type of separate parameterization which may carry incorrect ansatzes or assumptions (biases).
This can be done while still rendering a high-quality numerical model encoding all information available in the data. These types of biases can be managed by training the model with the biased term included. The DNN term then becomes uniquely parameterized to account for the biases. Consequently, the resulting combination of terms in the model remains largely unbiased, even though each individual term may not be independently meaningful.  The schema can not only facilitate the exploration of the existing formalism found in the literature but also provides a means to derive and test novel phenomenology using a technique that can directly manage and study the biases in individual terms. This will be discussed in more detail in Section \ref{sys}.

The phenomenology used to interpret and analyze experimental data relies on TMD factorization \cite{JC1981, JC1983, JC1985, JC2011, Meng1992, Ji2004, Ji2005, XJi2005, Eche2012, Arnold_2009}, proven for single-spin asymmetries (SSAs) described in terms of convolutions of TMDs. The Sivers function has been previously extracted from SIDIS data by several groups, with generally consistent results \cite{Anselmino2005, Collins2006, Vogelsang_2005, Anselmino2009, bachetta2011, Anselmino2017}. All previous phenomenological fits of the Sivers function (and other TMDs) require an ansatz characterizing the shape of the distribution combined with an assumed form of the Bjorken-x dependence. This can lead to ambiguity in determining both the quantitative results from the fit and the qualitative features of the momentum distributions and their associated dynamics. The function form of the Bjorken-x dependence is usually offered only as a placeholder and is assumed to at least contain the appropriate ingredients to facilitate the extraction. This is undoubtedly a considerable oversimplification but one that has permitted significant progress. In the following analysis, we perform a global fit with the goal of illustrating a method that can also permit significant progress even with data limitations, and with assumptions, and ansatzes in place.  We focus on testing the extraction ability of a DNN representing a single term to maximize information extraction and minimize both the fit error and the analytical ambiguity associated with the interpretation of a generic Bjorken-x dependence, $\mathcal{N}_q(x)$.  There are, of course, broader implications of how these tools can be used to accelerate the field, some of which we will also touch on through our extraction of the Sivers function.

The exceptional capacity of DNNs to be ideal for function approximation is rigorously provable through the Universal Approximation Theorem \cite{UAT89, UAT01}. This is the advantage of DNNs over other machine learning (ML) approaches. In this regard, even the mere existence of a function implies that DNNs can be used to represent it and work with it without actually knowing the function form. With such a high-level abstraction, one can make use of the available data and make assessments not otherwise possible, even given an arbitrary degree of complexity. The complexity can be contained in the data relationships as well as in the experimental uncertainties. In order to make optimal use of experimental uncertainties, a detailed analysis must be provided on their estimated scale and correlations.

DNNs are also Turing-complete, implying the potential to simulate the computational aspects of any real-world general-purpose computing process. The implications are that there is potential for a type of generalizable framework that can be utilized and further developed over time without the knowledge of the exact rigorous details of the underlying mechanisms. Provided appropriately detailed global models, higher-level symbolic regressions can then be performed to infer the strict mathematical form. However, such an approach requires access to a significant amount of experimental data that holds the necessary information. Without the necessary amount of quality data, no matter the number of nodes or sophistication of architecture, DNNs are limited in what useful information they can extract, along with any other technique. Even with this constraint, DNNs can make considerable advancements with the use of sparse data with large experimental uncertainties. Generally, fitting with large errors or performing computational tasks with inherent fuzzy logic are tasks that are difficult to make optimal use of modern computational resources. DNNs are uniquely suited for such challenges.

In the remainder of this article, a first-level DNN extraction of $\mathcal{N}_q(x)$ is performed to deduce the Sivers function from a global analysis using HERMES and COMPASS data. This investigation is exploratory, with the intention of developing tools and techniques that minimize error and maximize utility, which we hope to expand upon in further work. The choice in this article is to focus on the Sivers function, rather than the unpolarized TMD $f_{q/N}(x,k_{\perp})$, is to more clearly demonstrate the power of the method even given the limitations in data.  Our global fit of $f_{q/N}(x,k_{\perp})$ will be presented in later work.  The examination of $\mathcal{N}_q$ in relation to $x$ is prioritized over other possibilities only to illustrate how generating functions can be utilized to measure and analyze systematics arising from the extraction produced. As we demonstrate in this work, this technique enables a systematic enhancement of information extraction.

In Section \ref{kinematics}, we present the formalism of the Sivers function and the kinematics for both SIDIS and DY. In Section \ref{fitting}, a discussion of the fitting techniques of $\mathcal{N}_q(x)$ is presented with a focus on the methodology of the DNN approach. Section \ref{extraction} explains in detail the extraction technique, starting with model testing. We perform a baseline fit using the classical Minuit $\chi^2$ minimization algorithm and then perform the DNN fit, demonstrating with pseudodata the fidelity of the procedure. We then walk through the final DNN fit to experimental data for the polarized proton and the deuteron separately. The results of the fits are presented in Section \ref{results}.  Some further analysis and systematic studies are reported in Section \ref{sys}.  Then in Section \ref{tom} we present the implication of the results in terms of the 3D tomography of the proton.  In Section \ref{evol}, we perform a preliminary analysis on TMD evolution and finally, in Section \ref{conclusion}, some concluding remarks are provided.

\section{Kinematics and Formalism}
\label{kinematics}
With the spin of the proton perpendicular to the transverse plane, the Sivers function is expected to reflect an anisotropy of quark momentum distributions for the up and down quarks, indicating that their motion is in opposite directions \cite{Sivers_1990,Sivers_1991}. This is manifestly due to quark orbital angular momentum (OAM). The most interesting and relevant aspects of the OAM, such as magnitude and partonic distribution shape as a function of the proton's state, cannot be determined by the Sivers effect alone. However, systematic studies can be performed to investigate the full 3D momentum distribution of the quarks in a transversely polarized proton, which can be used in concert with other information to exploit multi-dimensional partonic degrees of freedom using a variety of hard processes. Here, we focus specifically on SIDIS and DY, but it should be noted that there is significant potential for broader model development that can come from combining all available data from multiple processes with additional constraints using the simultaneous DNN fitting approach presented here.

The Sivers function describes a difference in probabilities, which implies that it may not be positive definite. Making a comparison between the Sivers function from the DY process and the SIDIS process is still the focus of much experimental and theoretical effort. Under time reversal, the future-pointing Wilson lines are replaced by past-pointing Wilson lines that are appropriate for factorization in the DY process. This implies that the Sivers function is not uniquely defined and cannot exhibit process universality, as it depends on the contour of the Wilson line. This feature of the Sivers function is directly tied to the QCD interactions between the quarks (or gluons) active in the process, resulting in a conditional universality, as shown in \cite{JC2002},
\begin{equation}
\left.\Delta^N f_{q / N^{\uparrow}}\left(x, k_{\perp}\right)\right|_{\text{SIDIS}}=-\left.\Delta^N f_{q / N^{\uparrow}}\left(x, k_{\perp}\right)\right|_{\text{DY}}.
\label{eq:conditional_universality}
\end{equation}
This fundamental prediction still needs to be tested. Direct sign tests \cite{Kang2009, Kang2010, Anselmino2017} can be performed, but experimental proof would require an analysis over a broad phase space of both SIDIS and DY, with consideration given to flavor and kinematic sensitivity for both valence and sea quarks. Our analysis will, in part, rely on this relationship rather than making direct tests of the validity of the sign change.
\subsection{\label{sec1-1}SIDIS process}
\begin{figure}[h]
    \centering
    \includegraphics[width=87mm]{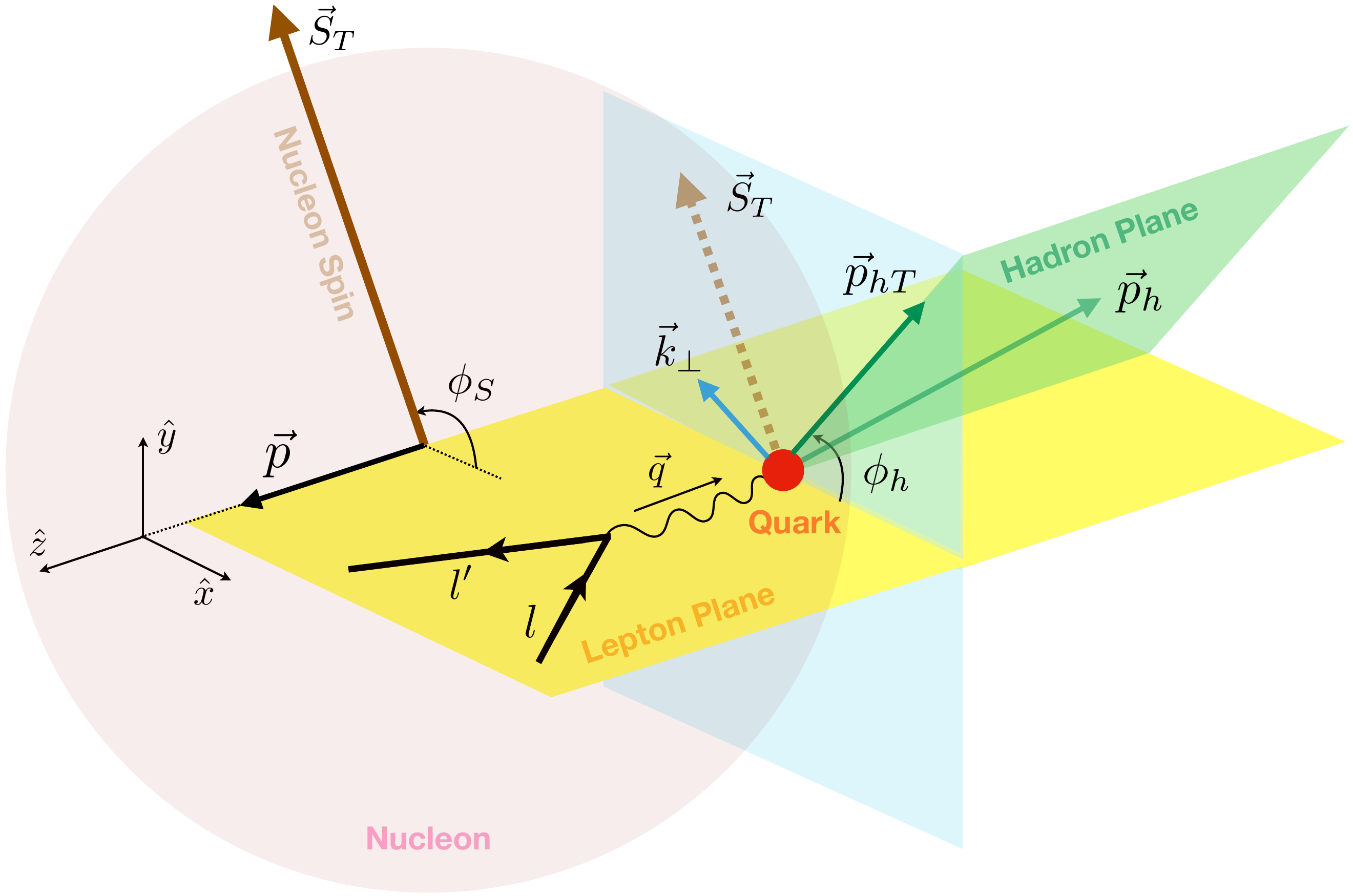}
    \caption{Kinematics of the SIDIS process in the nucleon-photon center-of-mass frame.    }
    \label{fig:sidis}
\end{figure}

The Semi Inclusive Deep Inelastic Scattering (SIDIS) process involves scattering a lepton off of a polarized nucleon and measuring the scattered lepton and a fragmented hadron. In the nucleon-photon center of mass frame, the nucleon three-momentum $\vec{p}$ is along the $z$-axis and its spin-polarization $\vec{S}_T$ is on the plane perpendicular (transverse) to the $\hat{z}$-axis. In Fig. \ref{fig:sidis} the struct-quark, virtual-photon (with four-momentum $\vec{q}$), and the lepton belong to a plane called ``Lepton Plane" (represented in yellow). The fragmented-hadron with momentum $\vec{p}_h$ and its projection onto the $\hat{x}-\hat{y}$ (i.e. $\vec{p}_{hT}$) belong to so-called ``Hadron Plane" (represented in transparent-green). Thus, the transverse momentum $\vec{k}_{\perp}$ of the struck quark and $\vec{p}_{hT}$ are falling onto the transverse plane (represented in transparent-blue) perpendicular to both lepton plane and hadron plane. The azimuthal angle $\phi_h$ of the produced hadron $h$ is the angle between the lepton plane and the hadron plane \cite{bachetta2004}. 

To perform a global analysis using a generalized DNN approach, it is not necessary to postulate an expression or function form for the shape of the partonic distributions or any aspects of the nonperturbative contribution which encodes information about the intrinsic quark-gluon correlations within the nucleon. But to provide a more careful demonstration we do not generalize and focus directly on $\mathcal{N}_q(x)$ to explore using quality metrics in combination with techniques to improve the overall extraction.  It is useful to demonstrate this using a very standard approach that is already well-developed in the literature. With this in mind the
TMDs (FFs) $x$ and $k_\perp$ ($z$ and $p_{\perp}$) dependence are decoupled. 
It is common to apply a Gaussian parametrization for the transverse
momentum dependence but this imposes a bias implying that the transverse momentum is nonperturbative and primarily driven by the intrinsic properties of the colliding hadrons rather than
hard gluon radiation.  Such biases can be managed by the technique itself which we expect to come clear through the examples to come.
There are several options for the choice of formalism and there is still much ongoing debate on how formalism should correctly be implemented for single-spin asymmetries to best respect factorization theory. Our choice of formalism is purely illustrative.  There remains much ongoing debate about this and we intend no favoritism in one strategy over another.  The focus of the present work is the demonstration of how the methods and tools can be implemented for arbitrary formalism.  There are however some very useful characteristics in one of the traditional descriptions of the TSSA \cite{Anselmino_DY_2003} that offer a clear means of detailing the necessary steps.  However, we stress that the type of parameterization used in \cite{Anselmino2017,Anselmino_2015}  does not have the complete features of TMD evolution but is comparable to full TMD evolution at next-to-leading logarithmic (NLL) accuracy.

The differential cross-section for the SIDIS process depends on both collinear parton distribution functions (PDFs) $f_{q/N}(x;Q^2)$ and fragmentation functions $D_{h/q}(z;Q^2)$, where $q$ is the quark flavor, $N$ represents the target nucleon, $h$ is the hadron type produced by the process, and $z$ is the momentum fraction of the final state hadron with respect to the virtual photon. A simplified version of the SIDIS differential cross-section can be written up to $\mathcal{O} (k_{\perp}/Q)$ as \cite{Anselmino_2005_April,Anselmino2009},
  
\begin{align}
    \frac{d^5 \sigma^{lN \rightarrow lh X}}{dxdQ^2dzd^2p_{\perp}} = \sum_q e_q^2 \int d^2 {\bf k_\perp} \; \left(  \frac{2\pi \alpha^2}{x^2 s^2} \frac{\hat{s}^2+\hat{u}^2}{Q^4} \right) \nonumber \\
    \times \hat{f}_{q/N^{\uparrow}} (x, k_\perp) D_{h/q} (z, p_\perp) + \mathcal{O} (k_{\perp}/Q) \;,
\end{align}  
where $\hat{s}, \hat{u}$ are partonic Mandelstam invariants, and $\hat{f}_{q/p^{\uparrow}} (x, k_\perp)$ is the unpolarized quark distribution, 
\begin{align}
    \label{Siv_dist}
    \hat{f}_{q/N^{\uparrow}} (x, k_\perp) &= f_{q/N} (x,k_\perp) + \frac{1}{2}\Delta^N f_{q/N^{\uparrow}} (x,k_\perp) \vec{S}_T\cdot(\hat{p} \times \hat{k}_{\perp}) \nonumber \\
    &= f_{q/N} (x,k_\perp) - \frac{k_\perp}{m_p} f_{1T}^{\perp q } (x,k_\perp) \vec{S}_T\cdot(\hat{p} \times \hat{k}_{\perp})
\end{align}
with transverse momentum $k_{\perp}$ inside a transversely polarized (with spin $\vec{S}_T$) proton with three-momentum $\vec{p}$, where $\Delta^N f_{q/p^{\uparrow}} (x,k_\perp)$ denotes Sivers functions that carry the nucleon's spin-polarization effects on the quarks which can be considered as a modulation to the unpolarized quark PDFs \cite{Anselmino2017},
\begin{align}
    \Delta^N f_{q/N^{\uparrow}} (x,k_\perp) &= 2 \mathcal{N}_q(x)h(k_{\perp})f_{q/N}(x,k_{\perp})
    \label{Siv_func}
\end{align}
where, 
\begin{align}
    f_{q/N}(x,k_{\perp}) &= f_q(x) \frac{1}{\pi \langle k_\perp^2 \rangle} e^{-k_\perp^2/\langle k_\perp^2 \rangle},\\
        h(k_{\perp}) &= \sqrt{2e}\frac{k_{\perp}}{m_1}e^{-k_{\perp}^2/m_1^2}.
\end{align}
Here $\mathcal{N}_q(x)$ is considered a factorized $x$-dependent function with a form that has yet to be formally established.  In fact the form of $f_{q/N}(x,k_{\perp})$ and $h(k_{\perp})$ are also assumed and these expressions should be considered embedded biases. For the best quality model of the Sivers function, the form of $f_{q/N}(x,k_{\perp})$ could be determined and parameterized separately using unpolarized data however, it is instructive to preserve these original forms for this exercise.  The parameter $m_1$ allows the $k_\perp$ Gaussian dependence of the Sivers function to be different from that of the unpolarized TMDs \cite{Anselmino2017}. $f_{q}(x;Q^2)$ is the collinear PDF for flavor $q$ that is obtained from CTEQ6l \cite{CTEQ6l_2003} grid through LHAPDF \cite{LHAPDF6} initially to be consistent with \cite{Anselmino2017} during method testing and later improved using NNPDF4.0 \cite{NNPDF40_2022} for the real extraction, whereas the fragmentation functions for $\pi^{\pm,0}$ are from \cite{DSS14}, and for $K^{\pm}$ are from \cite{DSS17} (DSS \textit{formalism}), from recent global analyses of fragmentation functions at next-to-leading-order (NLO) accuracy in QCD.

In terms of the cross-section ratios, the Sivers asymmetry in the SIDIS process can be written as,
\begin{align}
   A_{UT}^{\sin(\phi_h - \phi_S)}(x,y,z,p_{hT}) &= \frac{d\sigma^{l \uparrow N \rightarrow hl X} - d\sigma^{l\;\downarrow N \rightarrow l h X}}{d\sigma^{l\; \uparrow N \rightarrow hl X} + d\sigma^{l\; \downarrow N \rightarrow hl X}},
\end{align}
and can be parameterized \cite{Anselmino2017} and further re-arranged as,
\begin{align}
\label{eq:SivSIDIS}
    &A_{UT}^{\sin(\phi_h - \phi_S)}(x,z,p_{hT}) \nonumber \\
    &= \mathcal{A}_0(z,p_{hT},m_1) \left(\frac{\sum_q \mathcal{N}_q(x) e_q^2 f_{q}(x)D_{h/q}(z) }{\sum_q e_q^2 f_{q}(x)D_{h/q}(z)} \right),
\end{align}
where, 
\begin{align}
\label{eq:A0}
\mathcal{A}_0 & (z,p_{hT},m_1) \nonumber \\
&= \frac{\sqrt{2e}zp_{hT}}{m_1} \frac{[z^2 \langle k_\perp^2 \rangle + \langle p_\perp^2 \rangle  ]\langle k_S^2 \rangle^2}{[z^2 \langle k_S^2 \rangle + \langle p_\perp^2 \rangle ]^2\langle k_\perp^2 \rangle} \nonumber \\
& \times \exp \left[- \frac{p_{hT}^2z^2 \left(\langle k_S^2 \rangle - \langle k_\perp^2 \rangle \right)}{\left(z^2 \langle k_S^2 \rangle + \langle p_\perp^2 \rangle \right)\left(z^2 \langle k_\perp^2 \rangle + \langle p_\perp^2 \rangle  \right)} \right],
\end{align}

\begin{align}
     \langle k_S^2 \rangle &= \frac{m_1\langle k_\perp^2 \rangle}{m_1^2 + \langle k_\perp^2 \rangle},
\end{align}
and fragmentation functions $D_{h/q}(z,p_\perp)$ (before $p_\perp$-integration),
\begin{align}
    D_{h/q}(z,p_\perp) &= D_{h/q}(z)\frac{1}{\pi\langle p_\perp^2 \rangle}\exp^{-p_\perp^2 /\langle p_\perp^2 \rangle},
\end{align}
with $\langle k_\perp^2 \rangle = 0.57\pm 0.08$ GeV$^2$ and $\langle p_\perp^2 \rangle = 0.12 \pm 0.01$ GeV$^2$ from the fits \cite{Signori_2013,Anselmino_2014} to HERMES multiplicities \cite{HERMES_2013}. Note that we use the shorthand notation for the PDFs, FFs as well as TMDs by omitting $Q^2$ in the expressions for the sake of convenience as is done in the literature.

Through this azimuthal asymmetry, the SIDIS process provides information about the correlations between the transverse momentum of the partons leaving through the fragmented target and the spin of the target itself. In this regard, SIDIS allows one to study the structure of individual hadrons by selecting these decay fragments at the detection level. In general, SIDIS provides access to a wide range of TMDs, and allows for studying TMDs of hadrons carrying different flavors and polarizations.

For our present analysis, HERMES and COMPASS have the best-polarized proton target data for SIDIS, while COMPASS has the best-polarized neutron target data.  In the COMPASS data, the neutron target is actually a polarized deuteron but the neutron carries over 90\% of the deuteron polarization when polarized in solid-state form.  The JLab data on polarized $^3$He is of a different class of experiments and will not be combined with the polarized deuteron data from COMPASS.  It is worth noting that the uncertainties in the experimental data can greatly differ depending on the choice of polarized target.

\begin{figure}[h]
    \centering
    \includegraphics[width=88mm]{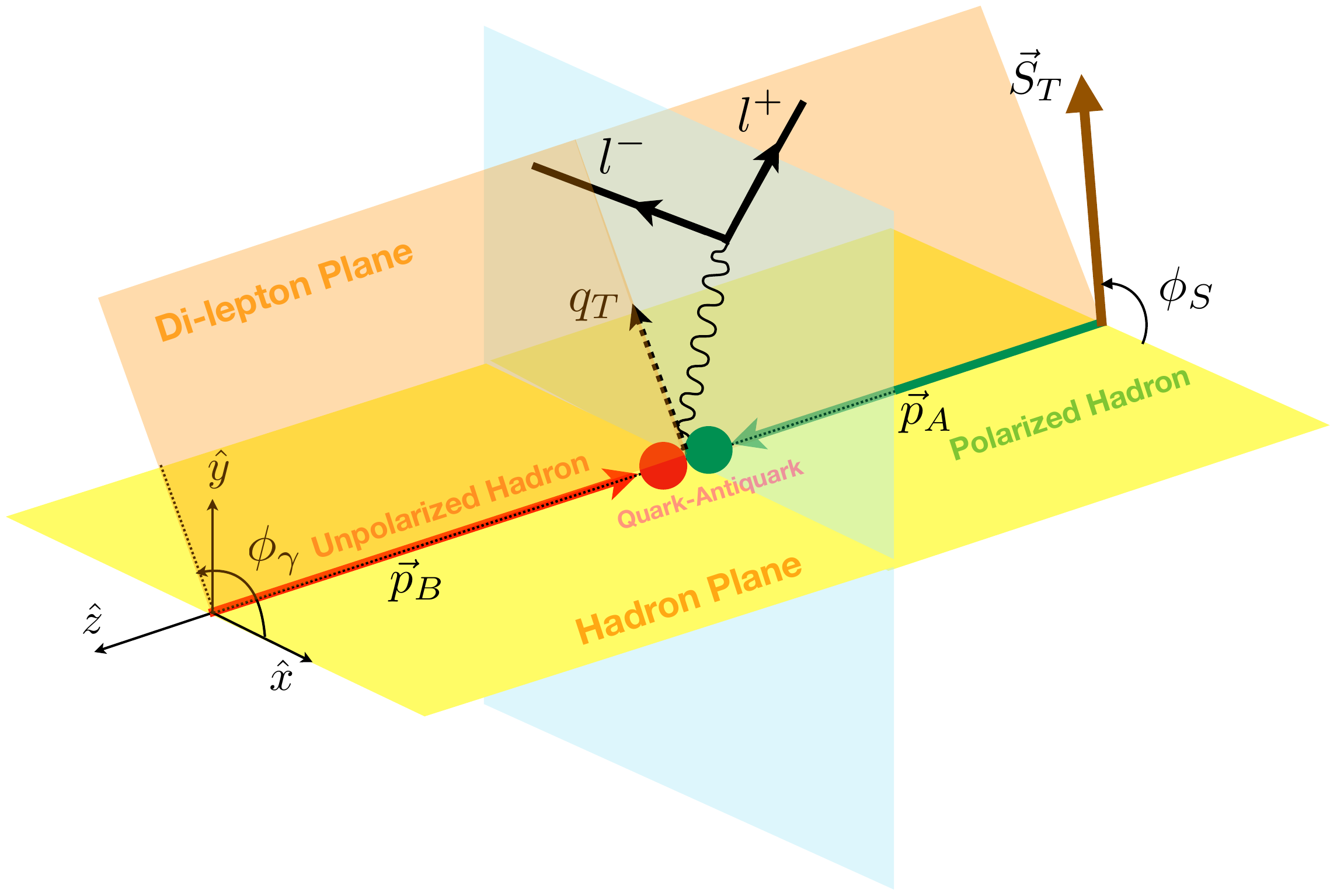}
    \caption{Kinematics of the DY process in the hadronic center-of-mass frame.}
    \label{fig:my_label}
\end{figure}

\subsection{\label{sec1-2} DY process}
Consider the Drell-Yan process $A^{\uparrow}B \rightarrow l^+l^- X$, where $A^{\uparrow}$ is a transversely polarized target, and $B$ is the hadron beam. In the hadronic c.m frame, the 4-momentum $q$ and the invariant mass squared ($Q_{M}$) of the final-state di-lepton pair, Feynman-$x$ ($x_F$) and the Mandelstam variable $s$ are related as,

\begin{align}
    q =  \left(q_0, q_T, q_L \right), \qquad q^2 &= Q_{M}, \qquad x_F = \frac{2q_L}{\sqrt{s}}, \nonumber \\ s&=\left(p_A + p_B \right)^2.
\end{align}

In the kinematical region of,

\begin{align}
    q_T^2 \ll Q_{M}, \quad \quad k_{\perp} \simeq q_T,
\end{align}

at order $\mathcal{O}{\left(k_{\perp}/Q_{M}\right)}$, and in the hadronic c.m frame, the Sivers Single Spin Asymmetry can be given as \cite{Anselmino_DY_2003,Anselmino_DY_2009},

\begin{widetext}
\begin{align}
     A_{N}^{\sin(\phi_{\gamma} - \phi_S)}&(x_F,Q_{M},q_T) = \frac{\int_0^{2\pi} d\phi_\gamma\left(d\sigma^{A^{\uparrow}B \rightarrow l^+l^- X} - d\sigma^{A^{\downarrow}B \rightarrow l^+ l^- X}\right)\sin(\phi_\gamma - \phi_S)}{\frac{1}{2}\int_0^{2\pi} d\phi_\gamma\left(d\sigma^{A^{\uparrow}B \rightarrow l^+l^- X} + d\sigma^{A^{\downarrow}B \rightarrow l^+ l^- X}\right)}, \\
   &= \frac{\int_0^{2\pi} d\phi_\gamma\left( \sum_q \int d^2k_{\perp2} d^2k_{\perp1} \delta^2(k_{\perp1}+k_{\perp1}-q_T) \Delta^N f_{q/A^{\uparrow}} (x_1,k_{\perp1})f_{\bar{q}/B} (x_2,k_{\perp2})\hat{\sigma}_0^{q\bar{q}}\right)\sin(\phi_\gamma - \phi_S)}{\int_0^{2\pi} d\phi_\gamma\left(\sum_q \int d^2k_{\perp2} d^2k_{\perp1} \delta^2(k_{\perp1}+k_{\perp1}-q_T) f_{q/A} (x_1,k_{\perp1})f_{\bar{q}/B} (x_2,k_{\perp2})\hat{\sigma}_0^{q\bar{q}}\right)}
   \label{eq:SivDY_Asym}
\end{align}
\end{widetext}

where,
\begin{align}
    \hat{\sigma}_0^{q\bar{q}} &= e_q^2 \frac{4\pi\alpha^2}{9 Q_{M}}, \\
    x_{1,2} &= \frac{\pm x_F + \sqrt{x_F^2 +4Q_{M}/s}}{2}.
\end{align}

Note that here we follow the same convention as in \cite{JC_2006_PRD73-094023,Anselmino_DY_2009,EFREMOV2005233,Vogelsang_2005} for the azimnuthal angle in the $A-B$ center-of-mass frame with the hadron $A^{\uparrow}$ moving along the positive $z$-axis and hadron $B$ along negative $z$-axis. Thus the mixed product $\vec{S}_T\cdot(\hat{p} \times \hat{k}_{\perp i})$ upon integration in $k_{\perp_i}$ (where $i=\{1,2\}$) yields a $\sin(\phi_\gamma - \phi_S) = \cos\phi_\gamma$ (when $\phi_S=\pi/2$) dependence for the Sivers asymmetry, which implies an overall $[-\sin^2(\phi_\gamma - \phi_S)]$ in Eq.(\ref{eq:SivDY_Asym}). For the case in which polarized hadron $A^{\uparrow}$ moves along the $-\hat{z}$ axis (i.e. for the processes $BA^{\uparrow} \rightarrow l^+l^-X$), the corresponding overall factor is $[+\sin^2(\phi_\gamma - \phi_S)]$. The analytical integration of the numerator and denominator of Eq. (\ref{eq:SivDY_Asym}) can be written as,

\begin{align}
   A_{N}^{\sin(\phi_{\gamma} - \phi_S)}&(x_F,Q_{M},q_T) \nonumber \\
   &= \frac{\int d\phi_{\gamma} \; \mathcal{C}(x_F,Q_{M},q_T,\phi_{\gamma})\sin (\phi_{\gamma}-\phi_S)}{\int d\phi_{\gamma}\; \mathcal{D}(x_F,Q_{M},q_T)},
\end{align}

where,

\begin{align}
    \mathcal{C}&(x_F,Q_{M},q_T,\phi_{\gamma}) \nonumber \\
    & \equiv \frac{d^4 \sigma^{\uparrow}}{dx_FdQ_{M}d^2q_T} - \frac{d^4 \sigma^{\downarrow}}{dx_FdQ_{M}d^2q_T}
    \\ &= \frac{4 \pi \alpha^2}{9 s Q_{M}} \left( \frac{q_T}{m_1}\sqrt{2e} \frac{\langle k_S^2 \rangle^2 \exp [-q_T^2/\left( \langle k_{S}^2 \rangle + \langle k_{\perp 2}^2 \rangle \right)]}{\pi \left( \langle k_{S}^2 \rangle + \langle k_{\perp 2}^2 \rangle \right)^2 \langle k_{\perp 2}^2 \rangle }  \right) \nonumber \\
    & \times \sin (\phi_{\gamma} - \phi_{S})\sum_q \frac{e_q^2}{x_1 + x_2} \Delta^N f_{q/A^{\uparrow}}(x_1)f_{{\bar{q}}/B}(x_2),
\end{align} 
and 
\begin{align}
    \mathcal{D}&(x_F,Q_{M},q_T)  \nonumber \\ 
    & \equiv \frac{1}{2} \left [ \frac{d^4 \sigma^{\uparrow}}{dx_FdQ_{M}d^2q_T} +  \frac{d^4 \sigma^{\downarrow}}{dx_FdQ_{M}d^2q_T} \right ] \nonumber \\
    &= \frac{4 \pi \alpha^2}{9 s Q_{M}} \left(\frac{ \exp [-q_T^2/\left( \langle k_{\perp 1}^2 \rangle + \langle k_{\perp 2}^2 \rangle \right)]}{\pi \left( \langle k_{\perp 1}^2 \rangle + \langle k_{\perp 2}^2 \rangle \right) }  \right) \nonumber \\
    & \times \sum_q \frac{e_q^2}{x_1 + x_2} f_{q/A}(x_1)f_{{\bar{q}}/B}(x_2),
\end{align}

and it can be further simplified as,
\begin{align}
   A_{N}&^{\sin(\phi_{\gamma} -  \phi_S)}(x_F,M,q_T) \nonumber \\
 &= \mathcal{B}_0(q_T,m_1)\frac{\sum_q \frac{e_q^2}{x_1 + x_2} \mathcal{N}_q(x_1)f_{q/A}(x_1)f_{{\bar{q}}/B}(x_2)}{\sum_q \frac{e_q^2}{x_1 + x_2} f_{q/A}(x_1)f_{{\bar{q}}/B}(x_2)}
   \label{eq:SivDY}
\end{align}

where, 
\begin{align}
    \label{eq:B0}
    \mathcal{B}_0(q_T,m_1) &= \frac{q_T \sqrt{2e}}{m_1} \frac{Y_1(q_T, k_S,k_{\perp 2})}{Y_2(q_T,k_{\perp 1},k_{\perp 2}) }
\end{align}
and, 
\begin{align}
    Y_1(q_T, k_S,k_{\perp 2}) &= \left(\frac{ \langle k_S^2 \rangle^2}{ \langle k_{\perp 2}^2 \rangle \left( \langle k_S^2 \rangle +  \langle k_{\perp 2}^2 \rangle \right) ^2} \right) \nonumber \\
    & \times \exp{\left( \frac{-q_T^2}{\langle k_S^2 \rangle + \langle k_{\perp 2}^2 \rangle} \right)},
\end{align}

\begin{align}
    Y_2(q_T,k_{\perp 1},k_{\perp 2}) &= \left(\frac{1}{\langle k_{\perp 1}^2 \rangle +  \langle k_{\perp 2}^2 \rangle } \right) \nonumber \\
    & \times \exp{\left( \frac{-q_T^2}{ \langle k_{\perp 1}^2 \rangle + \langle k_{\perp 2}^2 \rangle} \right)},
\end{align}

\begin{align}
    \frac{1}{\langle k_S^2 \rangle} &= \frac{1}{m_1^2} + \frac{1}{\langle k_{\perp 1}^2 \rangle},
\end{align}

with the assumption $\langle k_{\perp 1}^2 \rangle = \langle k_{\perp 2}^2 \rangle = \langle k_{\perp}^2 \rangle = 0.25$ GeV$^2$ as in \cite{Anselmino_DY_2009}.

Through this azimuthal asymmetry, the Drell-Yan process allows one to preferentially probe from the target and beam hadrons to create the quark anti-quark annihilation process of interest resulting in a dimuon pair in the detector. SIDIS only permits the measurement of a convolution of the TMDs function with a fragmentation function, whereas Drell-Yan allows the direct measurement of the TMDs without the complications of fragmentation functions and final state interactions. Coupled with its innate sensitivity to sea quarks, Drell-Yan is a critical process for determining the TMDs of the sea quarks.

\section{Fitting $\mathcal{N}_q(x)$}
\label{fitting}
To obtain accurate three-dimensional tomographic information on quarks and gluons inside the nucleon, it is critical to extract TMDs with minimal model dependence and little to no unknown biases. Fitting with statistical analysis tools such as MINUIT \cite{iminuit_ref} rely on $\chi^2$ minimization or log-likelihood functions to compute the best-fit parameter values and uncertainties, including correlations between parameters. This class of algorithms has well-established statistical methods that have been used for decades in various scientific fields, making them a reliable and trusted tool. In frequentist statistics, the reliability of a $\chi^2$ minimization fitting method can be evaluated through the concept of hypothesis testing. The fitting method minimizes the difference between the observed data and the expected theoretical model, expressed through a $\chi^2$ statistic. The $\chi^2$ statistic follows a known distribution and the probability of obtaining a value as extreme as the observation can be directly calculated. The reliability of the $\chi^2$ minimization fitting method depends greatly on the ability to accurately estimate the theoretical uncertainties and the degree to which the model approximates the observed data. When these conditions are met, the method can provide a reliable estimate of the parameters that describe the model and its uncertainties. However, chi-square fits can be sensitive to the choice of initial parameter values and may not always converge to the correct solution. Fitting with DNNs can provide considerable advantages and does not inherently sacrifice the statistical framework provided by chi-square fits but it's worth touching on some key attributes needed in the method in order to best maintain quality statistical relevance and interpretation of resulting fits.

To preserve the statistical robustness and reliability of traditional $\chi^2$ minimization fitting when using a DNN, it is important to carefully consider the data quality, model selection, validation, interpretation, and testing criteria. The quality of the data used to train the DNN should be as high as possible to ensure the DNN learns the correct relationships between inputs and outputs. This is crucial because the reliability of the DNN is only as good as the quality of the data it is trained on. Quantifying differences between the training data and the real data used in the fit can be challenging and can lead to unknown biases and systematic errors. In the method used here, Monte Carlo data, which has been tuned and matched to the experimental data, is utilized. This is done by successively extracting information from the experimental data to impose into the generated Monte Carlo data and then using the improved Monte Carlo data to further refine the extraction technique.

The choice of DNN architecture, activation functions, regularization techniques, and other hyperparameters should be carefully selected to minimize over-fitting and maximize generalization performance. Cross-validation techniques can be used to tune these hyperparameters and ensure the best possible fit to the data.  The quality of the fit should be quantified with a metric that is well-defined and can be interpreted statistically. This could still be the $\chi^2$ statistic but may also be a variety of possible loss functions. The trained DNN should be validated on an independent test dataset to ensure that it generalizes well to new data and that it does not over-fit the training data. This is critical because over-fitting can lead to an unreliable and unstable model.

When interpreting the results of the DNN fit, it's important to carefully examine the relationships between the inputs and outputs. To better understand how the DNN makes its predictions, techniques such as feature importance and attention mechanisms can be used. While DNNs can have a reliable statistical interpretation, it requires a more detailed analysis than traditional algorithms like MINUIT.  

Directed testing of the model predictions and verification of reliability through multiple trials is crucial.  Studies to test accuracy and precision are useful along with quantifying the robustness of the extraction method itself once a DNN architecture has been chosen.  In this regard, it's important to prove that the method can be flexible as well as correct.  It is not normally useful to have a model and method that yields highly accurate results but only for a fine slice of phase space under particular conditions.

The conventional chi-square minimization routines are limited in their flexibility and applicability to more complex problems because they assume a specific functional form of the relationship between inputs and outputs. In contrast, DNNs can learn complex and nonlinear relationships, making them suitable for tasks that require some degree of abstraction and where there is no known specific functional form of the relationship. DNNs are proven to be universal approximators and can handle large amounts of data while generalizing well to new data, which improves their accuracy and robustness. This is especially helpful in the present application, where DNNs can be used to build better models with new experimental data as it becomes available.

The ambiguity in the literature around $\mathcal{N}_q(x)$ and the accompanying factorized terms in the Sivers function ($h (k_\perp)$ and $f_{q/N}(x,k_\perp;Q^2)$) makes it a good candidate for a DNN extraction. In most Sivers function extractions \cite{Anselmino_2005_April,Vogelsang_2005,Anselmino2005,Collins_2006,Anselmino2009,Anselmino2012,bachetta2011,Sun_2013,Echevarria_2014,Luo_2020,Cammarota_2020,bachetta2022}, $\mathcal{N}_q(x)$ differs either by its parameterization or by the treatment of $q$ in $\mathcal{N}_q(x)$. For example, in \cite{Anselmino2009, bachetta2011, Bury2021_PRL, Bury2021_JHEP}, all anti-quarks ($\bar{u}$, $\bar{d}$, and $\bar{s}$) were treated the same (or combined) and referred to as the `unbroken sea'.

Our first step is a generalization of the MINUIT fit parameterization of $\mathcal{N}_q(x)$ for all light quark flavors.  Section \ref{sec:minuit-fits} summarizes the corresponding MINUIT fit results using $iminuit$ (python version of MINUIT) \cite{iminuit_ref}. In these fits we use the same dataset as in \cite{Anselmino2017}, and obtained the fit parameters for $\mathcal{N}_q(x)$ defined as,
\begin{align}
\label{eq:Nq}
    \mathcal{N}_q(x) &= N_q x^{\alpha_q}(1-x)^{\beta_q}\frac{(\alpha_q + \beta_q)^{(\alpha_q + \beta_q)}}{\alpha_q^{\alpha_q} \beta_q^{\beta_q}}.
\end{align}
This expression is generalized for all light quark flavors, where $N_q$ is a scalar for quark flavor $q$. After our consistency check, this parameterization is used as a pseudodata generator to train and test the DNN model. We emphasize that our original MINUIT fit parameterization is used as a tool to demonstrate that the DNN model is capable of predicting (or confirming) the 19-parameter model used to generate pseudodata, as illustrated in Section \ref{sec:DNN-pseudo-fits}. It should be clarified that we do not rely on the form of Eq. (\ref{eq:Nq}) in any way and any function that can be used to generate quality\footnote{The term quality here refers to how well it can represent the real experimental data.} pseudodata could be used.  After building confidence in the DNN model from these preliminary tests, we move towards extracting the Sivers functions from experimental data from the SIDIS process with a polarized-proton target (see section \ref{sec:DNN-fits}). Previous work on global fits to SIDIS data considered the data from polarized-proton targets, polarized-deuteron targets, and polarized $^3$He gas targets as a combined dataset. This is usually motivated only by the overall lack of separate polarized target data.  In \cite{Bury2021_JHEP}, isospin symmetry was assumed for $f_{1T,u}^{\perp}$ and $f_{1T,d}^{\perp}$ for COMPASS2009 \cite{COMPAS_Data_Alekseev_2009} and COMPASS2017 \cite{COMPASS_DY_2017} datasets. However, these are very different polarized targets with very different results so we explore if our method will be sensitive to flavor and target dependence by fitting each polarized target data independently. As the nuclear effects on the Sivers functions are not very well understood, separately extracting the same observable with different types of polarized targets that contain different flavor dominance and different nuclear effects could provide valuable insight. Significant data would be required for each target type so there are clear limitations at the moment and our fits can only be considered preliminary in this regard.

The DNN's unique capacity to manage abstraction allows for the capture of additional complexity in a semi-model-independent way.  To obtain the most information from the data and fit results we attempt to decompose some of this abstraction using the following conditions,
\begin{itemize}
    \item DNN fits to SIDIS data from proton and deuteron targets are performed independently to obtain separate models. 
    \item No kinematic cuts are applied to take full advantage of the available data and to allow the DNN to build implicit inclusion of the necessary corrections related to TMD factorization.
    \item A Sivers function for each light-quark flavor is obtained to ensure the $SU(3)_{\text{flavor}}$ breaking effects in QCD are also contained.
\end{itemize}
The technique to achieve the extraction under the aforementioned conditions is somewhat novel, so explicit details are provided step by step for clarity in the next section.

\begin{figure}[h]
    \centering
    \includegraphics[width=75mm]{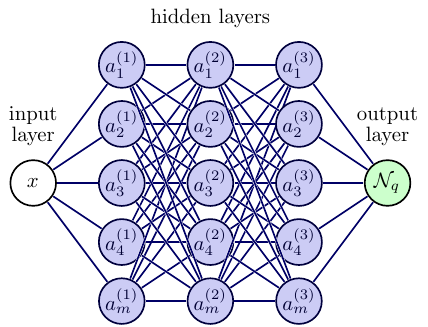}
    \caption{A generic representation of the DNN architecture for $\mathcal{N}_q(x)$, where $q=\{u,d,s,\bar{u},\bar{d},\bar{s}\}$, and $a^{(n)}_{m}$ represent the node $m$ in the hidden layer $n$. The figure represents only up to $n=3$ for demonstration purposes.}
    \label{fig:NN_Architecture}
\end{figure}

\section{The Extraction Technique}
\label{extraction}
The factorized form, $\mathcal{N}_q(x)h(k_{\perp})f_{q/N}(x,k_{\perp})$ assumes the validity of Eq. (\ref{Siv_func}), leading to model dependencies associated with this assertion and the Gaussian interpretation for the $k_\perp$ distributions of the partons. This introduces a bias to the analysis, but it is a known bias and can be managed.  Within the construction of a DNN model in the fit function,
\begin{align}
\label{eq:Nq1}
    \mathcal{N}^{\text{DNN}}_q(x)h(k_\perp)f_{q/N}(x,k_\perp),
\end{align}
there could be considerable variation in definitions (or assumptions) of $h(k_\perp)$ or $f_{q/N}(x,k_\perp)$ still leading to a quality fit of $\mathcal{N}^{\text{DNN}}_q(x)$ to the data due to its considerable flexibility and high number of parameters with respect to either of the other two terms in the function. This makes the final $\mathcal{N}^{\text{DNN}}_q(x)$ highly dependent on the choice of $h(k_\perp)$ and $f_{q/N}(x,k_\perp)$ so the resulting $\mathcal{N}^{\text{DNN}}_q(x)$ model must always be used with the same definitions of $h(k_\perp)$ and $f_{q/N}(x,k_\perp)$ it was trained with.  However, no particular definition of $h(k_\perp)$ or $f_{q/N}(x,k_\perp)$ is required to make the combination of $\mathcal{N}^{\text{DNN}}_q(x)h(k_\perp)f_{q/N}(x,k_\perp)$ non-biased and perform well in the fit. From this standpoint, we do not consider this to be a fully model-independent extraction, but every attempt is made to make it minimally model-dependent while still preserving the original assumptions relevant to the phenomenology for transparency of the schema. The DNN treatment of $\mathcal{N}_q(x)$ enables the flexibility of handling all the light quark-flavors $q=\{u,\bar{u},d,\bar{d},s,\bar{s}\}$ independently. The method presented is intended to be somewhat analogous to the approach of treating $\mathcal{N}_q(x)$ as a fit function to be parameterized for the sake of illustrating the advantages. It is natural to extend this study towards a symbolic regression of $\mathcal{N}_q(x)$ or the full expression of the Sivers function but that is not the scope of the present work.

\begin{figure}[h]
    \centering
    \includegraphics[width=75mm]{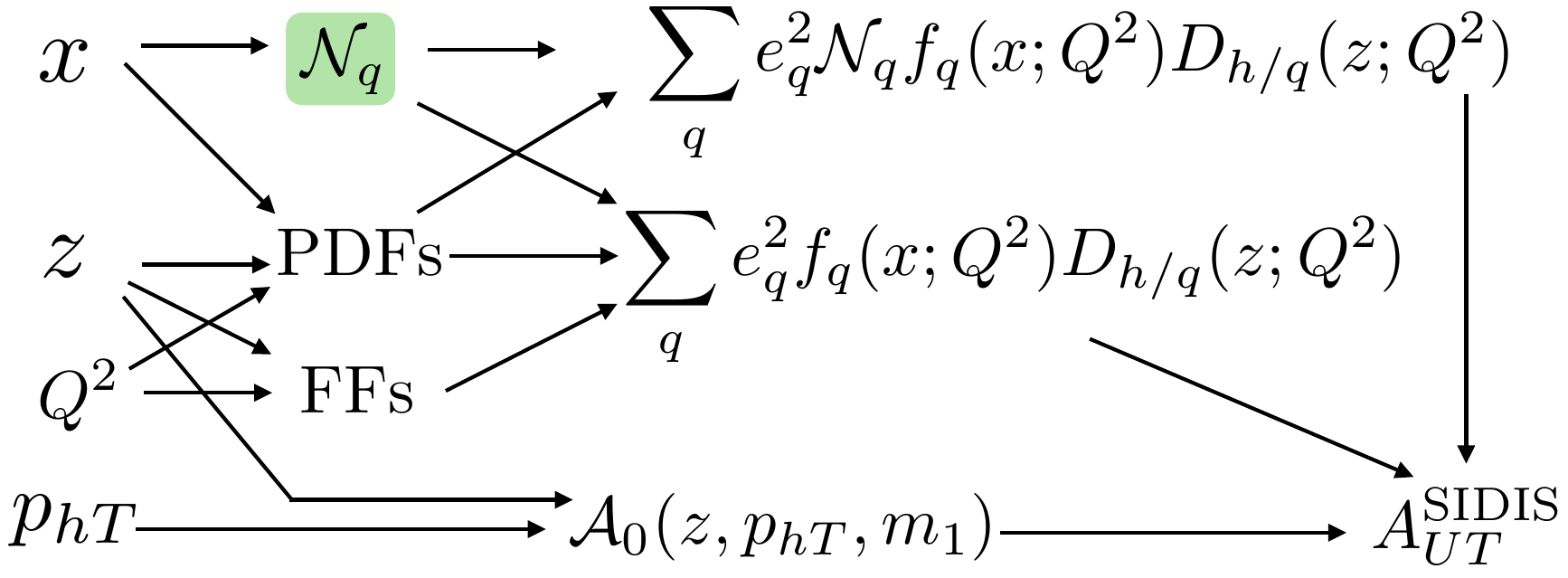}
    \caption{The block-diagram for SIDIS Sivers Asymmetry (see Eq.(\ref{eq:SivSIDIS}) ): $\mathcal{N}_q$ denote the DNN models for each quark-flavor $q=\{u,d,s,\bar{u},\bar{d},\bar{s}\}$, and $h=\{\pi^+,\pi^-,\pi^0,K^+,K^-\}$.}
    \label{fig:Block-Diagram-SIDIS}
\end{figure}

\begin{figure}[h!]
    \centering
    \includegraphics[width=78mm]{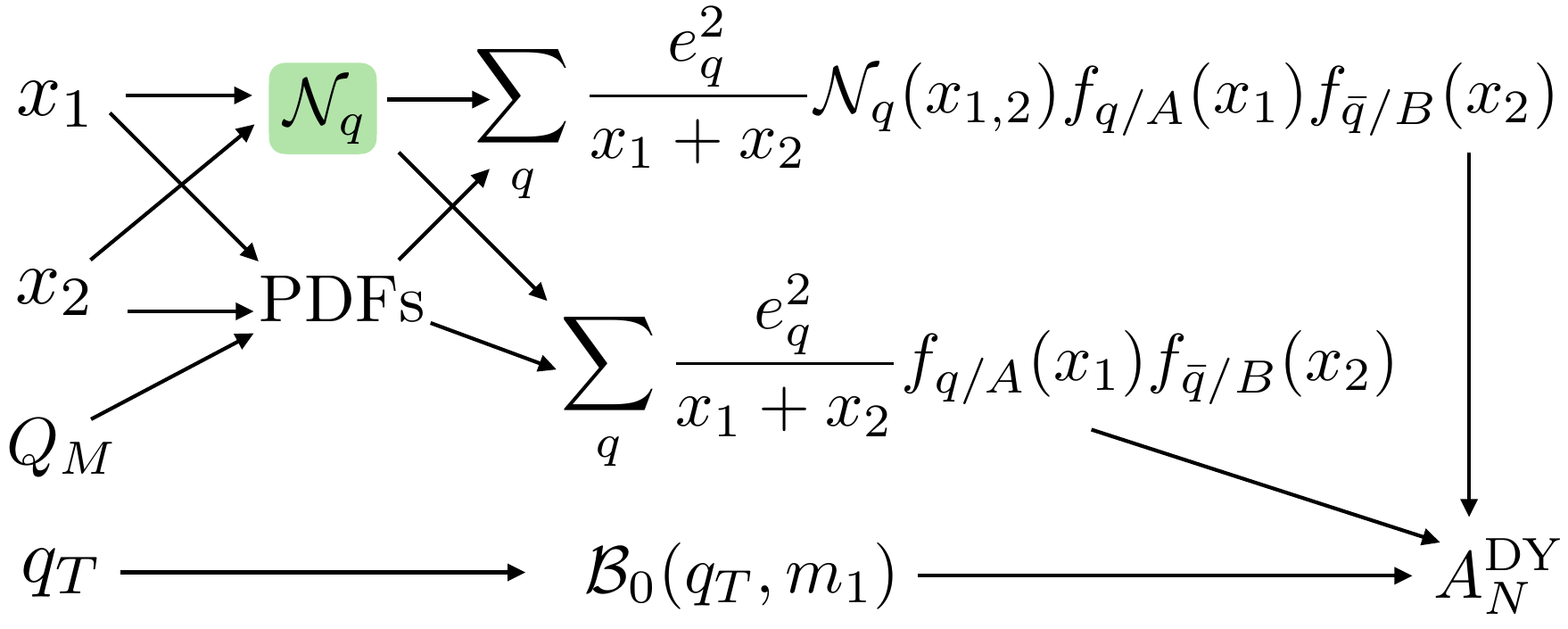}
    \caption{The block-diagram for DY Sivers asymmetries (see Eq. (\ref{eq:SivDY})). $\mathcal{N}_q$ denotes the DNN models for each quark-flavor $q=\{u,d,s,\bar{u},\bar{d},\bar{s}\}$ obtained by training to SIDIS Sivers asymmetries.}
    \label{fig:Block-Diagram-DY}
\end{figure}

We use a relationship for $\mathcal{N}_q(x)$, similar to the seminal work \cite{Anselmino2017}, as a tool to generate our pseudodata for testing accuracy and reproducibility only.  This definition is not used in any aspect of the final DNN fit results. The generic feedforward DNN structure for $\mathcal{N}_q(x)$ that we use in this work is represented in Fig. \ref{fig:NN_Architecture}. As we consider the $SU(3)_{\text{flavor}}$ symmetry breaking in QCD, we have $\mathcal{N}_u(x)$, $\mathcal{N}_{\bar{u}}(x)$, $\mathcal{N}_d(x)$, $\mathcal{N}_{\bar{d}}(x)$, $\mathcal{N}_s(x)$, and $\mathcal{N}_{\bar{s}}(x)$ to handle the six light-quark-flavors independently. Bjorken-$x$ is the only input as the initial layer of each $\mathcal{N}_q(x)$, and the final layer is a single-node output. The $m_1$ in $\mathcal{A}_0(z,p{hT},m_1)$ as defined in Eq. (\ref{eq:A0}) is treated as a free parameter, with the initialization obtained from our first chi-square minimization fit (discussed in Section \ref{sec:minuit-fits}), and then allowed to vary throughout the DNN training process with SIDIS data, as shown in Fig. \ref{fig:Block-Diagram-SIDIS}. The DNN model results are then used to infer the projections for both SIDIS kinematics and DY kinematics (see Fig. \ref{fig:Block-Diagram-DY}).

There are many open-source software libraries for machine learning and artificial intelligence.  For this particular extraction, we use TensorFlow \cite{tensorflow2015-whitepaper}.
A deep feedforward architecture is used with the hidden layers expanded with multiple numbers of nodes which are initialized randomly with Gaussian sampling of weights around $zero$ with a standard deviation of $0.1$.  The degree of potential Non-linearity is introduced into the network by the choice of the activation function. The selection of the activation function can have a substantial impact on DNN performance and training dynamics. We chose the $Relu6$ activation function. This activation is a variant of the Rectified Linear Unit ($ReLU$) function. The $ReLU6$ activation function has been shown to empirically perform better under low-precision conditions by encouraging the model to learn sparse features earlier, which is beneficial for learning complex patterns and relationships from the experimental data.  We also use \textit{Least Absolute Shrinkage and Selection Operator Regression} which is a regularization technique used to prevent overfitting and improve the model's performance and generalization ability, while also encouraging sparsity and feature selection.  We also use L1 regularization. L1 regularization encourages sparsity in the activation by adding a penalty term to the loss function that is proportional to the absolute value of the weights \cite{Geron_2019}. By adding this regularization term, the most important inputs are weighted the most, so that noisy or redundant information is discarded. The strength of the regularization is controlled by the magnitude of the regularization coefficient, which is set to $10^{-12}$. Additionally, we use a dynamically decreasing learning rate.  The learning rate is automatically reduced by $10\%$ if the {\it{training loss}} has not decreased within the last $200$ epochs\footnote{An epoch is a complete cycle of the passing of training data through the algorithm.} (i.e. {\it{patience}} = $200$).  The optimizer used was \textit{Adam} while the loss function used was \textit{Mean Squared Error}.  During the hyperparameter optimization process, there were slight deviations in the number of layers, nodes per layer, initial learning rate, batch size, and the number of epochs, but the basics of the scheme just described remain consistent for all DNNs used.

Data shuffling is used to randomly organize the data into batches that pass through the training process.  When training, it is crucial to expose the DNN to a diverse range of training samples. Shuffling the training data ensures that the DNN doesn't learn patterns based on the sequential order of data. Instead, it promotes a more randomized and representative distribution of the data, preventing the model from becoming biased toward any specific subset or order. This randomization helps the model generalize better and makes it more robust to variations in the input data, ultimately improving its overall performance and accuracy.  Batch size is a critical component of successful training as well.  A batch size too large for a small data set, even well-shuffled, can also lead to a biased model. In what follows, we train two dedicated DNN models: one for the SIDIS data with a polarized proton target and one for the data with a polarized deuteron (neutron) target. Since the number of available data points for the deuteron target is significantly smaller than the amount of data for the proton target, care is taken to find the optimum configuration of shuffling and batch size for the case of deuteron whereas for the proton the shuffling and batch size's impact is nearly negligible.

Our strategy is to first perform an exercise using only pseudodata to verify the extraction method that will ultimately be used on the real experimental data. First, we devise a \textit{generating function} for the SIDIS Sivers asymmetry data using a conventional $\chi^2$-minimization routine (MINUIT in this case), without following the popular assumption of ``unbroken sea" \cite{Anselmino2017} in order to generalize the treatment of quarks and antiquarks. We perform a series of conventional MINUIT fits step-wise to obtain the final 19 parameters for the case of broken $SU(3)_{\text{flavor}}$ symmetry in QCD. Then, we produce pseudodata (or replicas) for the SIDIS asymmetry by sampling from the mock experimental errors using the \textit{generating function} with kinematics and binning in $x$, $z$ and $p_{hT}$ as in the experimental data. Then a DNN model is constructed with all hyperparameters tuned in order to achieve the highest possible accuracy and precision.  Here our nomenclature becomes quite specific and we refer to the resulting distribution of DNN fits as a \textit{DNN model}.  The first model obtained with the method for a particular set of data is referred to as the \textbf{First Iteration}. At this stage, we use the distribution of fit results to obtain the mean and the error band from the initial DNN model to re-parameterize the \textit{generating function} so that it produces more realistic pseudodata.  The DNN fits are performed again improving the quality (both accuracy and precision) of the resulting fits to result in a \textbf{Second Iteration} DNN model. One can repeat the number of iterations until the resulting model is no longer improving within the experimental uncertainties. In this way, the DNN model approaches the best approximation of the Sivers functions in comparison to the $\textit{true}$ values put into the \textit{generating function}.

After confirming that the method works well, an extraction of the Sivers function using the SIDIS experimental data is performed. We treat the data for a polarized proton-target and deuteron-target separately for two reasons. First, fitting these together would introduce bias that would need to be managed directly.  This is the case even assuming isospin symmetry in the $u$ and $d$-quarks' Sivers functions. Second, our approach leaves open the possibility to explore the nuclear dependence of the Sivers functions. The construction of the DNN models for proton and deuteron is analogous. To perform the fit another \textbf{First Iteration}, as previously described, is performed by developing and tuning a DNN model using the data from the real experiment rather than pseudodata from the generating function. In the proceeding iterations, we use the generating function in order to tune the hyperparameters to achieve the highest possible quality of fit in comparison with the results from the \textbf{First Iteration}. Once a tuned model is obtained, we perform an extended study for evaluating the algorithmic uncertainty\footnote{Algorithmic uncertainty is the degree of increase in the distribution of the resulting fits that is not directly from propagated experimental error.} as well as the systematic uncertainty of the DNN extraction method.

To elaborate on the pedagogy of this method, we organize the remainder of this section into the following subsections: (\ref{sec:data-selection}) Data selection, (\ref{sec:minuit-fits}) MINUIT fits for the case of $SU(3)_{\text{flavor}}$, (\ref{sec:DNN-pseudo-fits}) DNN model training with pseudodata, (\ref{sec:DNN-fits}) DNN model training with real experimental data. 

\subsection{Data selection}
\label{sec:data-selection}
Generally, TMD factorization is considered valid only under the kinematics restriction of $\Lambda \ll Q$, and $p_{hT} \ll zQ$.  The first of these requirements is important because it is only in this limit one can apply massless-hadron kinematics resulting in near-negligible contamination from power corrections work where $Q > 2$GeV is common.  The second of these requirements provides the kinematic range a traditional factorization scheme will work.  Additionally, the transverse mass scale should generally be preserved around $(m/Q/z)^2$.  We do not strictly adhere to any of these guidelines but instead, study the consequence of using progressively less restrictive cuts.  For global data selection, we focus our attention on the fixed target SIDIS and DY data.  For the proton DNN fits, kinematically 3D binned data from HERMES2020 \cite{HERMES_Data_Airapetian_2020} is left out for validation and only used 1D binned HERMES2020 data for the training.  For the neutron DNN fits, the polarized $^3$He data from Jefferson Lab \cite{JLab2011,JLab2014} is left out and used to test the new projections of the DNN model trained on the deuteron COMPASS data only.

Table \ref{tab:data-sets} summarizes the kinematic coverage, the number of data points, and reaction types of the datasets that are considered in this work. In addition to the SIDIS datasets that are used in the fits, the polarized DY dataset from the COMPASS experiment is also listed in Table \ref{tab:data-sets} as we demonstrate the predictive capability of the DNN model by comparing the projections with the real data points. The DY projections are made using the trained SIDIS DNN model assuming a sign change expected from \textit{conditional} universality. 
For the case of training the DNN model related to proton-target we use HERMES2009 \cite{HERMES_Data_Airapetian_2009}, COMPASS2015 \cite{COMPAS_Data_Adolph_2015} and HERMES2020~  \cite{HERMES_Data_Airapetian_2020} data points associated with 1D kinematic binning, leaving the HEREMES2020\cite{HERMES_Data_Airapetian_2020} data associated with the 3D kinematic binning to compare with the projections from the trained model. The COMPASS2009 \cite{COMPAS_Data_Alekseev_2009} dataset with a polarized-deuteron target is used for the neutron Sivers extraction as a separate DNN model.

\begin{table}[h]
    \centering
    \begin{tabular}{cccc}
    Dataset  & Kinematic & Reaction & Data \\
      & coverage &  &  points\\
    \hline
        HERMES2009   & $0.023 < x < 0.4 $   &   $p^{\uparrow}+\gamma^* \rightarrow \pi^+$ & 21\\
          (SIDIS)     &    $0.2 < z < 0.7$    &   $p^{\uparrow}+\gamma^* \rightarrow \pi^-$ & 21\\
        \cite{HERMES_Data_Airapetian_2009}       &   $0.1 < p_{hT}< 0.9$   &   $p^{\uparrow}+\gamma^* \rightarrow \pi^0$ & 21\\
               &   $Q^2>$ 1 GeV$^2$    &   $p^{\uparrow}+\gamma^* \rightarrow K^+$ & 21\\
               &       &   $p^{\uparrow}+\gamma^* \rightarrow K^-$ & 21\\
        \hline
        HERMES2020    & $0.023 < x < 0.6 $   &   $p^{\uparrow}+\gamma^* \rightarrow \pi^+$ & 27, \textbf{64} \\
          (SIDIS)     &    $0.2 < z < 0.7$    &   $p^{\uparrow}+\gamma^* \rightarrow \pi^-$ & 27, \textbf{64}\\
           \cite{HERMES_Data_Airapetian_2020}    &  $0.1 < p_{hT}< 0.9$      &   $p^{\uparrow}+\gamma^* \rightarrow \pi^0$ & 27\\
               &     $Q^2>$ 1 GeV$^2$   &   $p^{\uparrow}+\gamma^* \rightarrow K^+$ & 27, \textbf{64}\\
               &       &   $p^{\uparrow}+\gamma^* \rightarrow K^-$ & 27, \textbf{64}\\
        \hline
        COMPASS2015   & $0.006 < x < 0.28 $   &   $p^{\uparrow}+\gamma^* \rightarrow \pi^+$ & 26\\
        (SIDIS)       &    $0.2 < z < 0.8$    &   $p^{\uparrow}+\gamma^* \rightarrow \pi^-$ & 26\\
        \cite{COMPAS_Data_Adolph_2015}       &    $0.15 < p_{hT}< 1.5$   &   $p^{\uparrow}+\gamma^* \rightarrow K^+$ & 26\\
               &  $Q^2>$ 1 GeV$^2$     &   $p^{\uparrow}+\gamma^* \rightarrow K^-$ & 26\\
        \hline
        COMPASS2009   & $0.006 < x < 0.28 $   &   $d^{\uparrow}+\gamma^* \rightarrow \pi^+$ & 26\\
        (SIDIS)       &    $0.2 < z < 0.8$    &   $d^{\uparrow}+\gamma^* \rightarrow \pi^-$ & 26\\
        \cite{COMPAS_Data_Alekseev_2009}       &   $0.15 < p_{hT}< 1.5$  &   $d^{\uparrow}+\gamma^* \rightarrow K^+$ & 26\\
               &    $Q^2>$ 1 GeV$^2$     &   $d^{\uparrow}+\gamma^* \rightarrow K^-$ & 26\\
        \hline \hline
        JLAB   & $0.156 < x < 0.396 $   &   $^3He^{\uparrow}+\gamma^* \rightarrow \pi^+$ & 4\\
        2011,2014       &    $0.50 < z < 0.58$    &   $^3He^{\uparrow}+\gamma^* \rightarrow \pi^-$ & 4\\
         (SIDIS)   &    $0.24 < p_{hT}< 0.43$             &        $^3He^{\uparrow}+\gamma^* \rightarrow K^+$        &   4      \\
           \cite{JLab2011,JLab2014}    &     $1.3 < Q^2 < 2.7$       &     $^3He^{\uparrow}+\gamma^* \rightarrow K^-$    &     1    \\            
        \\
        COMPASS2017   & $0.1 < x_N < 0.25 $   &   $p^{\uparrow}+\pi^- \rightarrow l^+l^- X$ & 15\\
        (DY) \cite{COMPASS_DY_2017}      &    $0.3 < x_{\pi} < 0.7$    &  \\ 
         &       $4.3 < Q_M < 8.5$          &                      &              \\
         &       $0.6 < q_T < 1.9$       &                      &              \\
        \hline
    \end{tabular}
    \caption{The SIDIS and DY datasets considered in the fits include the DY data, which is used to demonstrate the predictive capability of the DNN model. Specifically, we make projections using the trained SIDIS DNN model, assuming a sign change to predict the real experimental DY data points. For the HERMES2020 dataset, data is available with both 1D and 3D kinematic bins. The 3D bin numbers are indicated in bold font.}
    \label{tab:data-sets}
\end{table}

For the initial $\chi^2$-minimization fit with MINUIT the same datasets are used as in \cite{Anselmino2017} for consistency which is HERMES2009  \cite{HERMES_Data_Airapetian_2009}, COMPASS2009 \cite{COMPAS_Data_Alekseev_2009}, and COMPASS2015 \cite{COMPAS_Data_Adolph_2015}. This fit is described in the next subsection. The error in Fit 1 of Table \ref{table:SIDIS_MINUIT_Fits} is quite large and many of the parameters are consistent with zero. This highlights the challenges of reproducing this type of analytical fit with a standard $\chi^2$ minimization fit. The last column, Fit 5, is a fit from our generating function to the proton-DNN model which provides the most accurate parameterization of $\mathcal{N}_q$ to date. Again, we point out we do not use this parameterization for any physics extraction but only to generate pseudodata.

\subsection{MINUIT fits for $SU(3)_{\text{flavor}}$}
\label{sec:minuit-fits}

The analysis begins with a $\chi^2$-minimization fit with MINUIT similar to the approach in \cite{Anselmino2017} except we expand the number of parameters to treat each of the light-quark flavors separately.  The results of the MINUIT fits are shown in Table \ref{table:SIDIS_MINUIT_Fits}. Fit 1 is from the original fit results from Anselmino \textit{et al} directly from \cite{Anselmino2017}.  Here the $\mathcal{N}_q(x)$ for the $u$ and $d$ quark used is Eq. (\ref{eq:Nq}) but $\mathcal{N}_{\bar{q}}(x)=N_{\bar{q}}$ for anti-quarks.  In this fit, there are three parameters $\alpha_q$ and $\beta_q$ and $N_q$ for each quark flavor, and for each anti-quark it's just $N_{\bar{q}}$ plus $m_1$. This results in a 9-parameter fit. Fit 2 is a test to reproduce the same parameterization as in Fit 1.  We note that in Fit 2 non of the 9 parameters are fixed or have bounds imposed.  Both of these first columns only consider $u$ and $d$ quarks and antiquarks. The Fit 1 parameters were used as the initial values to perform Fit 2.  The difference in these two sets of fit parameters demonstrates the challenge of systematic consistency with this method though some parameters match reasonably well. For fit 3 we use the same convention but add in the strange quark so there is an additional four parameters $N_s$, $\alpha_s$, $\beta_s$, and $N_{\bar{s}}$ which leads to a 13-parameter fit. In order to initialize the 13 parameters in Fit 3, we use the corresponding values for those parameters from Fit 2 and zeros for the rest. Fit 4 uses Eq. (\ref{eq:Nq}) for both quarks and antiquarks so that the treatment of all three light-quark-flavors is the same.  In addition to the parameters from Fit 3, Fit 4 contains six more parameters for the antiquarks. The result of Fit 4 leads to a larger $N_{\bar{q}}$ value to compensate for the fact that $\alpha_{\bar{s}}$ and $\beta_{\bar{s}}$ are now present in the fit.
However, the motivation behind performing Fit 4 in this way is to generalize the $\mathcal{N}_q(x)$ in a flavor-independent fashion for both quarks and anti-quarks. Fit 4 is the final fit that we will use to generate pseudodata for testing the DNN fits and for calculating the model's accuracy. 
The last column, Fit 5, uses the analytical generating function to fit the proton-DNN model in fine bins.  The error of the model is propagated to each bin and a MINUIT fit is performed providing a highly accurate parameterization of $N_q$ represented in this particular function form. We provide this can as a basis for comparison.

\begin{table}[h!]
    \centering
    \begin{tabular}{c|ccccc}
      Parameter  & Fit 1 & Fit 2 & Fit 3 & Fit 4 & Fit 5\\
      \hline
     $m_1$ & 0.8 & 3.9 & 7.0 & 7 & 3.63 \\
      $\delta m_1$ & $\pm$0.9 & $\pm$0.3  & $\pm$0.6 & $\pm$4 & $\pm$0.03 \\ \hline
     $N_u$  &  0.18 & 0.48 & 0.89 & 0.89 & 1.08\\
     $\delta N_u$ & $\pm$0.04 & $\pm$0.03  & $\pm$0.05 & $\pm$0.06 & $\pm$0.02 \\ \hline
     $\alpha_u$  & 1.0  & 2.41 & 2.78  & 2.75 & 2.85\\
     $\delta \alpha_u$ & $\pm$0.6 & $\pm$0.16  & $\pm$0.17 & $\pm$0.11 & $\pm$0.02 \\ \hline
     $\beta_u$  & 6.6  & 15.0 & 19.4 & 20 & 12.4\\
     $\delta \beta_u$ & $\pm$5.2 & $\pm$1.4  & $\pm$1.6 & $\pm$2 & $\pm$0.1 \\ \hline
     $N_{\bar{u}}$  & -0.01  & -0.032 & -0.07 & -0.12 & 11.8\\
     $\delta N_{\bar{u}}$ & $\pm$0.03 & $\pm$0.017  & $\pm$0.06 & $\pm$0.60 & $\pm$0.3 \\ \hline
     $\alpha_{\bar{u}}$  & -  & - & - & 0.4  & 1.90\\
     $\delta \alpha_{\bar{u}}$ & -  &  -  & - & $\pm$0.5 & $\pm$0.03 \\ \hline
     $\beta_{\bar{u}}$  &  - & - & - & 20  & 1.28\\
     $\delta \beta_{\bar{u}}$ & -  &  -  & -  & $\pm$16 & $\pm$0.07 \\ \hline
     $N_d$  & -0.52  & -1.25 & -2.33 & -2.4 & -2.81\\
     $\delta N_d$ & $\pm$0.20 & $\pm$0.19  & $\pm$0.31 & $\pm$0.4 & $\pm$0.07 \\ \hline
     $\alpha_d$  & 1.9  & 1.5 & 2.5 & 2.7 & 1.21\\
     $\delta \alpha_d$ & $\pm$1.5 & $\pm$0.4  & $\pm$0.4 & $\pm$0.6 & $\pm$0.02 \\ \hline
     $\beta_d$  & 10  & 7.0 & 15.8 & 17 & 3.58\\
     $\delta \beta_d$ & $\pm$11 & $\pm$2.6  & $\pm$3.2 & $\pm$4 & $\pm$0.11 \\ \hline
     $N_{\bar{d}}$  & -0.06  & -0.05 & -0.29 & -0.7 & -32.9\\
     $\delta N_{\bar{d}}$ & $\pm$0.06 & $\pm$0.11  & $\pm$0.27 & $\pm$0.5 & $\pm$0.7 \\ \hline
     $\alpha_{\bar{d}}$  & -  & - & - & 1.5 & 5.2 \\
     $\delta \alpha_{\bar{d}}$ & -  &  -  & -  & $\pm$0.6 & $\pm$0.7 \\ \hline
     $\beta_{\bar{d}}$  & -  & - & - & 20 & 12.9 \\
     $\delta \beta_{\bar{d}}$ &  - &  -  & -  & $\pm$17 & $\pm$0.3 \\ \hline
     $N_s$  & -  & - & -14 & -20 & -15.2 \\
     $\delta N_s$ & -  &  -  & $\pm$10  & $\pm$40 & $\pm$0.8 \\ \hline
     $\alpha_s$  & - & - & 4.9 & 4.7 & 2.9 \\
     $\delta \alpha_s$ & -  & -   & $\pm$3.3  & $\pm$3.0 & $\pm$0.1 \\ \hline
     $\beta_s$  & -  & - & 3 & 2.3  & 3.2\\
     $\delta \beta_s$ &  - &  -  & $\pm$4  & $\pm$3.1 & $\pm$0.3 \\ \hline
     $N_{\bar{s}}$  & -  & - & -0.1 & 20 & -19.90 \\
     $\delta N_{\bar{s}}$ &   &    & $\pm$0.2  & $\pm$5 & $\pm$0.09 \\ \hline
     $\alpha_{\bar{s}}$  & -  & - & - & 9.5 & 7.35 \\
     $\delta \alpha_{\bar{s}}$ & -  &  -  &  -  & $\pm$1.4 & $\pm$0.05 \\ \hline
     $\beta_{\bar{s}}$  & -  & - & - & 20 & 19.80 \\
     $\delta \beta_{\bar{s}}$ &  - &  -  &  -  & $\pm$14 & $\pm$0.02 \\ 
     \hline
     $\chi^2/N_{data}$  & 1.29  & 1.59 & 1.69 & 1.66 & - \\
    \end{tabular}
    \caption{Collection of MINUIT fit results. Fit 1 is from Anselmino et al \cite{Anselmino2017}, Fit 2: Re-fit as similar to \cite{Anselmino2017}, Fit 3: fit results including strange-quarks, Fit 4: fit results with the same treatment for all three light-quark-flavors.}
    \label{table:SIDIS_MINUIT_Fits}
\end{table}

\subsection{DNN Method Testing}
\label{sec:DNN-pseudo-fits}
We develop a systematic method of constructing, optimizing, and testing the DNN fits by using pseudodata to ensure a quality extraction from the experimental data. Our approach uses a combination of Monte Carlo sampling and synthetic data generation. The pseudodata points are randomly generated by sampling within multi-Gaussian distributions centered around each experimental data point, with variance given by the experimental uncertainty.  Many pseudodata DNN fits (instances) are performed together to obtain the uncertainty of the resulting DNN model (mean and distribution). The general approach is to use existing experimental data to parameterize a fit function and then use it to generate new synthetic data (replicas) with similar characteristics.  The pseudodata is generated with a known Sivers function so that the extraction technique can be explicitly tested.  An error bar is assigned to each new data point which is taken directly from the experimental uncertainties reported for the complete set of kinematic bins.  This approach aims to produce pseudodata that simulates the experimental data as closely as possible with particular sensitivity to phase space. It does this so that the test metrics are also relevant for the real experimental extraction. To do this the pseudodata generator must be very well-tuned to the kinematic range of the experimental data. Hence, the \textit{generating function} contains as much feature space information as possible. It's important to emphasize here that the metrics that we use to quantify the improvement in the \textbf{Second Iteration} compared to the \textbf{First Iteration} are sensitive to phase space.  The \textit{accuracy} (proximity of the mean of the DNN fits to the \textit{true} Sivers) is defined as,
\begin{align}
    \label{eq:accuracy}
    \epsilon_q(x,k_{\perp}) = \left( 1 - \frac{|\Delta^N f_{q / p^{\uparrow}}^{\text{(true)}}-\Delta^N f_{q / p^{\uparrow}}^{\text{(mean)}}|}{\Delta^N f_{q / p^{\uparrow}}^{\text{(true)}}} \right)\times 100 \%,
\end{align}
and \textit{precision} (the standard deviation of replicas), as
\begin{align}
    \label{eq:precision}
    \sigma_q (x,k_\perp) =  \sqrt{\frac{\sum_{i} \left(\Delta^N f_{q / p^{\uparrow}}^{(i)} - \Delta^N f_{q / p^{\uparrow}}^{(\text{mean})} \right)^2}{N}}.
\end{align}

The \textit{generating-function} used to produce the \textit{true} value of the Sivers is improved in the process of optimizing the DNN hyperparameters.  This approach improves the \textit{generating-function} and the DNN fit with each iteration.  As a result, more realistic data can be generated in each iteration, which enables better hyperparameter optimization and testing for the experimental data in the subsequent iteration.  Note that experimental data still refers to pseudodata replicas that are generated using the real experimental data rather than the \textit{generating-function}.

In the pseudodata test, the same number of replicas are used in the \textbf{First Iteration} and in the \textbf{Second Iteration}.  The number of replicas should be kept the same across consecutive iterations to control statistical error variation from the replicas. To most accurately propagate the experimental uncertainty using the replica approach requires a sufficient amount of replicas so that only negligible statistical error from the replicas is added. For the present study, our pseudodata uncertainty is simplified and is represented by a single error bar which contains the experimental statistical error and systematic error provided by the data publication.\footnote{A more detailed analysis could be performed if a full covariance matrix from the experimental analysis was provided but this type of information is often difficult to obtain after the experiment results are published.}

These are the steps to perform the DNN \textit{Method Testing} with pseudodata using the \textit{generating function}
\begin{enumerate}
    \item Generate pseudodata for the SIDIS Sivers asymmetry using the \textit{generating function} (The initial function used is from Fit 4 in Table \ref{table:SIDIS_MINUIT_Fits}).
    \item \textbf{First Iteration}: construct a DNN fit and tune its hyperparameters by training with the pseudodata from Step 1. Use 10\% of the pseudodata for the validation in each epoch. 
    \item Determine the optimum number of epochs by analyzing the training and validation losses for each replica in Step 2.
    \item Improve the \textit{generating function}: 
    \begin{enumerate}
        \item Perform an intermediate DNN fit to the experimental data. This fit is performed using the optimized hyperparameters from the above Steps 2 and 3.
        \item Use the trained DNN fit from (a) to infer the asymmetry over the 3D kinematics ($x,z,p_{hT}$) in fine bins.\footnote{When generating fine-binned kinematics for the Sivers asymmetry data for a given kinematic variable, the averaged values are used for other variables including $Q^2$ as the $Q^2-$evolution is near negligible over the considered kinematics.}
        \item Perform a MINUIT fit on the fine-binned data to obtain the new \textit{generating function} parameterization.
    \end{enumerate}
    \item Perform Step 1 again using the improved \textit{generating function}.
    \item \textbf{Second Iteration}: Perform Step 2 again with the pseudodata generated from Step 5.
    \item Perform a comparison of the Sivers functions extracted at the \textbf{First Iteration} vs \textbf{Second Iteration} in terms of the \textit{accuracy}, \textit{precision} and the magnitude of the \textit{loss-function} at the final epoch.
\end{enumerate}

The accuracy metric is a critical part of the testing and is used to understand the resulting extraction methods' strengths and weaknesses. A truly robust method would be consistently able to make accurate predictions for any \textit{generating function} parameterization over the relevant kinematic range.  To verify the robustness of our extraction method, we test it with many artificial Sivers functions and their corresponding pseudodata at the final stage.

The hyperparameters optimization process is a systematic process of trial and error with architecture changes like adding or subtracting layers and the number of nodes per layer.  The initial learning rate varies as does the batch size. Care is taken to achieve the lowest training and validation loss.  We monitor the stability of the loss by examining the magnitude and frequency of fluctuations.  A more moderate and even trend downward is optimal.  After the \textit{generating function} is improved, in principle, it holds more information.  The number of layers and nodes is increased to make a better fit to the new, more realistic, pseudodata in the \textit{Second Iteration}.  If there are too few hidden layers and nodes, it becomes apparent when the training loss does not decrease compared to previous fits.  Too many additional layers and nodes result in instability in the training loss.  Tuning continues until little improvement in accuracy and precision can be determined.

\begin{figure}[h!]
    \centering
    \includegraphics[width=85mm,left]{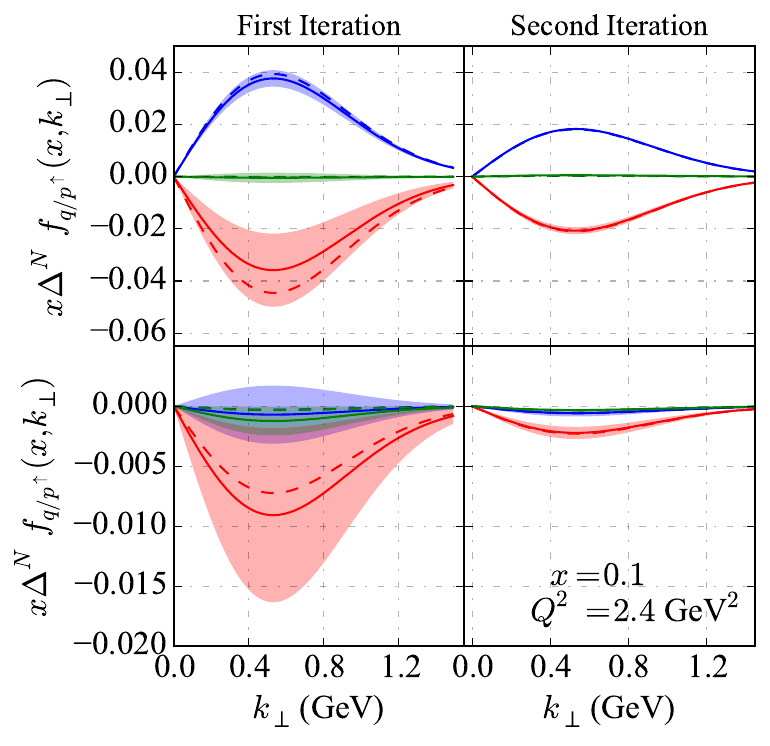}
    \caption{A comparison of the DNN model before (left-column) and after (right-column) hypertuning with the improved \textit{generating-function}. In the \textbf{First Iteration}, the DNN model is based on Fit 4, whereas in the \textbf{Second Iteration}, the model is based on the hyper-tuned DNN fit. Additionally, hypertuning optimization leads to higher accuracy and precision with the same number of epochs.  The dotted line in each case represents generating function or the true value while the solid line represents the mean of DNN fits.  In the top two plots, $u$ is blue, $s$ is green and $d$ is red.  Similarly for the anti-quark in the lower two plots.}
    \label{fig:Siv_Pseudo}
\end{figure}

\begin{table}[ht]
    \centering
    \begin{tabular}{ccccccc}
        Hyperparameter & $\mathcal{C}_0^i$ & $\mathcal{C}_0^f$  & $\mathcal{C}_p^i$ & $\mathcal{C}_p^f$ & $\mathcal{C}_d^i$ & $\mathcal{C}_d^f$ \\ \hline
        Hidden Layers & 5 & 7 & 5 & 7 & 5 & 8\\
        Nodes/Layer & 256 & 256 & 550 & 550 & 256 & 256\\
        Learning Rate  & 1 & 0.125 & 5 & 1 & 10 & 1 \\
        Batch Size & 200 & 256 & 300 & 300 & 100 & 20\\
        Number of epochs & 1000 & 1000 & 300 & 300 & 200 & 200\\
        Training Loss  & 0.6 & 0.05 & 1.5 & 1 & 2 & 1 \\ \hline
        $\varepsilon_{u}^{max}$   & 95.67 & 99.27 & 55.21 & 94.04 & 56.80 & 93.02\\
        $\varepsilon_{\bar{u}}^{max}$ & 42.62 & 98.09 & 52.57 & 96.70 & 34.83 &  91.40 \\
        $\varepsilon_d^{max}$  & 80.46 & 98.89 & 55.69 & 93.13 & 52.44 & 89.27\\
        $\varepsilon_{\bar{d}}^{max}$  & 74.59 & 97.08 & 55.37 & 95.04 & 46.60 & 92.58\\
        $\varepsilon_s^{max}$  & 45.53 & 79.27 & 49.54 & 90.64 & 36.34 & 93.41\\
        $\varepsilon_{\bar{s}}^{max}$   & 59.27 & 91.13 & 33.89 & 82.51 & 65.57 & 91.45\\ \hline
        $\sigma_{u}^{max}$   & 3 & 0.1 & 5 & 2 & 2 & 0.4\\
        $\sigma_{\bar{u}}^{max}$  & 2 & 0.2 & 6 & 2 & 8 & 2\\
        $\sigma_d^{max}$  & 10 & 1 & 20 & 6 & 2 & 1\\
        $\sigma_{\bar{d}}^{max}$  & 7 & 4 & 20 & 8 & 7 & 1 \\
        $\sigma_s^{max}$  & 2 & 0.2 & 4 & 1 & 6 & 2 \\
       $\sigma_{\bar{s}}^{max}$  & 1 & 0.1 & 4 & 2 & 6 & 3 \\ \hline
    \end{tabular}
    \caption{The summary of the optimized sets of hyperparameters: The indications in the table are $\mathcal{C}_0^i$ and $\mathcal{C}_0^f$ for results from the pseudodata from the generating function, $\mathcal{C}_p^i$, and $\mathcal{C}_p^f$ for results from SIDIS data from experiments associated with polarized-proton target, $\mathcal{C}_d^i$ and $\mathcal{C}_d^f$ for results from SIDIS data from experiments associated with polarized-deuterium target, where $i$ and $f$ indicates the \textbf{First Iteration} and \textbf{Second Iteration} respectively. The initial learning rate is also listed ($\times10^{-4}$) as is the final training loss ($\times10^{-3}$). The accuracy and precision in each case are the maxima over the phase space.}
    \label{tab:h-params}
\end{table}

Figure (\ref{fig:Siv_Pseudo}) illustrates the results from the \textbf{First Iteration} (\textit{two plots on the left-hand side}) and the \textbf{Second Iteration} (\textit{two plots on the right-hand side}). The set of hyperparameters in the \textbf{First Iteration} and the \textbf{Second Iteration} are given in the columns for each DNN model in Table \ref{tab:h-params}. The indications in the table are $\mathcal{C}_0^i$ and  $\mathcal{C}_0^f$ for results from the pseudodata from the generating function, $\mathcal{C}_p^i$, and $\mathcal{C}_p^f$ for results from SIDIS data from experiments associated with polarized-proton target, $\mathcal{C}_d^i$ and $\mathcal{C}_d^f$ for results from SIDIS data from experiments associated with polarized-deuterium target. Here $i$ and $f$ indicate the \textbf{First Iteration} and \textbf{Second Iteration} respectively. The listed learning rate (multiplied by $10^{-4}$) is the initial learning rate as a dynamically decreasing learning rate is used. (This is explained in Sec. \ref{extraction}). The accuracy $\varepsilon_q(x,k_\perp)$ is defined in Eq. (\ref{eq:accuracy}), and the results in this table correspond to the maximum deviation of the mean of the replicas from the true values $\varepsilon_q^{max}$; whereas the precision $\sigma_q(x,k_\perp)$ is defined in Eq. (\ref{eq:precision}) and the results are the maximum standard deviations of the replicas $\sigma_q^{max}$, and are in the units of $\times 10^{-3}$.

It is worth noting that the improvement in both \textit{accuracy} $\varepsilon_q^{max}$ and \textit{precision} $\sigma_q^{max}$ can be observed from the closeness of the solid line (mean of the 1000 DNN replicas) to the dashed line (\textit{generating function}) for quarks (upper plots) and antiquarks (lower plots). The improvement of the DNN model in each case is significant to the point where it is difficult to distinguish the solid line and the corresponding dashed line, indicating a high degree of accuracy and precision. Also, we observed that the training loss in the \textbf{Second Iteration} is about one order of magnitude less than the one from the \textbf{First Iteration}.

\subsection{DNN model from real data}
\label{sec:DNN-fits}
In contrast to the testing of the DNN model with pseudodata from the \textit{generating function}, we now describe the steps to apply the DNN fit method to real experimental data.  The pseudodata test from Sec. \ref{sec:DNN-pseudo-fits} is using the combined proton and deuteron data as in all previous work on global fits of the Sivers function.  In the following extraction with real experimental data, the proton and deuteron data are fitted separately.  To take full advantage of the information provided by the model testing in the previous section, the steps from Sec. \ref{sec:DNN-pseudo-fits} are performed again separately for proton and deuteron data.  The starting hyperparameters for the first DNN fit in this section including the architecture, initial learning rate, batch size, as well as optimal number of epochs, are all determined based on the best accuracy and precision using the well-tuned pseudodata from the generating function first in each case.  The provide more information on the initial hyperparameter so that even the \textbf{First Iteration} is a quality fit.  In the following steps, two distinct DNN models are developed, one for the \textit{proton} quarks Sivers asymmetry and one for \textit{neutron} quark Sivers asymmetry.  The following procedure is common to both.

\begin{enumerate}
    \item \textbf{First Iteration}: construct a DNN fit and tune its hyperparameters by training with the experimental data. Use 10\% of the data for the validation in each epoch.\footnote{For this section we list this as \textbf{First Iteration} though the real experimental data has already been fit to improve the \textit{generating function} for the pseudodata test.  With every subsequent iteration, the pseudodata test and the DNN fit to the experimental data improve.}
    \item Identify the optimum number of epochs when the validation loss exceeds the training loss (see Fig. (\ref{fig:Model_Improve_Demo_loss}) as an example). 
    \item Perform a DNN fit, with the optimized hyperparameters from Step 1 and the number of epochs determined from Step 2, using all the data without leaving any for validation.
    \item Improve the \textit{generating function}:
    \begin{enumerate}
        \item Use the tuned DNN model in step 3 to infer the asymmetry over the 3D kinematics ($x,z,p_{hT}$) in fine bins.
        \item Perform a MINUIT fit to obtain the new \textit{generating function}.
        \item Produce pseudodata for the SIDIS asymmetry using the \textit{generating function} in the previous step.
    \end{enumerate}
    \item Perform Step 1 and Step 2 with the pseudodata from Step 4 (from the improved \textit{generating function}).
    \item \textbf{Second Iteration}: perform a DNN fit with the optimized hyperparameters from Step 5 using experimental data without leaving any data for validation.
    \item Perform a comparison of the Sivers functions extracted at the \textbf{First Iteration} vs \textbf{Second Iteration} in terms of the \textit{accuracy}, \textit{precision} and the magnitude of the \textit{loss-function} at the final epoch.
\end{enumerate}

\begin{figure}[ht]
    \centering
    \includegraphics[width=65mm]{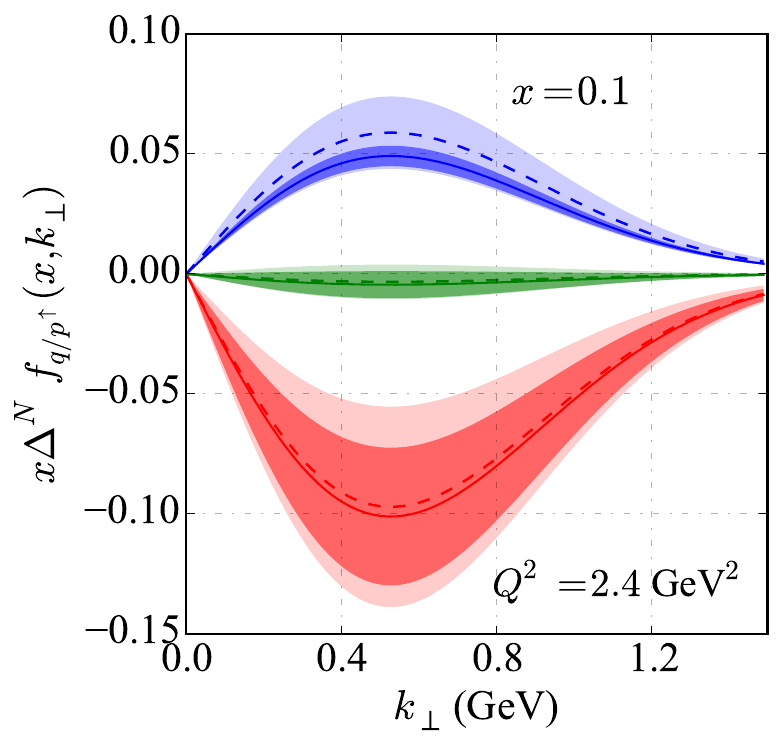}
    \caption{The qualitative improvement of the extracted Sivers functions for $u$ (blue), $d$ (red), and $s$ (green) quarks at $x=0.1$ and $Q^2$=2.4 GeV$^2$ using the optimized \textit{proton}-DNN model at the \textbf{Second Iteration} (solid-lines with dark-colored error bands with 68\% CL), compared to the \textbf{First Iteration} (dashed-lines with light-colored error bands with 68\% CL).}
    \label{fig:Model_Improve_Demo}
\end{figure}

\begin{figure}[ht]
    \centering
    \includegraphics[width=87mm]{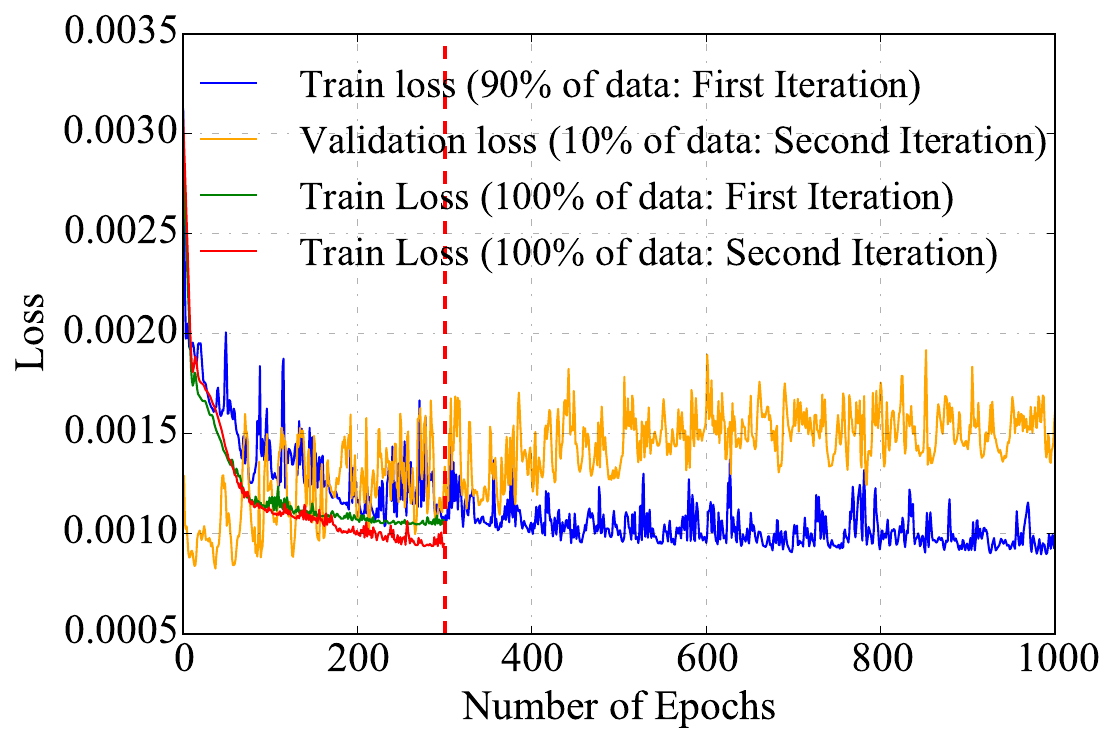}
    \caption{An example for the determination of the optimum number of epochs (about 300 in this case: marked by the vertical red dashed-line) at the \textbf{First Iteration} based on cross-over coordinates between the train (blue curve) and the validation (orange curve) losses. Then, the train-loss behaviors until the optimum number of epochs with all data points at the \textbf{First Iteration} and at the \textbf{Second Iteration} are represented by the green-curve and the red-curve respectively. }
    \label{fig:Model_Improve_Demo_loss}
\end{figure}

Although the architectural specifics such as the number of hidden layers, nodes per layer, and learning rate may be modified in the hyperparameter optimization step, the number of epochs and the number of replicas remain the same.  The overall feedforward architecture structure remains consistent as well for simplicity. Table \ref{tab:h-params} shows the optimized set of hyperparameter configurations are represented with $\mathcal{C}_{\{0,p,d\}}^{\{i,f\}}$ where $i,f$ represent the \textbf{First Iteration} and the \textbf{Second Iteration}, and $\{0,p,d\}$ represent with \textit{pseudodata}, with \textit{proton} data and with \textit{deuteron} data respectively.

Clearly, the pseudodata from the \textit{generating function} still plays a critical role in the tuning and testing of the fit of the real experimental data.  Accuracy is necessarily determined using pseudodata so that the mean of the final DNN model can be compared directly to the \textit{true} Sivers asymmetry.  This procedure for determining accuracy assumes that the \textit{true} Sivers asymmetry from the experiment can be approximated by the well-tuned \textit{generating function} after the final iteration.  There is a systematic error associated with analyzing accuracy this way but this type of error can be estimated. 

After completing the extraction method described above, we perform a systematic uncertainty assessment on the overall method of extraction.
To test the reliability of the extraction, we adjust the parameters of the \textit{generating function} and repeat the extraction process by again following the full set of steps for several variations of pseudodata. By using the Sivers asymmetry generated from the optimized \textit{generating function}, the absolute differences between the mean of the DNN model and the true value over $k_\perp$ was used to estimate the systematic uncertainties of the final DNN model from this extraction technique. 

\section{Results}
\label{results}
In this section, we present the results from two separate DNN models: the \textit{proton}-DNN and the \textit{deuteron}-DNN, along with their optimized hyperparameters. The number of parameters in the optimized proton-DNN model is 11 million wheras for deuteron-DNN model is 2.8 million. Only SIDIS data was used to train the DNN models in this exploratory AI-based extraction technique. The optimized hyperparameter configurations are provided for both models in columns $\mathcal{C}_p^{i(f)}$ and $\mathcal{C}_d^{i(f)}$, with the subscripts $p$ and $d$, respectively, in Table \ref{tab:h-params}. To quantitatively represent the improvement made by performing the steps mentioned in the previous section, we present our accuracy and precision results in the lower part of Table \ref{tab:h-params}.

\subsection{DNN fit to SIDIS data}

We now explore the results and compare some of our final fits and projections with those of other global fits. Note that the $\chi^2$ values are calculated values for each kinematic bin based on Pearson's reduced $\chi^2$ statistic, and indicated in our plots are calculated after the analysis is complete, rather than as a part of the minimization process.

\begin{figure*}
    \centering
        \begin{tabular}{cc}
            \includegraphics[width=80mm]{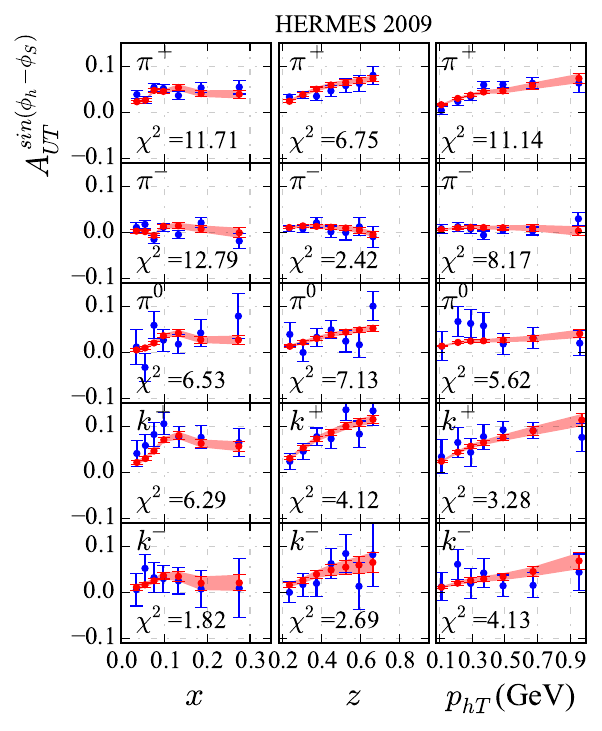} &  
            \includegraphics[width=72.5mm]{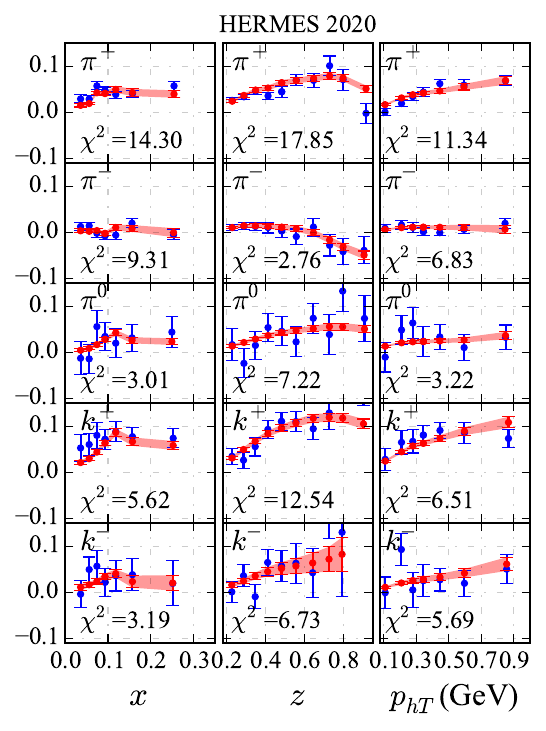} \\
            \includegraphics[width=80mm]{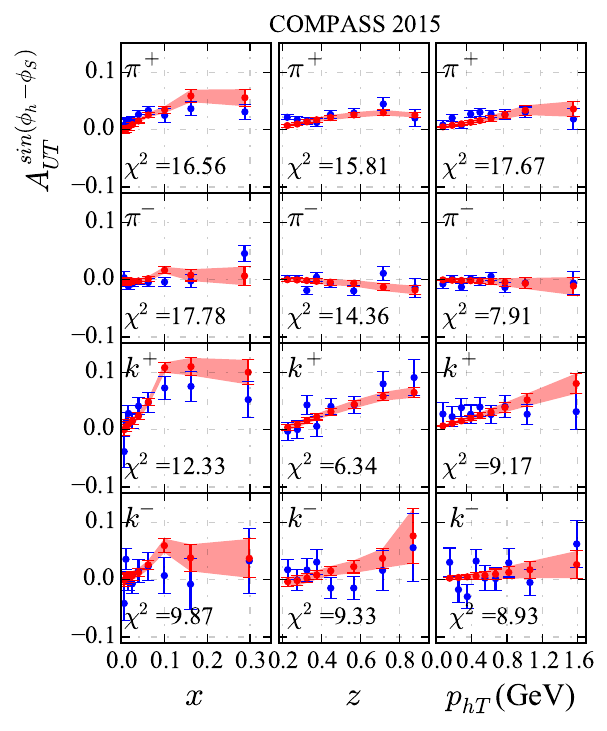} &
            \includegraphics[width=72.8mm]{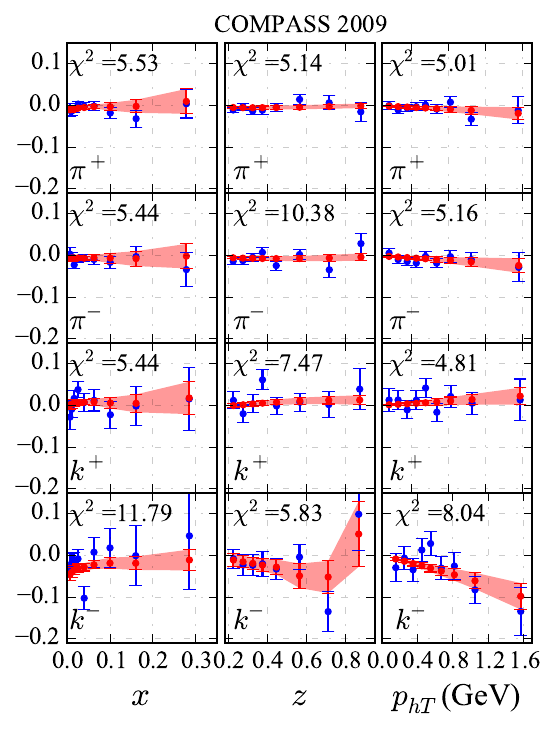}
        \end{tabular}
        \caption{The DNN fit results of the SIDIS Sivers asymmetries (red) accompanied by 68\% CL error bands in comparison with the actual data (blue). The \textit{proton}-DNN model is trained with HERMES2009, HERMES2020, and COMPASS2015, whereas the \textit{deuteron}-DNN model is trained with the COMPASS2009 data. The calculated partial $\chi^2$ values are provided as quantitative assessments for all kinematic bins.}
        \label{fig:SIDIS_Asym_from_DNNs}
\end{figure*}
The plots of SIDIS Sivers asymmetry data and our resulting DNN models (for \textit{proton} and \textit{deuteron}) are shown in Fig. \ref{fig:SIDIS_Asym_from_DNNs}.  Each plot includes the partial $\chi^2$ values for the particular $x,z$, and $p_{hT}$ bins for each hadron type. HERMES2009 \cite{HERMES_Data_Airapetian_2009} (\textit{top-left}), HERMES2020 \cite{HERMES_Data_Airapetian_2020} (\textit{top-right}), and COMPASS 2015 \cite{COMPAS_Data_Adolph_2015} (\textit{bottom-left}) are described with the \textit{proton}-DNN model with Root-Mean-Square-Error (RMSE) of 0.0225, whereas the COMPASS2009 \cite{COMPAS_Data_Alekseev_2009} (\textit{bottom-right}) dataset is described with the \textit{deuteron}-DNN with RMSE of 0.0220. In comparison with \cite{Anselmino2017}, there are some improvements in describing the proton SIDIS data on $\pi^{\pm}$ and $K^+$ in HERMES 2009, which can be noticed quantitatively based on the partial $\chi^2$ values from the DNN model. This indicates that the possible effects attributed to the TMD evolution \cite{Anselmino2012,Echevarria_2014,Aybat_2012} and were assumed to be the cause of the larger $\chi^2$ values in \cite{Anselmino2017} for proton data on $\pi^+$ may have been somewhat integrated into the DNN model. Although HERMES 2020 \cite{HERMES_Data_Airapetian_2020} reported SIDIS data in 1D kinematic bins as well as with 3D kinematic bins, in our fits we use the data in the form of 1D kinematic bins to be consistent with the rest of the datasets in our fits.

The \textit{deuteron}-DNN model's description of COMPASS2009 \cite{COMPAS_Data_Alekseev_2009} data is shown in the bottom-left sub-figure of Fig. \ref{fig:SIDIS_Asym_from_DNNs}. Without applying any cuts on the data, the DNN model yields a RMSE of 0.0220, covering the full range in $x$, $z$, and $p_{hT}$ kinematic projections from the COMPASS2009 dataset. This is in contrast to the limited kinematic coverage considered in \cite{Bury2021_JHEP,Echevarria_2021,Cammarota_2020,bachetta2022}; notably, the data points at $p_{hT}>$ ~1 GeV are described somewhat better by the \textit{deuteron}-DNN model compare to the fits in \cite{Anselmino2009,Anselmino2017}. This suggests that performing dedicated fits to data specific to polarized nucleon targets enables better information extraction, which is true for both DNN and other fitting approaches. The advantage of DNNs in this case is to perform well even with limited data. 

We did not include JLab \cite{JLab2011,JLab2014} data in our \textit{deuteron}-DNN model fits to use it as a projection test for the neutron Sivers asymmetry. Our projection indicates good agreement with the $^3$He data, but both the data and the projection are largely consistent with zero. It is also important to note that in this work we are \textbf{\textit{not}} imposing any isospin symmetry condition ($f_{1T}^{\perp u}=f_{1T}^{\perp d}$ and/or $f_{1T}^{\perp \bar{u}}=f_{1T}^{\perp \bar{d}}$) for the SIDIS data with the deuteron target as was done in \cite{Bury2021_JHEP}.  The successful construction of the two different proton and neutron Sivers functions may indicate that our DNN approach can be particularly useful for analyzing data from polarized nucleons in different nuclei, potentially opening up a new way of exploring the nuclear effects associated with TMDs.   

\subsection{Sivers in Momentum Space}
\begin{figure}[h!!!]
    \centering
    \includegraphics[width=87mm]{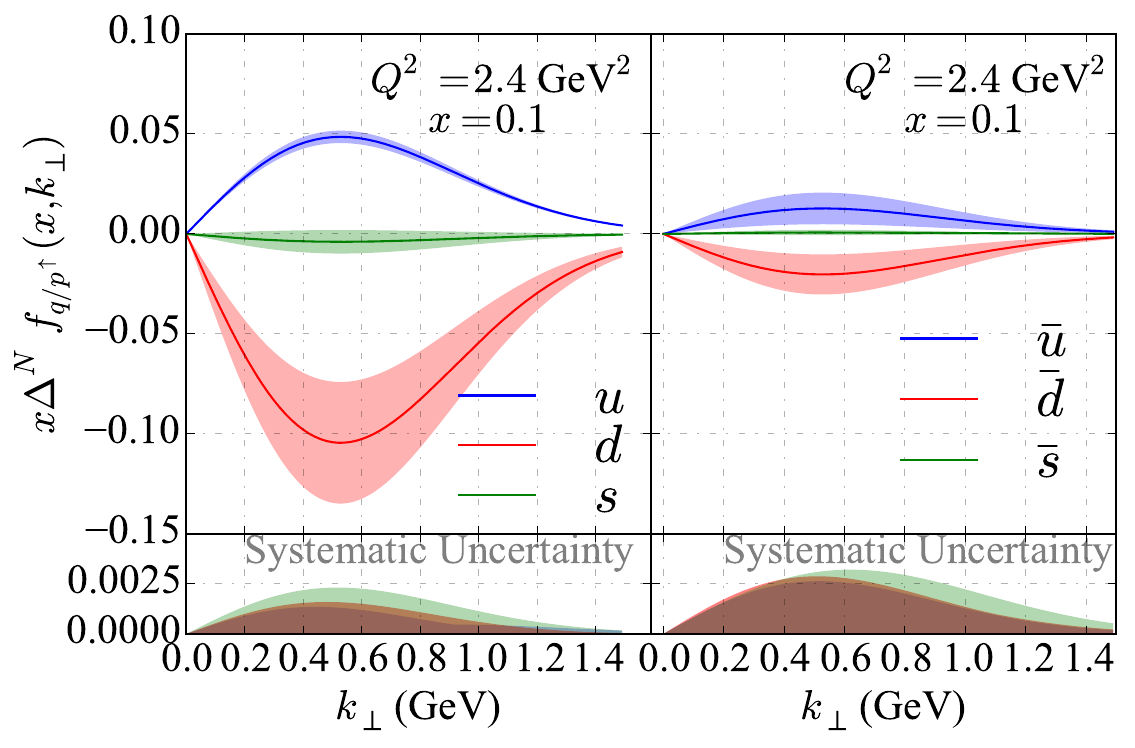} \\
    \includegraphics[width=87mm]{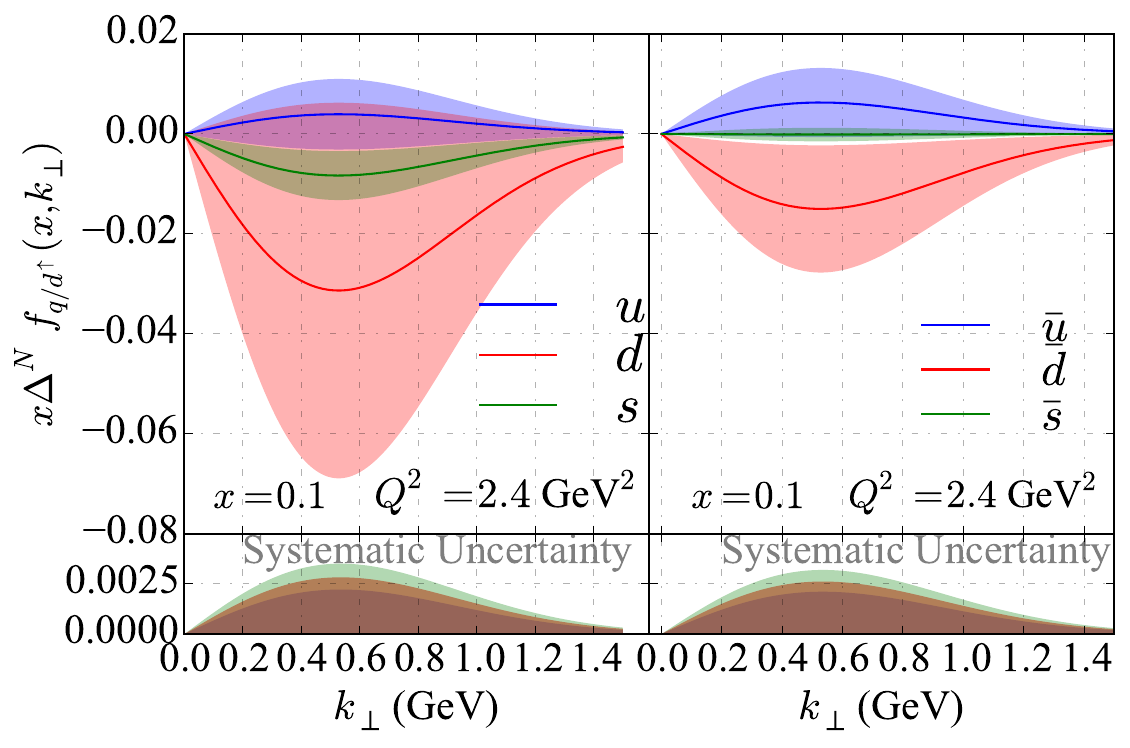}
    \caption{The extracted Sivers functions from the \textit{proton}-DNN model (upper) and \textit{deuteron}-DNN model (lower)  at $x=0.1$ and $Q^2=2.4$ GeV$^2$ with 1-$\sigma$ (68\%) CL \textit{error-bands}, including systematic uncertainties.}
    \label{fig:Siv_SIDISpd}
\end{figure}

The extracted Sivers functions, including the systematic uncertainties from the DNN models at $x=0.1$ and $Q^2=2.4$ GeV, are shown in Fig. \ref{fig:Siv_SIDISpd} represented by the \textit{mean} with 68\% Confidence Level (CL) \textit{error-bands}. The corresponding optimized hyperparameter configurations for the \textit{proton}-DNN model and \textit{deuteron}-DNN model are $\mathcal{C}_p^f$ and $\mathcal{C}_d^f$, respectively, as given in Table \ref{tab:h-params}. The Sivers functions extracted using the \textit{deuteron}-DNN model show consistency with zero, considering the accompanying systematic uncertainties.  However, this is still a significant result, given the limitation in statistics from the SIDIS data with a deuteron target. The Sivers functions extracted using the \textit{proton}-DNN model have small systematic uncertainties. Note that we use $\Delta^N f_{q/p^{\uparrow}} (x,k_\perp)$ notation, as in \cite{Anselmino2017}, to represent the Sivers functions in our plots, and one can use Eq. (\ref{Siv_dist}) to convert to $f_{1T}^{\perp q } (x,k_\perp)$ notation.

Comparing the extracted Sivers functions in $k_\perp$-space to other extractions in the literature can also be useful, although we have not included those curves in our plots. In summary, the \textit{proton}-DNN model extractions are relatively precise, with narrower error bands compared to those in \cite{Anselmino2017,Anselmino2009,Anselmino_DY_2009,Echevarria_2014,Bury2021_PRL,Bury2021_JHEP,Echevarria_2021}.

\subsection{Sivers First Transverse Moment}

The first transverse moment of the Sivers functions can be obtained through $d^2k_\perp$-integration of the Sivers functions \cite{Anselmino2017},
\begin{align}
    \label{eq:Siv_moment}
    \Delta^N & f_{q/p^{\uparrow}}^{(1)}(x) = \int d^2k_\perp \frac{k_\perp}{4m_p}\Delta^N f_{q/p^{\uparrow}}(x,k_\perp)  \nonumber \\
    &= - f_{1T}^{\perp(1)q}(x) = \frac{\sqrt{\frac{e}{2}}\langle k_\perp^2 \rangle m_1^3}{m_p(\langle k_\perp^2 \rangle + m_1^2)^2}\mathcal{N}_q(x)f_q(x;Q^2)
\end{align}

\begin{figure}[h!!!]
    \centering
    \includegraphics[width=87mm]{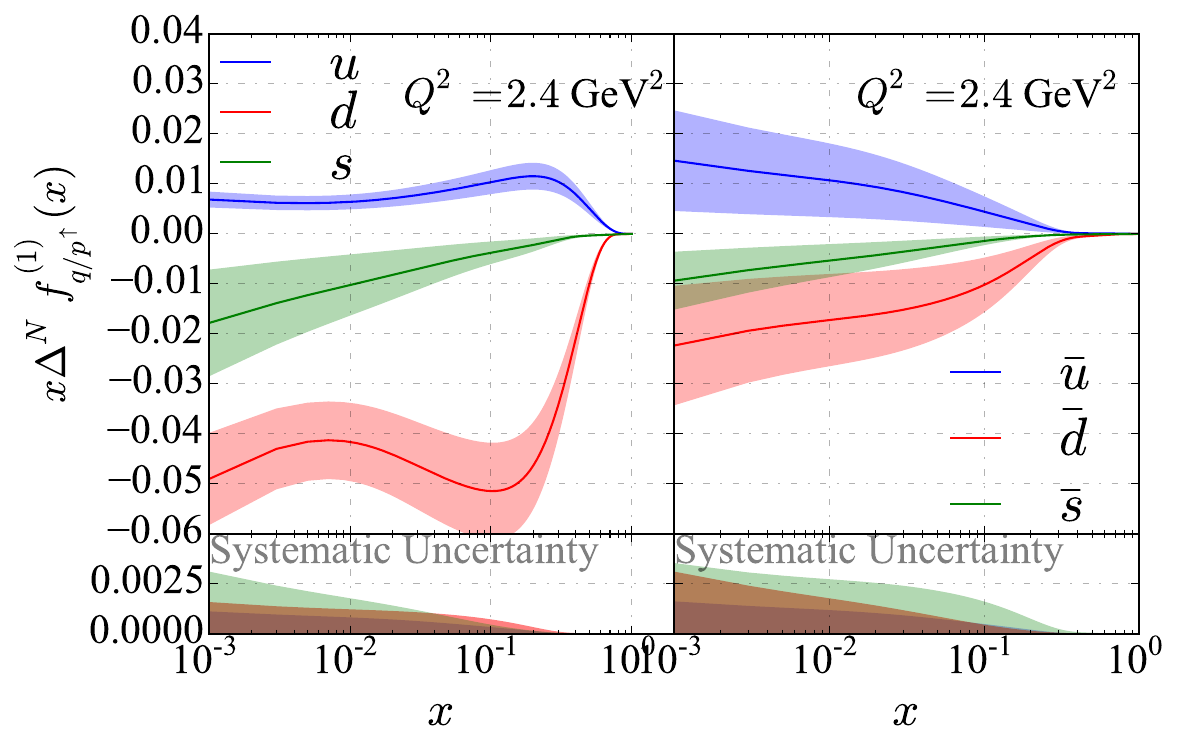} \\
    \includegraphics[width=87mm]{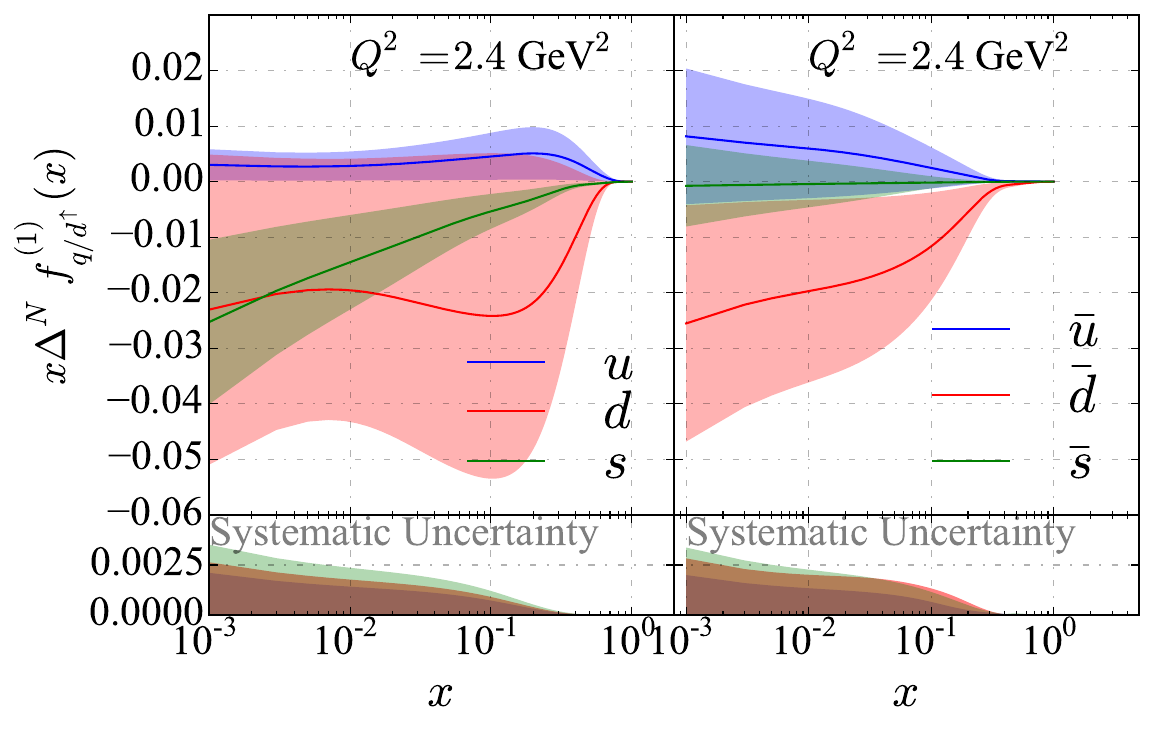}
    \caption{The extracted first transverse moments of Sivers functions from the \textit{proton}-DNN model (upper) and \textit{deuteron}-DNN model (lower)  at $x=0.1$ and $Q^2=2.4$ GeV$^2$ with 68\% CL \textit{error-bands}, including systematic uncertainties.}
    \label{fig:Siv_Moments_SIDISpd}
\end{figure}

\begin{figure}[h!!!]
    \centering
    \includegraphics[width=87mm]{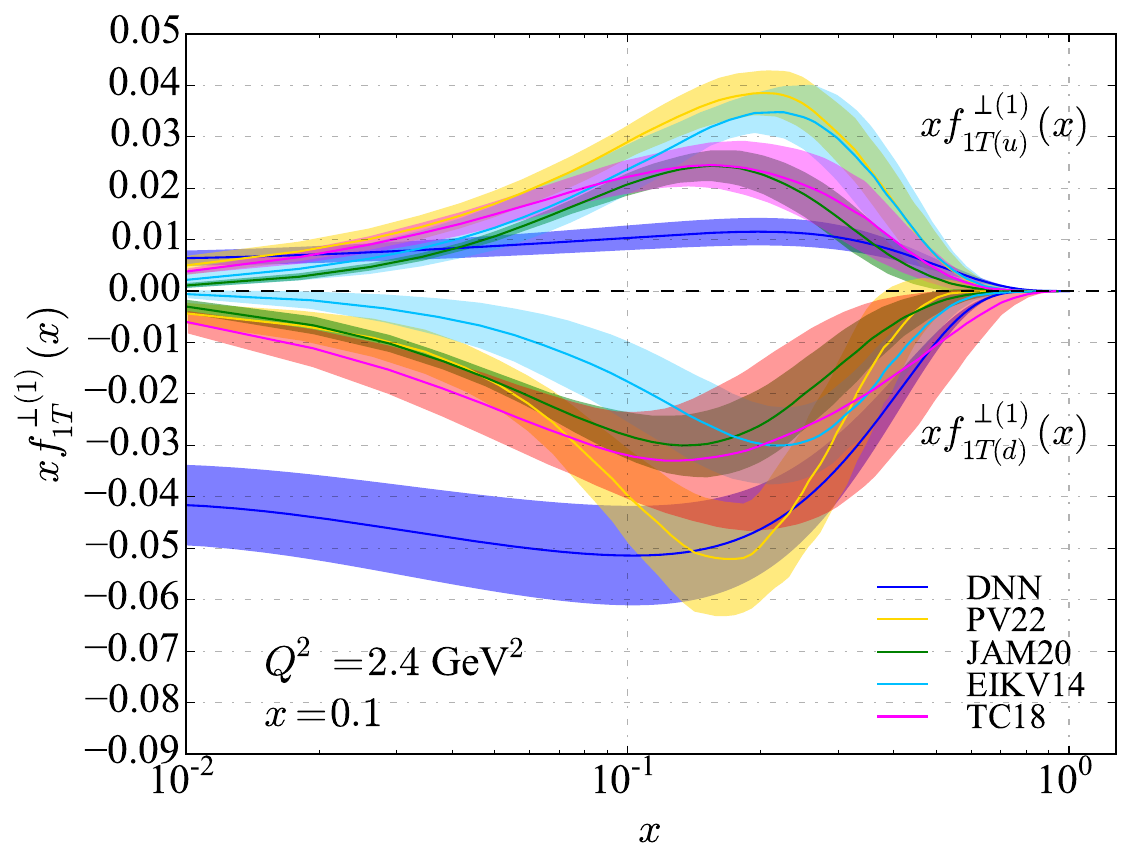} \\
    \caption{The extracted Sivers functions for valence $u (d)$ quarks from the \textit{proton}-DNN model represented in upper (lower) half of the figure; with the results from: PV22 \cite{bachetta2022}, JAM20 \cite{Cammarota_2020}, EIKV \cite{Echevarria_2014}, TC18 \cite{Boglione_2018}.}
    \label{fig:Siv_ud_comparison}
\end{figure}

The extracted first transverse moments of the Sivers functions including the systematic uncertainties from the DNN models are given in Fig. \ref{fig:Siv_Moments_SIDISpd} with 68\% CL \textit{error-bands} using the optimized hyperparameter configurations $\mathcal{C}2$ and $\mathcal{C}3$ in Table \ref{tab:h-params} respectively for \textit{proton}-DNN model and \textit{deuteron}-DNN model. The calculated moments using the \textit{deuteron}-DNN model are consistent with zero, based on the systematic uncertainties. 

Comparing the results in Fig. 1 of \cite{bachetta2022} as shown in Fig. \ref{fig:Siv_ud_comparison}, we see that the $x f_{1T}^{\perp (1)u}$ from the DNN model is more consistent with \cite{Echevarria_2014,Cammarota_2020} in the vicinity of $x=0.1$, although it is consistent with \cite{bachetta2011} at $x=0.01$. The $x f_{1T}^{\perp (1)d}$, in general, is consistent with the extractions from \cite{Anselmino2009,Anselmino2017,Echevarria_2014,Cammarota_2020,bachetta2022,bachetta2011,Boglione_2018}. Additionally, the extracted behavior of $x f_{1T}^{\perp (1)u}$ and $x f_{1T}^{\perp (1)d}$ is consistent with the qualitative observation in \cite{Anselmino2009},
\begin{align}
\label{eq:fu_fd}
    \Delta^N f_{u/p^{\uparrow}}^{(1)}(x) &= - \Delta^N  f_{d/p^{\uparrow}}^{(1)}(x) \nonumber \\
    \text{or} \; \; \; f_{1T}^{\perp(1)u}(x) & = - f_{1T}^{\perp(1)d}(x)
\end{align}
which was originally a prediction from the large-$N_c$ limit of QCD \cite{Pobylitsa_2003}. Most importantly, the DNN model is able to capture the feature of the $u$ and $d$ quarks orbiting in opposite directions without imposing this constraint directly as done in \cite{EFREMOV2005233}. In terms of the quantitative assessment, Eq. (\ref{eq:fu_fd}) could be accurate at the large-$N_c$ limit, if the isospin breaking effects are also included at the next to leading order in $\mathcal{O}(1/N_c)$.

In regards to the light sea-quarks, the \textit{proton}-DNN model extracts the features such as $\Delta^N f_{\bar{u}/p^{\uparrow}}^{(1)}(x)>0$ and $\Delta^N f_{\bar{d}/p^{\uparrow}}^{(1)}(x)<0$, even considering the scale of the uncertainties. Additionally, the \textit{proton}-DNN model is consistent with 
\begin{align}
    \Delta^N f_{\bar{u}/p^{\uparrow}}^{(1)}(x)&= -\Delta^N f_{\bar{d}/p^{\uparrow}}^{(1)}(x)
\end{align}
which was also a similar observation from a theoretical calculation based on $SU(2)$ chiral Lagrangian \cite{He_2019} and the predictions at large-$N_c$ limit of QCD \cite{Pobylitsa_2003}. The central values extracted in \cite{bachetta2011} are qualitatively similar to the features seen in Fig. \ref{fig:Siv_Moments_SIDISpd} which are small but non-zero within the uncertainties. Additionally, the corresponding central values extracted in \cite{Anselmino2017} are both negative but consistent with zero.

\begin{figure*}
    \centering
        \begin{tabular}{cc}
            \includegraphics[width=92mm]{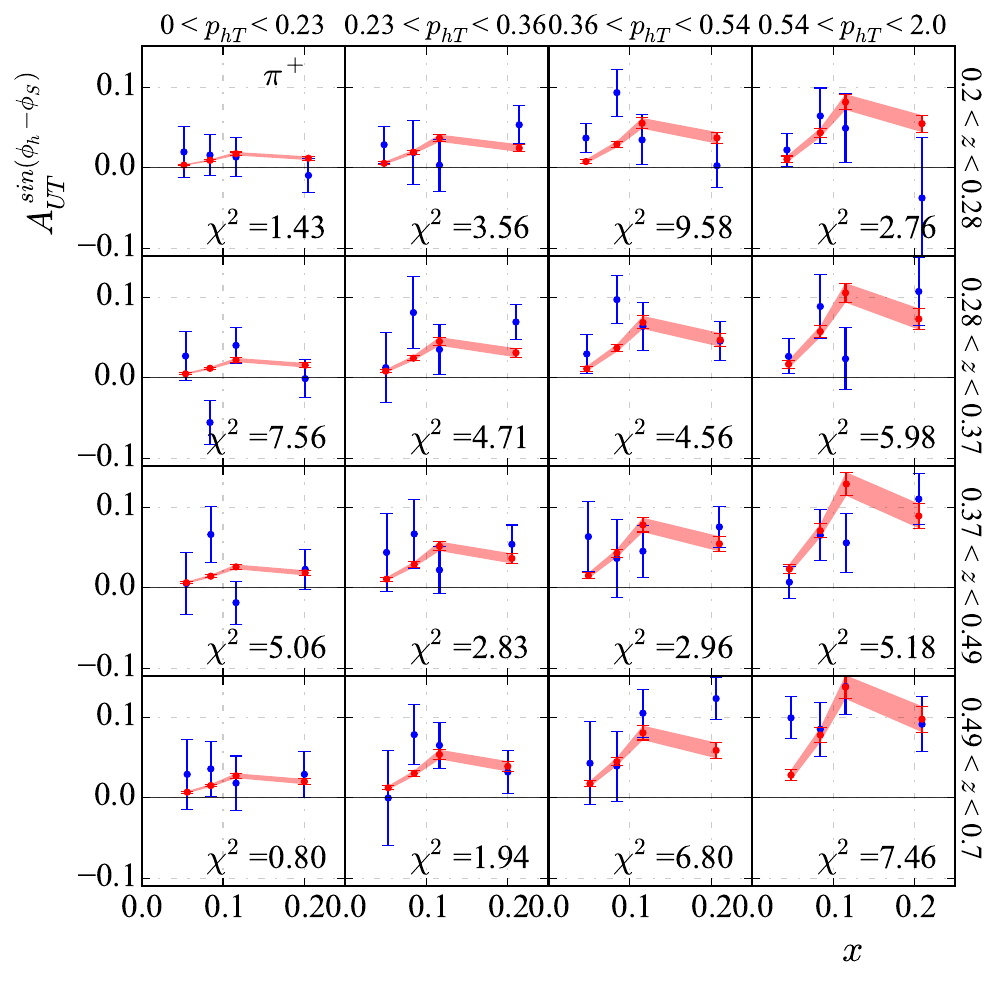} &  
            \includegraphics[width=86mm]{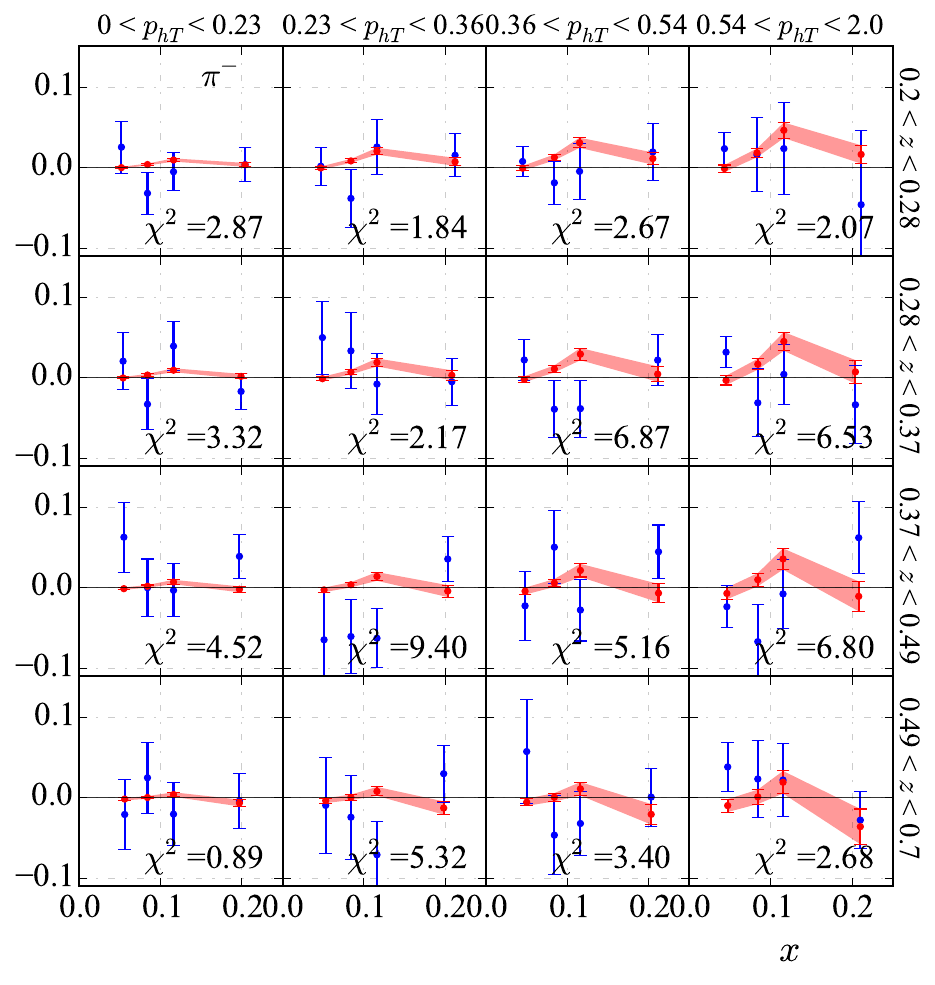}\\
            \includegraphics[width=92mm]{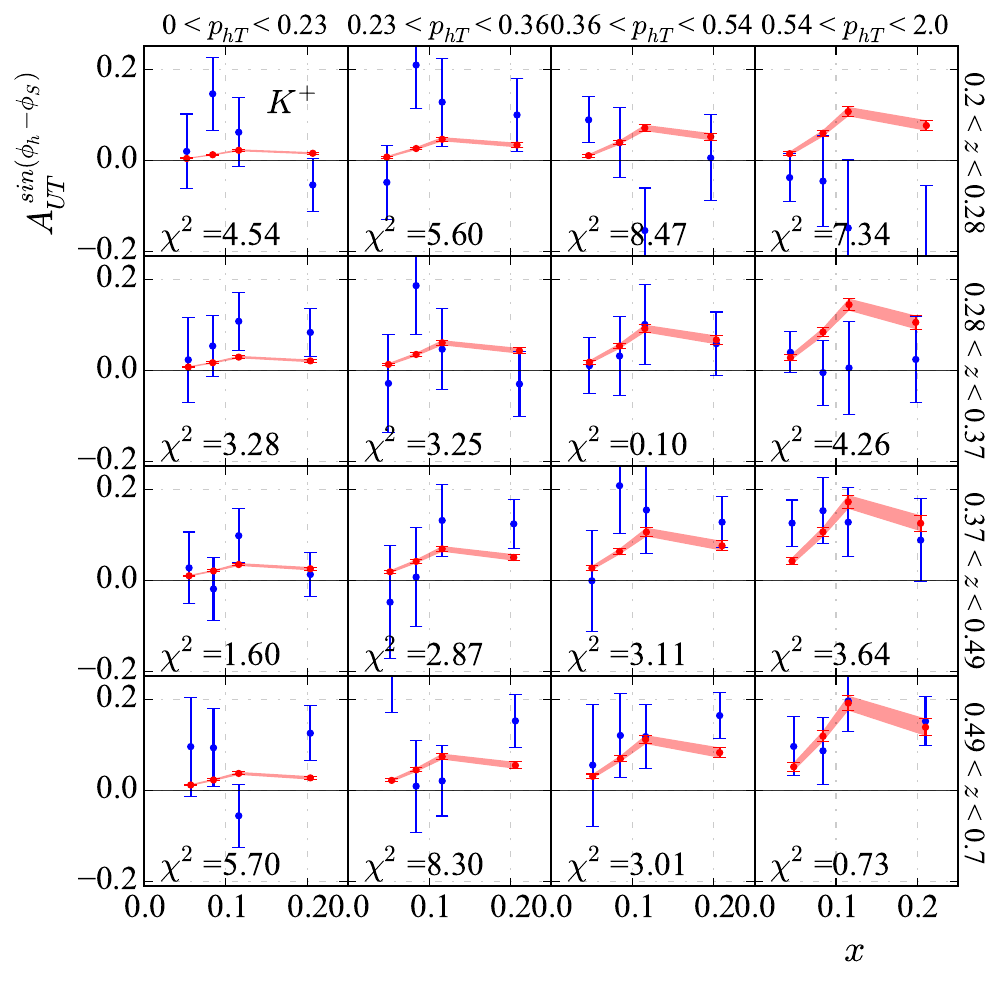} &
            \includegraphics[width=86mm]{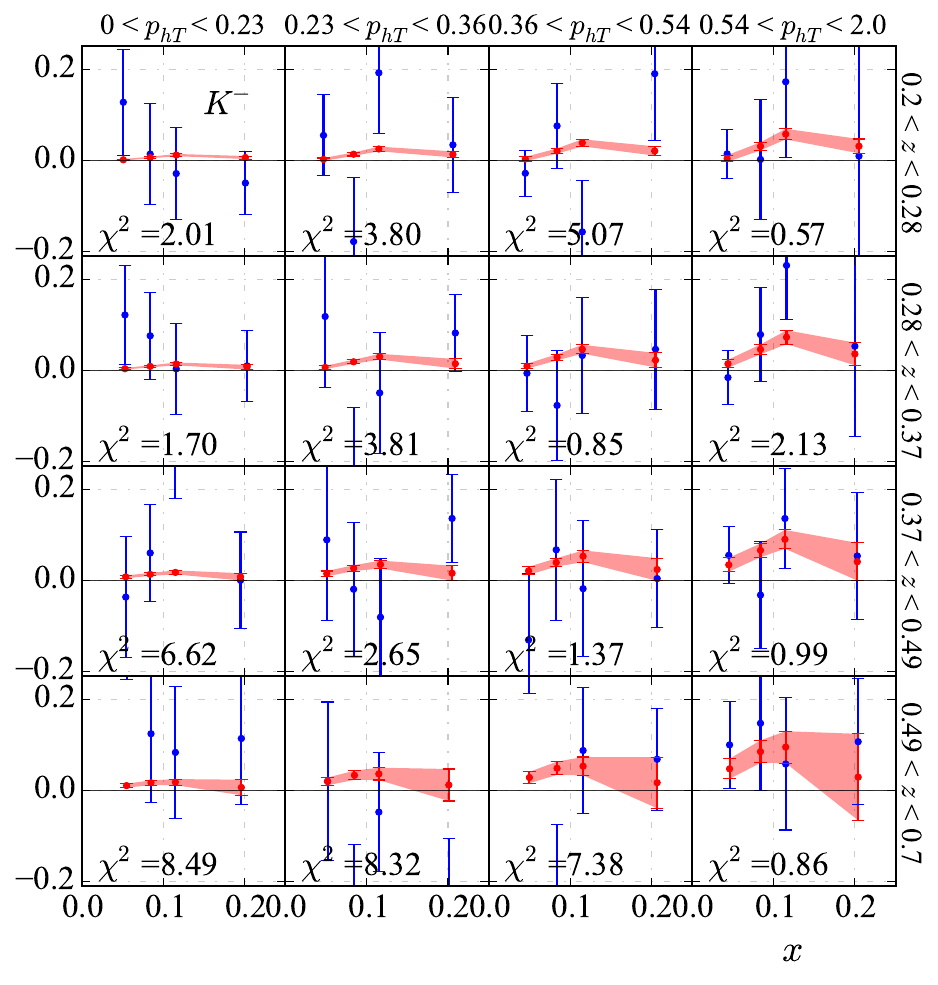}
        \end{tabular}
        \caption{Projections of the of HERMES 2020 data for 3D kinematic bins, using the \textit{proton}-DNN model including 68\% CL error bands (in red) in comparison with the actual data points (in blue).}
        \label{fig:HERMES2020_Asym_3D}
\end{figure*}

The first transverse moments $x f_{1T}^{\perp (1)q}(x)$, in the case of $SU(3)_{\textbf{flavor}}$, from our DNN result, are more precise (narrower error bands) than those in \cite{Anselmino2009,Anselmino2017,Anselmino2005}. However, the error bands are slightly larger than those in JAM20 \cite{Cammarota_2020}, which includes more data from SIDIS, DY and SIA, $pp$-collisions, and parameterizations for Sivers, Collins, and Transversity TMDs together.

\subsection{Projections}

\subsubsection{SIDIS Projections}
\begin{figure}[h!]
    \centering
    \includegraphics[width=80mm]{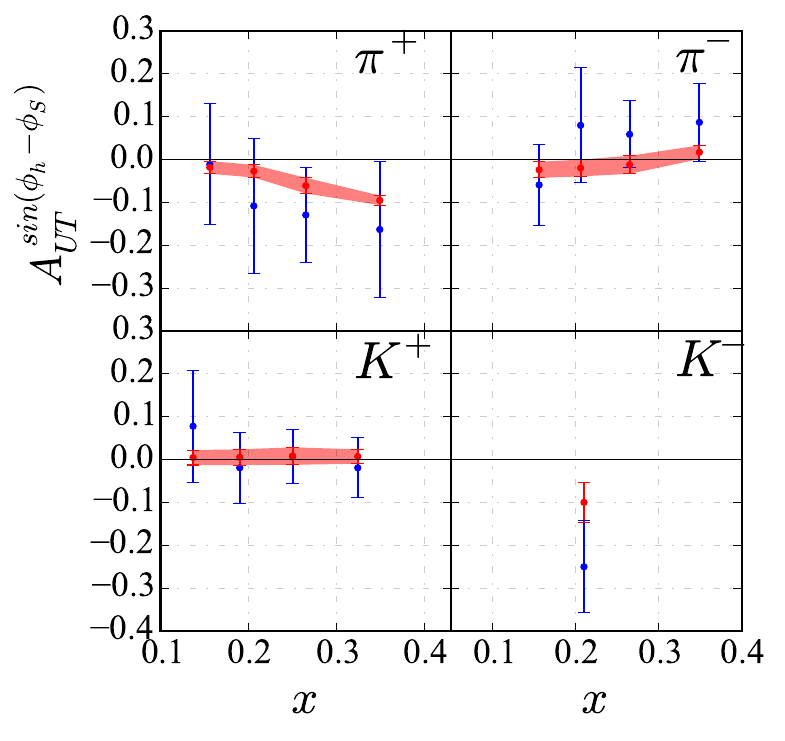} 
    \caption{The \textit{deuteron}-DNN model projections (red) with 68\% CL error-bands, for JLab kinematics \cite{JLab2011,JLab2014}  in comparison with measured data (blue) without the systematic uncertainty.}
    \label{fig:JLab_prediction}
\end{figure}

In Fig. \ref{fig:HERMES2020_Asym_3D}, we compare the SIDIS Sivers asymmetries (in red) projected onto the HERMES2020 3D kinematic bins with the experiment measurements (in blue). These results are obtained using our \textit{proton}-DNN model and are accompanied by 68\% CL error bands. We also provide calculated partial $\chi^2$ values for each kinematic bin as a quantitative assessment. Unlike in \cite{Bury2021_JHEP}, we have made projections for all the data points since we have not applied any data cuts. There are relatively larger partial $\chi^2/N_{pt}$ (as was also observed in \cite{Bury2021_JHEP}), but only in a couple of $K^+$ and $K^-$ bins. In Fig. \ref{fig:JLab_prediction}, we present the projected SIDIS Sivers asymmetries for the JLab kinematics \cite{JLab2011,JLab2014}, obtained using our \textit{deuteron}-DNN model. The figure includes 68\% CL error bands and a comparison with the JLab neutron Sivers asymmetry data. These results are consistent with those reported in \cite{Echevarria_2014, Anselmino2017, Boglione_2018, Echevarria_2021}.

\subsubsection{DY Projections}

The resulting DNN model based on the SIDIS Sivers asymmetries is capable of projecting the Sivers asymmetries in DY experiments which could be sensitive to either valence quarks or sea quarks depending on the relevant kinematic coverage. For example, the COMPASS2017-polarized DY Sivers asymmetry measurements \cite{COMPASS_DY_2017} are dominated by the valence quarks, and the upcoming SpinQuest (E1039) experiment's polarized DY asymmetry measurements \cite{SpinQuest_Proposal} will be dominated by the \text{sea} quarks. For these DY projections, we follow the block diagram represented in Fig. \ref{fig:Block-Diagram-DY}, which includes the assumption of the {\bf{\textit{sign-change}}} of the Sivers function in DY relative to the SIDIS process mentioned in Eq. (\ref{eq:conditional_universality}). Therefore, using the trained \textit{proton}-DNN model, we make projections for the DY Sivers asymmetries for the COMPASS2017 experiment \cite{COMPASS_DY_2017} with a proton target and a pion beam using CTEQ6l \cite{CTEQ6l_2003} and JAM21PionPDFnlo \cite{JAM21} for proton PDFs and pion PDFs respectively. Meanwhile, using both the \textit{proton}-DNN model and \textit{deuteron}-DNN model, we make predictions for the SpinQuest experiment\footnote{The kinematic bins from the SeaQuest experiment are used \cite{SeaQuest_2022}.}. The kinematic-inputs are $x_1$ (beam), $x_2$ (target), $x_F(=x_1-x_2)$, $q_T$ (transverse component of the virtual photon), and $Q_M$ (di-lepton invariant mass). 

\begin{figure}[h!]
    \centering
    \includegraphics[width=85mm]{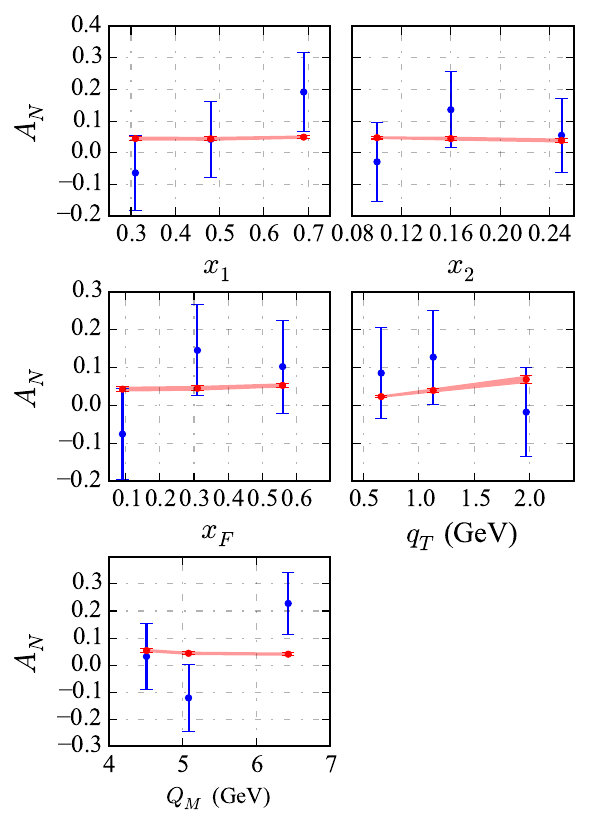}
    \caption{The \textit{proton}-DNN model's predictions (red) including 68\% CL error-bands, for Sivers asymmetries in $x_1, x_2, x_F, q_T$, and $Q_M$ kinematic projections for COMPASS DY kinematics \cite{COMPASS_DY_2017} in contrast with the measured data (blue).}
    \label{fig:DYp_COMPASS_Pred}
\end{figure}

The projected DY Sivers asymmetries for the COMPASS2017 kinematics using the trained \textit{proton}-DNN model in comparison with the data \cite{COMPASS_DY_2017} points are represented in the Fig. \ref{fig:DYp_COMPASS_Pred}. Although the projections are based on the assumption of conditional universality, it is worth noting that without this assumption, negative asymmetry projections were observed. However, for clarity, the projections without assuming conditional universality are not shown in Fig. \ref{fig:DYp_COMPASS_Pred}. When comparing the projections of the \textit{proton}-DNN model with the predictions from \cite{Anselmino2017,Sun_2013,Echevarria_2014}, it is evident that the mean of the \textit{proton}-DNN model projection, in terms of $x_F$, is more consistent with the measured mean Sivers asymmetry in the experiment (refer to Figure 6 in \cite{COMPASS_DY_2017}). At the same time, the \textit{proton}-DNN model projection has relatively smaller uncertainty.  It is important to note that the predictions mentioned in the cited works were based on different $Q^2-$ evolution schemes, while the \textit{proton}-DNN model incorporates DGLAP evolution through LHAPDF \cite{LHAPDF6}. In \cite{Bury2021_JHEP}, only two data points from the COMPASS2017 \cite{COMPASS_DY_2017} data were included in their fits, resulting in larger uncertainties in the projected asymmetry values for the remaining data points when compared to the projections generated by the \textit{proton}-DNN model. The increasing trend of the projected DY Sivers asymmetries with respect to the $q_T$ kinematic variable in \cite{Bury2021_JHEP} is consistent with the projections presented in the middle-right plot of Fig. \ref{fig:DYp_COMPASS_Pred} generated by the \textit{proton}-DNN model, while in \cite{Echevarria_2021}, the corresponding trend exhibits a very small negative slope in relation to $q_T$.
 
\begin{figure}[h!]
    \centering
    \includegraphics[width=88mm]{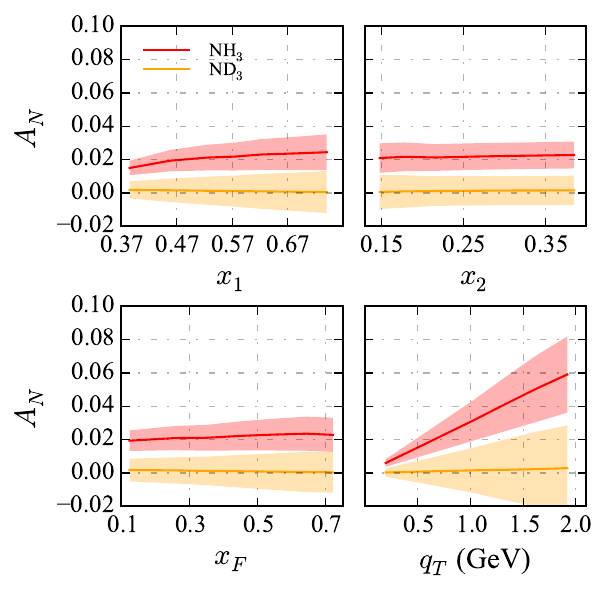}
    \caption{The \textit{proton}-DNN model (red) and the \textit{deuteron}-DNN model (orange) predictions including 68\% CL error-bands, for Sivers asymmetries in $x_1, x_2, x_F$, and $q_T$ kinematic projections for the SpinQuest DY kinematics \cite{SpinQuest_Proposal,SeaQuest_2022}.}
    \label{fig:DYpd_SQ_Pred}
\end{figure}

A non-zero sea-quark Sivers asymmetry is inferring that the sea quarks have non-zero orbital angular momentum. The \textit{proton}-DNN model predictions exhibit consistency with the non-zero Sivers asymmetry from the \textit{sea}-quarks, with higher precision compared to existing predictions \cite{Anselmino2017,Sun_2013,Echevarria_2014}, for the SpinQuest kinematics \cite{SpinQuest_Proposal,SpinQuest}. Additionally, in this work, we report our projections for the polarized Drell-Yan Sivers asymmetries for a deuteron target at the SpinQuest experiment, as shown in Figure \ref{fig:DYpd_SQ_Pred} by the orange-colored bands. The central lines in all kinematic projections of $x_1,x_2,x_F$, and $q_T$, yet consistent with zero. The \textit{proton}-DNN model predicts a positive slope with respect to $q_T$ for a \textit{proton} target as shown in the lower-right plot of Fig. \ref{fig:DYpd_SQ_Pred}. To date, with the exception of this work, no predictions have been made for the polarized DY Sivers asymmetry using a \textit{deuteron} target, which will be measured during the SpinQuest experiment. A noteworthy aspect of the forthcoming SpinQuest experiment is that, in addition to measuring the Sivers asymmetry from \textit{proton} and \textit{deuteron} targets, it will also ascertain the transversity distributions of both quarks and gluons, utilizing a tensor-polarized deuteron Spin 1 target, as proposed in \cite{Dustin_2022}.

\section{Systematic Studies}
\label{sys}
Using deep neural networks (DNNs) in global fits following a similar schema as laid out in this work may afford the loosening of some of the strict kinematic cuts to the experimental data while still preserving the validity of TMD factorization.  The methodology is still very new, but some distinct advantages are clear with our DNN approach compared to standard analytical fitting.

\subsection{Systematic Study of Data Cuts}

Neural networks, especially deep neural networks, have a profound capacity for learning complex patterns and relationships in data to model non-linear and high-dimensional functions without any prior information on the function dimensionality or form. When performing global fits the kinematic dependencies can be intricate and difficult to capture using conventional analytical approaches leading to the need to limit the data to achieve a decent fit. DNNs can adapt to these complexities and, in some cases, benefit from the lack of kinematic cuts, making the resulting model more robust and still sensitive to a range of kinematic variables.

The TMD factorization formula for the SIDIS hadronic tensor $W^{\mu\nu}$ was defined in \cite{JC2011,Aybat_2012},
\begin{align}
    W^{\mu\nu} &= \sum_f | \mathcal{H}_f(Q^2,\mu)|^{\mu \nu} \nonumber \\
    & \times \int d^2k_\perp p_\perp \delta^{(2)}\left(z_hk_\perp + p_\perp - p_{hT} \right) \nonumber \\
    & \times F_{f/N^{\uparrow}}(x,z_hk_\perp,S;\mu,\zeta_F) D_{h/f}(z_h,p_\perp;\mu,\zeta_D) \nonumber\\
    & + Y(p_{hT},Q^2),
    \label{TMD-factorization-formula}
\end{align}
where all non-perturbative information (the soft-part) is encoded in $F_{f/N^{\uparrow}}$ and the $D_{h/f}$ whereas the perturbatively calculable hard-part is denoted by $| \mathcal{H}_f(Q^2,\mu)|^{\mu \nu}$.
For preserving the validity of TMD factorization normally strict kinematic cuts are applied. However, applying stringent cuts to already limited experimental data reduces the statistical significance of the model and may lead to the loss of valuable information. We explore a range of possible cuts with the intention of preserving as much data as possible and taking care to not introduce bias by only selecting data that leads to better fits. The unique feature of DNNs to perform well even when trained on a broader range of data with complex correlations serves as a major advantage over fitting analytically. It then serves, whenever there is not a direct conflict with the necessary factorization theorem, to include data points from regions that might otherwise be excluded in traditional kinematic cuts allowing the DNN model to build implicit inclusion of the necessary corrections.

DNNs have the ability to implicitly capture higher-order effects that would be near impossible to obtain using a direct analytical fit unless the initial ansatz is very lucky or the function form has been proven to contain the necessary physics. TMD factorization relies on specific assumptions about the dominance of certain terms in the cross-section calculation.  One critical limitation is $q_{hT} \ll Q$ which is required for the derivation of the factorization property suitable for the case of relatively low transverse momentum.  In this factorization scheme approximations are made that have errors of order ($q_{hT}/Q$).  For $q_{hT}$ greater than $Q$, the conventional formalism, with integrated fragmentation functions, would no longer be valid.  This directly leads to the restriction of SIDIS data to a region where $p_{hT} \ll zQ$, which can severely reduce the available data. TMD factorization loses accuracy at large $q_{hT}$, with fractional errors characterized as $(q_{hT}/Q)^\alpha$.  The Collins and Soper approach \cite{JC2011} gives ($m/Q$) errors for the full range of $q_{hT}$ which treats the TMD term as a first approximation to the cross-section and allows for the application of a correction by applying an additive approximation from the ordinary collinear factorization.  Such corrections can be implicitly captured when training a DNN model over the full range of $p_{hT}$. This is the case even if the assumptions such corrections are based on are not relevant in all kinematic regions of the applicable data.  The only requirement is that TMD factorization does not break down at a rigid boundary but instead, remains valid but at a cost to the models' accuracy.  The scale of such errors can be numerically estimated so there is an advantage to using as much data as possible and then studying the systematic effects of certain limitations. This approach allows the implicit inclusion of higher-order effects in the DNN model, providing the means for a more comprehensive analysis that can handle a wider kinematic range without sacrificing factorization validity. However, careful consideration and validation are necessary to ensure the reliability of the model as well as accurate quantification of the systematic error as a function of $p_{hT}$.

The applicability of TMD factorization was ensured by applying cuts to SIDIS data based on various criteria in the literature. Although, the limit $q_T\ll Q$ \cite{JC1981,JC1985,Meng1996,JC2011,Scimemi_2018,Scimemi_2020,Bacchetta_2020,Bacchetta2022_unpolTMD} covers the conventional TMD factorization formalism, the large-$Q$ requirement is needed for suppressing the power corrections $\sim m^2/Q^2$ and $\sim \Lambda^2/Q^2$, where $\Lambda$ is a general nonperturbative scale of QCD. Since $m$ and $\Lambda$ are  $\sim 1$ GeV, $Q > 2$ GeV condition was applied on top of the $\delta = p_{hT}/(zQ) \leq 0.3$ condition in \cite{Bury2021_JHEP,Bury2021_PRL} although the phenomenological region for $\delta$ is $0.2-0.3$ in order for the TMD factorization to be valid. It is practically well known by the effort on global fits, that accommodating more data points is a bigger challenge when using a smaller lower bound for $\delta$. Therefore, a more conservative limit $q_T/Q < 0.75$ was used in \cite{Echevarria_2021} to retain a large enough data set to perform a meaningful fit. On the other hand, a combination of cuts: $Q^2 > 1.63$ GeV$^2$, $0.2 <z< 0.6$, $0.2<p_{hT}<0.9$ GeV was applied in \cite{Cammarota_2020}, and also discussed in detail in \cite{Boussarie_2023} with the standpoint that, data which satisfy $p_{hT}\ll Q$ may not satisfy $q_T \ll Q$ depending on the value of $z_h$ because $q_T \simeq p_{hT}/z_h$, and therefore be difficult to describe in a TMD approach. Not only in Sivers function extractions but this is also the case for other TMDs extractions \cite{Bhattacharya_2022}.

In this exploratory effort with DNNs, such later-mentioned power corrections are not directly imposed. In addition to the data basic cut $Q^2 > 1$ GeV$^2$ we performed $Q^2 > 2$ GeV$^2$ and $p_{hT} < zQ$ cuts separately with the proton-DNN model to understand the impact on the extracted Sivers functions. In addition, a dedicated DNN fit has been performed with the JAM20 \cite{Cammarota_2020} cuts: i.e. $Q^2 > 1.63$ GeV$^2$, $0.2 <z< 0.6$, $0.2<p_{hT}<0.9$ GeV as a demonstration of impact from such a selective combination of cuts on the extracted Sivers functions. 

\begin{figure}[h!!!]
    \centering
    \includegraphics[width=87mm]{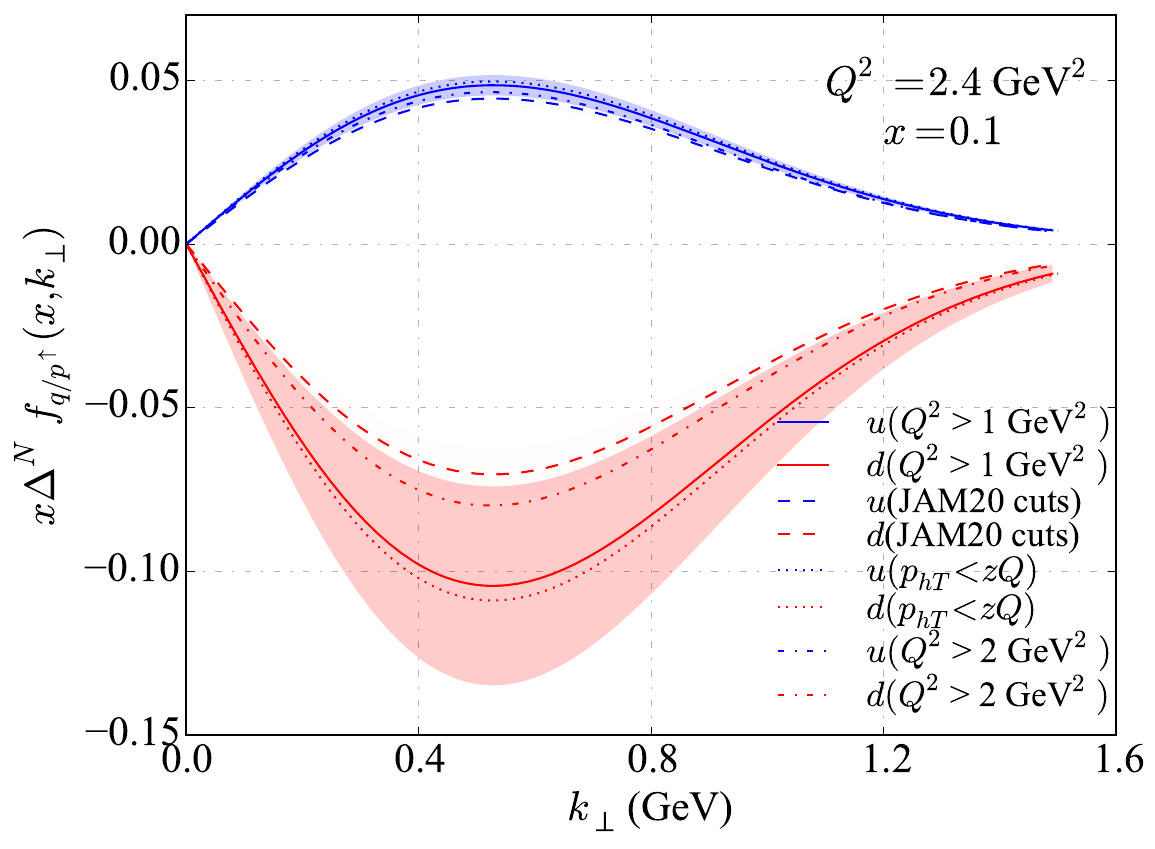} \\
    \caption{Solid lines with light-band represent the $u$ (in blue), $d$ (in red) Sivers functions using the cut $Q^2 > 1$ GeV$^2$.  These resulting DNN models made from the cuts from all tests are also shown.}
    \label{fig:Cuts_Comparison}
\end{figure}

\begin{figure}[h!!!]
    \centering
    \includegraphics[width=87mm]{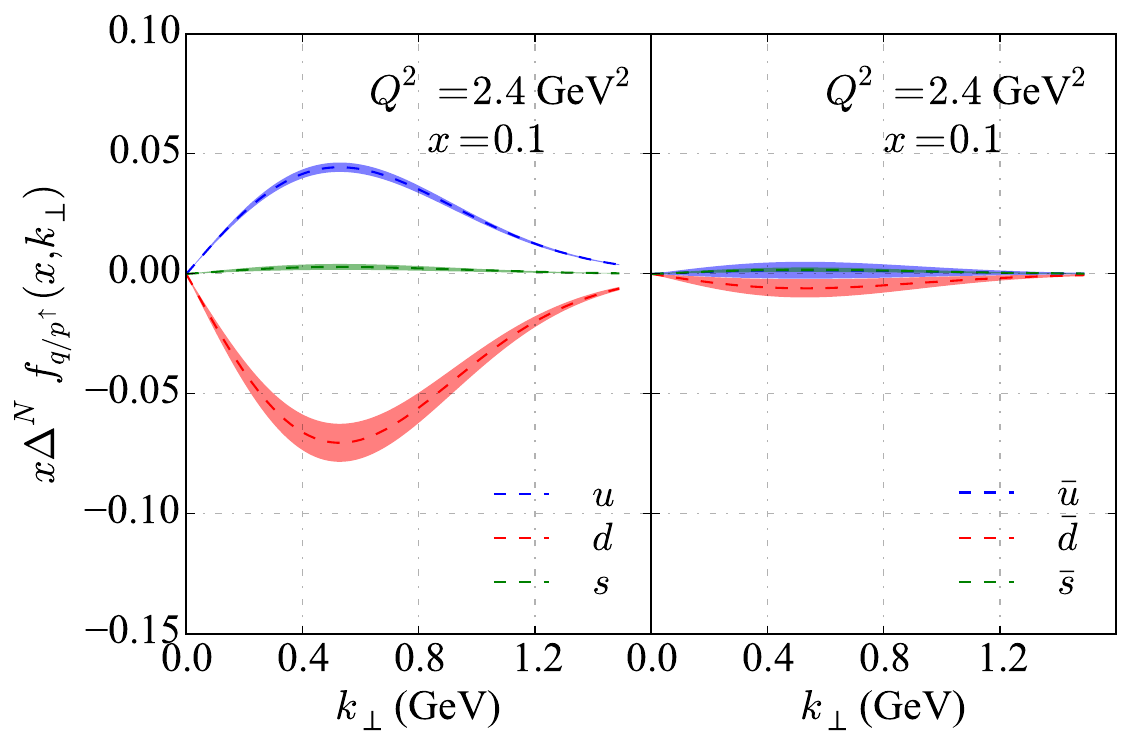} \\
    \caption{Sivers functions from a retrained DNN model using the cuts \cite{JAM21} to the data demonstrating that being selective with the data can reduce the error bands of the fit but may also add an unintentional bias.}
    \label{fig:Sivers_with_JAM_cuts}
\end{figure}

The results are plotted in Fig. \ref{fig:Cuts_Comparison} only representing the Sivers functions for $u,d$ flavors. In summary, all cuts analyzed which respect $Q^2 > 2$ GeV$^2$ and $p_{hT} < zQ$, except the JAM20 \cite{Cammarota_2020} cuts, are consistent with the DNN model that only contains the $Q^2 > 1$ GeV$^2$ cut which for this data set is no cuts at all.  Note that $Q^2 > 1$ is the recommended generic cut in \cite{JC2011} assuming the application of corrections and error estimates for increasing $q_T$.  The deviation measured in this study of cuts that is strictly dependent on $q_T$ is only of $\sim 2\%$ so it is clearly advantageous to incorporate a wide range of $q_T$ to ensure that there is implicit inclusion of these corrective contributions to the hadronic tensor from the $Y$-term built into the DNN model.  The Fig. \ref{fig:Sivers_with_JAM_cuts} shows the resultant Sivers functions from proton-DNN for all six light quark flavors with relatively small uncertainty bands caused by the selection of cuts in JAM20 \cite{Cammarota_2020}. Note that the uncertainty band represents the statistical component with 68\% CL from 1000 replica models. 

\subsection{Systematic Study on Choice of $h(k_\perp)$}

\begin{figure}[h!!!]
    \centering
    \includegraphics[width=87mm]{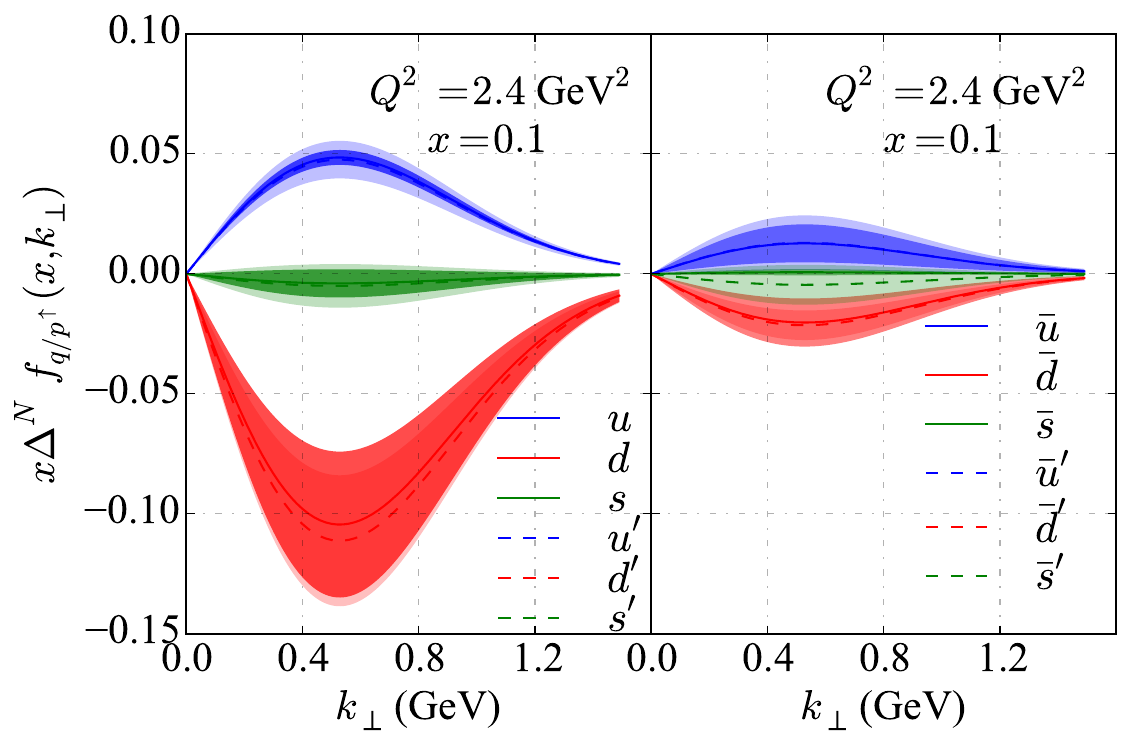} \\
    \caption{Using two different $h(k_\perp)$. Solid line with dark-band represents the Sivers functions with $h(k_{\perp}) = \sqrt{2e}\frac{k_{\perp}}{m_1}e^{-k_{\perp}^2/m_1^2}$, whereas the dashed line with light-band represents the Sivers functions with $h(k_{\perp}) = \frac{2 k_{\perp}m_1}{m_1^2 + k_\perp^2}$. }
    \label{fig:Siv_at_different_hk}
\end{figure}

In the original framework \cite{Anselmino2017}, the Sivers function is written as the factorized form, Eq. \ref{Siv_func}, where $h(k_\perp)$ is understood to be of an unknown form that is simply postulated by the authors (in \cite{Anselmino2017}) based on the assumed kinematic response. Indeed, the analytical expression of $h(k_\perp)$ has been treated with various types of ansatz and mostly with the Gaussian type parameterization \cite{Anselmino_2005_April,Anselmino2017,bachetta2022,Cammarota_2020,Bury2021_JHEP}. It's also entirely possible that this term has nothing to do with the proper theoretical treatment as suggested in \cite{Aybat_2012}.  When a term in the factorized TMD expression is required to manage some kinematic behavior but has not been formally derived, the DNN analysis presented can be particularly useful as it allows for a high-quality fit despite the dependence of $h(k_\perp)$.  After the determination of the other terms using the DNN, like $\mathcal{N}_q(x)$ in this case, the terms can then be separated and studied independently to determine the interpretation.  If $h(k_\perp)$ or any of the multiplicative terms are biases then the DNN can be used to compensate for the bias.  This is done by building an architecture that has orders of magnitude more parameters (usually thousands) than the expression in question, $h(k_\perp)$ in this case.  This effectively reduces the weight of $h(k_\perp)$ in the fit.  This can be done progressively and systematically with intentional control of how much each term contributes to the fit.  In the fit performed for this analysis, we build the DNN to optimize the loss directly so the contribution of $h(k_\perp)$ to the final fit is small though the DNN becomes specialized to the particular $h(k_\perp)$.  In this way, the DNN directly mitigates any type of incorrect ansatz if the final number DNN parameters is very large with respect to $h(k_\perp)$.
This can be done while still rendering a high-quality numerical model encoding all information available in the data.  The DNN term becomes uniquely parameterized to account for the biases while the resulting combination of terms in the model remains largely unbiased, even though each individual term may not be independently meaningful.  It should be noted that for our analysis the choice of having the DNN model $\mathcal{N}_q(x)$rather than $\mathcal{N}_q(x)h(k_\perp)$ or some other choice is arbitrary and done purely as an exercise to demonstrate the flexibility of the technique. To study the systematic variation of the choice of $h(k_\perp)$, we used the same candidate functions for $h(k_\perp)$ that were also used in \cite{Anselmino_2005_April}. Those are,
\begin{align}
    h(k_{\perp}) &= \sqrt{2e}\frac{k_{\perp}}{m_1}e^{-k_{\perp}^2/m_1^2},
\end{align}
and
\begin{align}
    h(k_{\perp}) = \frac{2 k_{\perp}m_1}{m_1^2 + k_\perp^2}.
\end{align}
The model must be trained separately with each $h(k_\perp)$ creating entirely different models for $\mathcal{N}_q(x)$ that result in the same Sivers function as shown in Fig. \ref{fig:Siv_at_different_hk}. The solid line with dark band represents the Sivers functions with $h(k_{\perp}) = \sqrt{2e}\frac{k_{\perp}}{m_1}e^{-k_{\perp}^2/m_1^2}$, whereas the dashed line with light-band represents the Sivers functions with $h(k_{\perp}) = \frac{2 k_{\perp}m_1}{m_1^2 + k_\perp^2}$. It is clear that the DNN $\mathcal{N}_q(x)$ is capable of incorporating both types of $h(k_\perp)$ without affecting the Sivers functions in the final model as well as the asymmetries (with deviation less than 1\%).

\section{The 3D Tomography of Proton}
\label{tom}

\begin{figure}[h!]
    \centering
    \includegraphics[width=94mm]{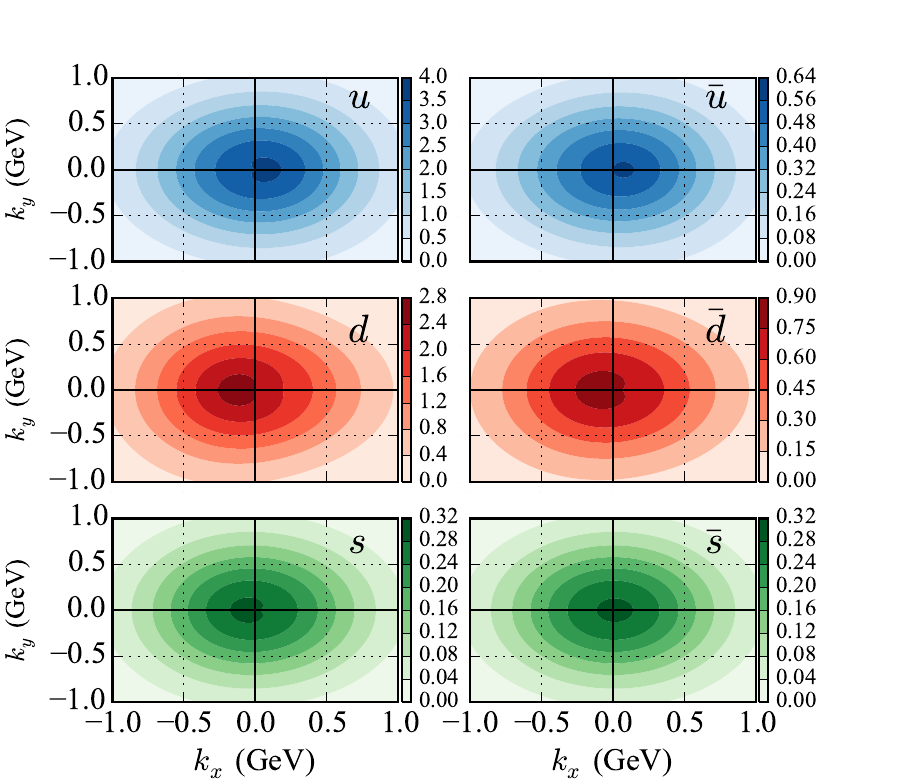}
    \caption{Quark density distributions $\rho^a_{p\uparrow}$ from the \textit{proton}-DNN model (average of 1000 replicas ) for the light quark flavor $a = \{u,\bar{u},d, \bar{d}, s, \bar{s} \}$ inside a proton polarized along the $+y$ direction and moving towards the reader, as a function of $(k_x,k_y)$ at $x=0.1$ and $Q^2=2.4$ GeV$^2$.}
    \label{fig:3D_density}
\end{figure}

The TMD density of unpolarized quarks inside a proton polarized in the $\hat{y}$-direction can be graphically represented using the relation \cite{bachetta2022,Bury2021_JHEP},

\begin{align}
    \rho^a_{p\uparrow}(x,k_x,k_y;Q^2) &= f_1^a (x, k_{\perp}^2; Q^2) - \frac{k_x}{m_p} f_{1T}^{\perp a} (x, k_{\perp}^2; Q^2),
    \label{eq:density_dist}
\end{align}

where $k_{\perp}$ is a two-dimensional vector $(k_x,k_y)$, and the unpolarized TMD and the Sivers function for quark-flavor $a$ are respectively represented as $f_1^a (x, k_\perp^2; Q^2)$, and $ f_{1T}^{\perp a} (x, k_{\perp}^2; Q^2)$. The corresponding quark density distributions from our \textit{proton}-DNN model for all light quark flavors in $SU(3)_{\text{flavor}}$ at $x=0.1$ and $Q^2=2.4$ GeV$^2$ are shown in Fig. \ref{fig:3D_density}. The observed shifts in each quark flavor are linked to the correlation between the OAM of quarks and the spin of the proton. The results shown in Fig. \ref{fig:3D_density} provide evidence of non-zero OAM in the wave function of the proton's valence and sea quarks. The \textit{proton}-DNN model calculations for the $u$ and $d$ quarks are similar to those reported in \cite{Bury2021_JHEP, bachetta2022}, where the distortion has a positive shift for the $u$-quark and a negative shift for the $d$-quark with respect to the $+x$ direction. From the results in Fig. \ref{fig:3D_density}, the \textit{proton}-DNN model demonstrates that a virtual photon traveling towards a polarized proton ``sees" an enhancement of the quark distribution, in particular more $u,\bar{u}$-quarks to its right-hand side and more $d,\bar{d}$-quarks to its left-hand side in the momentum space. Moreover, the resultant shifts for $\bar{u},s$ quarks from the \textit{proton}-DNN model are also in agreement with \cite{Bury2021_JHEP}. In the low-x region, the momentum space quark density becomes almost symmetric \cite{bachetta2022}, and it indicates that the Sivers effect becomes smaller and the corresponding experimentally observed asymmetry is small. 

The forthcoming data from Jefferson Lab at 12 GeV, Fermilab SpinQuest experiment, and the anticipated future data from the Electron-Ion Collider \cite{EIC_Boer_2011,EIC_Accardi_2016,EIC_Abdul_2022}, along with their extensive kinematic coverage, are expected to provide invaluable insights into the 3D structure of the nucleon. Obtaining a model-independent estimate of quark angular momentum requires  parton distributions that simultaneously depend on both momentum and position \cite{Ji_1997,Lorce_2012,Radici_2014,Stefanis_2017}. In addition to experimental observations, lattice QCD (LQCD) computations provide a valuable tool for QCD phenomenology from first principles. For instance, LQCD has been utilized to investigate the Sivers effect and other TMD observables at different pion masses \cite{Yoon_2017} as well as the generalized parton distribution at the physical pion mass \cite{Lin_2021}. Additionally, LQCD results on the Collins-Soper kernel over a range of $b_T$ (the Fourier transform of the transverse momentum) are useful for global fits of TMD observables from different processes \cite{Shanahan_2020}. In this way, LQCD could complement the experimental data and open up an avenue to enhance the DNN method to explore the 3D structure of nucleons more directly.

\section{Exploring Evolution}
\label{evol}

\begin{figure}[h!!!]
    \centering
    \includegraphics[width=87mm]{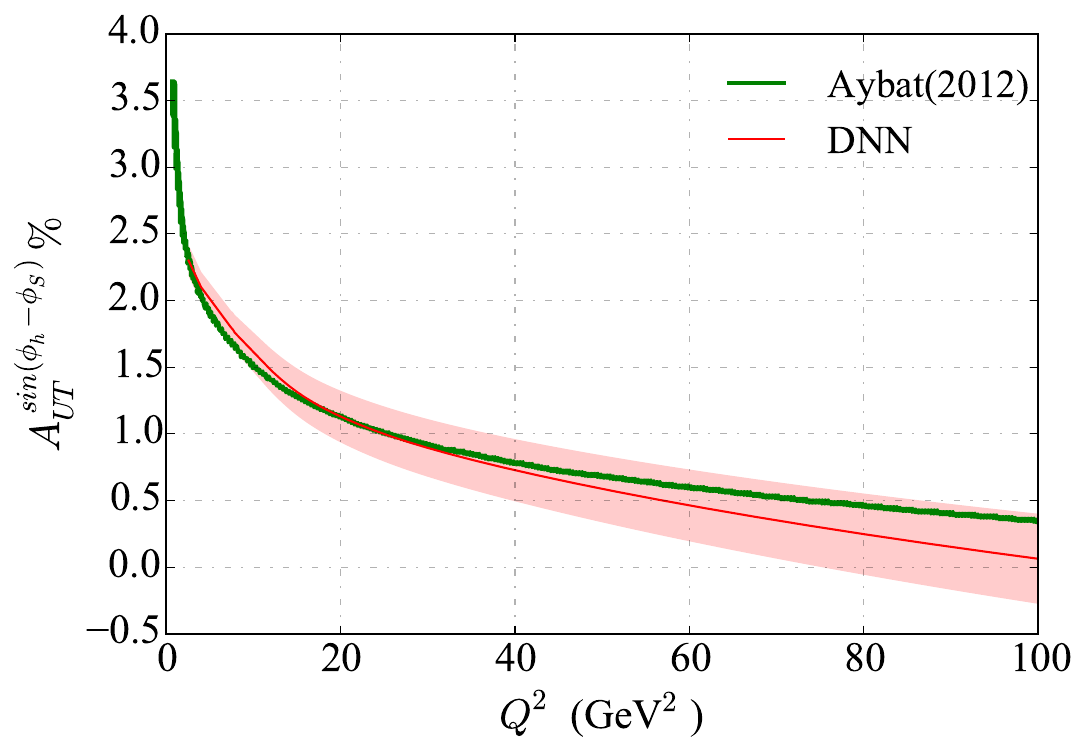} \\
    \caption{The Sivers asymmetry evolution in $Q^2$ (compared to the result from \cite{Aybat_2012}. The red-colored solid line and the band represent the mean and standard deviation of the $A_{UT}^{\sin (\phi_h - \phi_S)}$ from 1000 replica models of the proton-DNN at $x=0.12,\; z=0.32, \; p_{hT}=0.14$ GeV.}
    \label{fig:Q2_evolution}
\end{figure}

The solution of the TMD evolution equations \cite{JC2011,Scimemi_2020},
\begin{align}
    \mu^2 \frac{dF(x,b;\mu,\zeta)}{d\mu^2} &= \frac{\gamma_F(\mu,\zeta)}{2} F (x,b; \mu, \zeta) \\
    \zeta \frac{ F (x,b; \mu, \zeta)}{d\zeta} &= -\mathcal{D}(b,\mu) F (x,b; \mu, \zeta)
\end{align}
can be written as the following simplified form in terms of the Fourier transform of $k_\perp$ (i.e., $b$) \cite{Bury2021_JHEP} where $F$ can be any TMD distribution
\begin{align}
    F (x,b; \mu, \zeta) &= \left( \frac{\zeta}{\zeta_\mu (b)}\right)^{-\mathcal{D}(b,\mu)}F(x,b),
\end{align}
and $\mathcal{D}(b)$ is the nonperturbative Collins-Soper kernel. Also in the literature, these scales were generally selected as 
\begin{align}
    \mu \sim Q, \quad \quad \zeta_F\zeta_D\sim Q^4, \quad \quad \mu^2 = \zeta^2  = Q^2
\end{align}
\cite{Bury2021_JHEP,JC2011,Aybat_2012,Aybat_2012Feb,Echevarria_2014}, and the global fits have been performed using some form of evolution factor as a function of the Collin-Soper kernel. Although the full analysis of incorporating TMD evolution from the DNN fit is beyond the scope of this work, a preliminary DNN fit has been performed by modifying the $\mathcal{N}_q(x)$ as $\mathcal{N}_q(x,Q^2)$ by adding a separate input node for $Q^2$ in addition to $x$. The Fig. \ref{fig:Q2_evolution} shows the percentage of the Sivers asymmetry ($A_{UT}^{\sin (\phi_h - \phi_S)}$) vs $Q^2$ (GeV$^2$) in comparison with \cite{Aybat_2012}. The preliminary version of the TMD evolution from DNN is in agreement with the observation in \cite{Aybat_2012} within 68\% CL (with 1000 replica models) regarding the suppression of the full asymmetry faster than $\sim 1/\sqrt{Q}$, but slower than $\sim 1/Q^2$. Moreover, the dynamic trend and consistency with the evolution behavior seen in \cite{Aybat_2012} indicates that a complete evolution treatment at large $Q$ may be a collective effect and worthy of a deeper investigation with our DNN approach.

As presented the Sivers function satisfies DGLAP evolution but not a TMD evolution.  We have however introduced the first steps of how this analysis could be performed.  However, in order to perform the best evolution analysis it would be best to start with no prior analytical ansatz.

\section{Conclusion and Discussion}
\label{conclusion}

In conclusion, this article demonstrates the effectiveness of using specialized Deep Neural Networks (DNNs) as part of a fit function used to obtain a global extraction of the Sivers function from transverse single-spin asymmetries experimental data. It has been clearly shown that these tools are incredibly flexible and can be used to build accurate models even under the condition that the factorized terms contain biases. By training the model with these terms present, the DNN can account for these biases, resulting in a largely unbiased final model. This approach enables the exploration of existing formalism and the testing of new phenomenology while managing and studying biases in individual terms. The proposed method can provide a means to minimize errors and ambiguity associated with the ill-defined expressions normally constructed to meet the factorization requirements.

Our proposed method leverages artificial intelligence (AI) to perform global fits and extract the Sivers distributions of unpolarized quarks in both polarized protons and neutrons. We use a generating function to ensure robustness and accuracy in the extraction process. Progressive improvement in the extracted information can be achieved by optimizing architecture and the corresponding hyperparameters using pseudodata generated in the same kinematic bins as the real data and then translating that improvement to real experimental data extraction. Our schema handles complex and sparse data effectively, with the generating function providing additional quality control and the means to quantify accuracy and precision.

As the first attempt to extract the Sivers function using AI techniques, we chose the $\mathcal{N}_q(x)$ parameterization of the deep neural net to incorporate all x-dependent features. This initial DNN extraction of $\mathcal{N}_q(x)$ uses SIDIS data from HERMES and COMPASS to build a Sivers function model which can then be used to make predictions for the Sivers asymmetries for both SIDIS and DY processes. The fitting method successfully extracts the Sivers function for valence quarks and light quark flavors in a flavor-independent manner. This investigation is exploratory, with the intention of developing tools and techniques that minimize error and maximize utility, which we hope to expand upon in further work.
The trained model predicts results consistent with experimental data, demonstrating its predictive capability. Projections for the sea quark Siver function are made for the SpinQuest experiment with polarized proton and deuteron targets.

There are two key challenges when using data to extract TMDs. One is the applicability of TMD factorization and the other is achieving meaningful fit results. An exploratory effort has been performed to address this with our schema along with a variety of data cuts. As a result, the DNN shows the variation of $q_T$ dependent cuts, resulting in small deviations in the extracted Sivers functions compared to the result with a wide range of $q_T$, and simultaneously following the TMD factorization theorem. We also analyze the TMD evolution of the Sivers asymmetry, which aligns with previous work and indicates the potential of the DNN approach. Further analysis in this direction is warranted.

Our method and techniques are intended to be simple and reproducible, highlighting the potential of AI for data analysis and information extraction. Cooperation between experimentalists, theorists, and the growing computational efforts is crucial for accelerating advancements in the field.  A global effort to standardize data collection, organization, and storage in an unbinned format with detailed covariance information must be developed to take full advantage of the AI tools now available. AI and its emerging technologies will continue to accelerate the progress of data-driven physics, but the speed of that progress largely depends on the level of foresight and cooperation within the community.

\newpage
\begin{acknowledgments}
This work was supported by the U. S. Department of Energy (DOE) contract DE-FG02-96ER40950. The authors would like to thank Micheal Wagman, Andrea Signori, Alexei Prokudin, and Filippo Delcarro for their insightful discussions. Additionally, the authors would like to acknowledge Bakur Parsamyan for providing the COMPASS data, Luciano Pappalardo for providing the HERMES data, and Sassot Rodolfo for providing the Fragmentation Functions.
\end{acknowledgments}

\bibliography{Refs.bib}

\providecommand{\noopsort}[1]{}\providecommand{\singleletter}[1]{#1}%
\begin{thebibliography}{85}%
\makeatletter
\providecommand \@ifxundefined [1]{%
 \@ifx{#1\undefined}
}%
\providecommand \@ifnum [1]{%
 \ifnum #1\expandafter \@firstoftwo
 \else \expandafter \@secondoftwo
 \fi
}%
\providecommand \@ifx [1]{%
 \ifx #1\expandafter \@firstoftwo
 \else \expandafter \@secondoftwo
 \fi
}%
\providecommand \natexlab [1]{#1}%
\providecommand \enquote  [1]{``#1''}%
\providecommand \bibnamefont  [1]{#1}%
\providecommand \bibfnamefont [1]{#1}%
\providecommand \citenamefont [1]{#1}%
\providecommand \href@noop [0]{\@secondoftwo}%
\providecommand \href [0]{\begingroup \@sanitize@url \@href}%
\providecommand \@href[1]{\@@startlink{#1}\@@href}%
\providecommand \@@href[1]{\endgroup#1\@@endlink}%
\providecommand \@sanitize@url [0]{\catcode `\\12\catcode `\$12\catcode
  `\&12\catcode `\#12\catcode `\^12\catcode `\_12\catcode `\%12\relax}%
\providecommand \@@startlink[1]{}%
\providecommand \@@endlink[0]{}%
\providecommand \url  [0]{\begingroup\@sanitize@url \@url }%
\providecommand \@url [1]{\endgroup\@href {#1}{\urlprefix }}%
\providecommand \urlprefix  [0]{URL }%
\providecommand \Eprint [0]{\href }%
\providecommand \doibase [0]{https://doi.org/}%
\providecommand \selectlanguage [0]{\@gobble}%
\providecommand \bibinfo  [0]{\@secondoftwo}%
\providecommand \bibfield  [0]{\@secondoftwo}%
\providecommand \translation [1]{[#1]}%
\providecommand \BibitemOpen [0]{}%
\providecommand \bibitemStop [0]{}%
\providecommand \bibitemNoStop [0]{.\EOS\space}%
\providecommand \EOS [0]{\spacefactor3000\relax}%
\providecommand \BibitemShut  [1]{\csname bibitem#1\endcsname}%
\let\auto@bib@innerbib\@empty
\bibitem [{\citenamefont {Sivers}(1990)}]{Sivers_1990}%
  \BibitemOpen
  \bibfield  {author} {\bibinfo {author} {\bibfnamefont {D.}~\bibnamefont
  {Sivers}},\ }\href {https://doi.org/10.1103/PhysRevD.41.83} {\bibfield
  {journal} {\bibinfo  {journal} {Phys. Rev. D}\ }\textbf {\bibinfo {volume}
  {41}},\ \bibinfo {pages} {83} (\bibinfo {year} {1990})}\BibitemShut {NoStop}%
\bibitem [{\citenamefont {Sivers}(1991)}]{Sivers_1991}%
  \BibitemOpen
  \bibfield  {author} {\bibinfo {author} {\bibfnamefont {D.}~\bibnamefont
  {Sivers}},\ }\href {https://doi.org/10.1103/PhysRevD.43.261} {\bibfield
  {journal} {\bibinfo  {journal} {Phys. Rev. D}\ }\textbf {\bibinfo {volume}
  {43}},\ \bibinfo {pages} {261} (\bibinfo {year} {1991})}\BibitemShut
  {NoStop}%
\bibitem [{\citenamefont {Anselmino}\ \emph
  {et~al.}(2009{\natexlab{a}})\citenamefont {Anselmino}, \citenamefont
  {Boglione}, \citenamefont {D'Alesio}, \citenamefont {Melis}, \citenamefont
  {Murgia},\ and\ \citenamefont {Prokudin}}]{Anselmino_DY_2009}%
  \BibitemOpen
  \bibfield  {author} {\bibinfo {author} {\bibfnamefont {M.}~\bibnamefont
  {Anselmino}}, \bibinfo {author} {\bibfnamefont {M.}~\bibnamefont {Boglione}},
  \bibinfo {author} {\bibfnamefont {U.}~\bibnamefont {D'Alesio}}, \bibinfo
  {author} {\bibfnamefont {S.}~\bibnamefont {Melis}}, \bibinfo {author}
  {\bibfnamefont {F.}~\bibnamefont {Murgia}},\ and\ \bibinfo {author}
  {\bibfnamefont {A.}~\bibnamefont {Prokudin}},\ }\href
  {https://doi.org/10.1103/PhysRevD.79.054010} {\bibfield  {journal} {\bibinfo
  {journal} {Phys. Rev. D}\ }\textbf {\bibinfo {volume} {79}},\ \bibinfo
  {pages} {054010} (\bibinfo {year} {2009}{\natexlab{a}})}\BibitemShut
  {NoStop}%
\bibitem [{\citenamefont {Anselmino}\ \emph {et~al.}(2017)\citenamefont
  {Anselmino}, \citenamefont {Boglione}, \citenamefont {D’Alesio},
  \citenamefont {Murgia},\ and\ \citenamefont {Prokudin}}]{Anselmino2017}%
  \BibitemOpen
  \bibfield  {author} {\bibinfo {author} {\bibfnamefont {M.}~\bibnamefont
  {Anselmino}}, \bibinfo {author} {\bibfnamefont {M.}~\bibnamefont {Boglione}},
  \bibinfo {author} {\bibfnamefont {U.}~\bibnamefont {D’Alesio}}, \bibinfo
  {author} {\bibfnamefont {F.}~\bibnamefont {Murgia}},\ and\ \bibinfo {author}
  {\bibfnamefont {A.}~\bibnamefont {Prokudin}},\ }\href
  {https://doi.org/10.1007/JHEP04(2017)046} {\bibfield  {journal} {\bibinfo
  {journal} {Journal of High Energy Physics}\ }\textbf {\bibinfo {volume}
  {2017}},\ \bibinfo {pages} {46} (\bibinfo {year} {2017})}\BibitemShut
  {NoStop}%
\bibitem [{\citenamefont {Cammarota}\ \emph {et~al.}(2020)\citenamefont
  {Cammarota}, \citenamefont {Gamberg}, \citenamefont {Kang}, \citenamefont
  {Miller}, \citenamefont {Pitonyak}, \citenamefont {Prokudin}, \citenamefont
  {Rogers},\ and\ \citenamefont {Sato}}]{Cammarota_2020}%
  \BibitemOpen
  \bibfield  {author} {\bibinfo {author} {\bibfnamefont {J.}~\bibnamefont
  {Cammarota}}, \bibinfo {author} {\bibfnamefont {L.}~\bibnamefont {Gamberg}},
  \bibinfo {author} {\bibfnamefont {Z.-B.}\ \bibnamefont {Kang}}, \bibinfo
  {author} {\bibfnamefont {J.~A.}\ \bibnamefont {Miller}}, \bibinfo {author}
  {\bibfnamefont {D.}~\bibnamefont {Pitonyak}}, \bibinfo {author}
  {\bibfnamefont {A.}~\bibnamefont {Prokudin}}, \bibinfo {author}
  {\bibfnamefont {T.~C.}\ \bibnamefont {Rogers}},\ and\ \bibinfo {author}
  {\bibfnamefont {N.}~\bibnamefont {Sato}} (\bibinfo {collaboration} {Jefferson
  Lab Angular Momentum (JAM) Collaboration}),\ }\href
  {https://doi.org/10.1103/PhysRevD.102.054002} {\bibfield  {journal} {\bibinfo
   {journal} {Phys. Rev. D}\ }\textbf {\bibinfo {volume} {102}},\ \bibinfo
  {pages} {054002} (\bibinfo {year} {2020})}\BibitemShut {NoStop}%
\bibitem [{\citenamefont {Aybat}\ \emph
  {et~al.}(2012{\natexlab{a}})\citenamefont {Aybat}, \citenamefont {Prokudin},\
  and\ \citenamefont {Rogers}}]{Aybat_2012}%
  \BibitemOpen
  \bibfield  {author} {\bibinfo {author} {\bibfnamefont {S.~M.}\ \bibnamefont
  {Aybat}}, \bibinfo {author} {\bibfnamefont {A.}~\bibnamefont {Prokudin}},\
  and\ \bibinfo {author} {\bibfnamefont {T.~C.}\ \bibnamefont {Rogers}},\
  }\href {https://doi.org/10.1103/PhysRevLett.108.242003} {\bibfield  {journal}
  {\bibinfo  {journal} {Phys. Rev. Lett.}\ }\textbf {\bibinfo {volume} {108}},\
  \bibinfo {pages} {242003} (\bibinfo {year} {2012}{\natexlab{a}})}\BibitemShut
  {NoStop}%
\bibitem [{\citenamefont {Bury}\ \emph
  {et~al.}(2021{\natexlab{a}})\citenamefont {Bury}, \citenamefont {Prokudin},\
  and\ \citenamefont {Vladimirov}}]{Bury2021_JHEP}%
  \BibitemOpen
  \bibfield  {author} {\bibinfo {author} {\bibfnamefont {M.}~\bibnamefont
  {Bury}}, \bibinfo {author} {\bibfnamefont {A.}~\bibnamefont {Prokudin}},\
  and\ \bibinfo {author} {\bibfnamefont {A.}~\bibnamefont {Vladimirov}},\
  }\href {https://doi.org/10.1007/JHEP05(2021)151} {\bibfield  {journal}
  {\bibinfo  {journal} {Journal of High Energy Physics}\ }\textbf {\bibinfo
  {volume} {2021}},\ \bibinfo {pages} {151} (\bibinfo {year}
  {2021}{\natexlab{a}})}\BibitemShut {NoStop}%
\bibitem [{\citenamefont {Echevarria}\ \emph {et~al.}(2014)\citenamefont
  {Echevarria}, \citenamefont {Idilbi}, \citenamefont {Kang},\ and\
  \citenamefont {Vitev}}]{Echevarria_2014}%
  \BibitemOpen
  \bibfield  {author} {\bibinfo {author} {\bibfnamefont {M.~G.}\ \bibnamefont
  {Echevarria}}, \bibinfo {author} {\bibfnamefont {A.}~\bibnamefont {Idilbi}},
  \bibinfo {author} {\bibfnamefont {Z.-B.}\ \bibnamefont {Kang}},\ and\
  \bibinfo {author} {\bibfnamefont {I.}~\bibnamefont {Vitev}},\ }\href
  {https://doi.org/10.1103/PhysRevD.89.074013} {\bibfield  {journal} {\bibinfo
  {journal} {Phys. Rev. D}\ }\textbf {\bibinfo {volume} {89}},\ \bibinfo
  {pages} {074013} (\bibinfo {year} {2014})}\BibitemShut {NoStop}%
\bibitem [{\citenamefont {Echevarria}\ \emph {et~al.}(2021)\citenamefont
  {Echevarria}, \citenamefont {Kang},\ and\ \citenamefont
  {Terry}}]{Echevarria_2021}%
  \BibitemOpen
  \bibfield  {author} {\bibinfo {author} {\bibfnamefont {M.~G.}\ \bibnamefont
  {Echevarria}}, \bibinfo {author} {\bibfnamefont {Z.-B.}\ \bibnamefont
  {Kang}},\ and\ \bibinfo {author} {\bibfnamefont {J.}~\bibnamefont {Terry}},\
  }\href {https://doi.org/10.1007/JHEP01(2021)126} {\bibfield  {journal}
  {\bibinfo  {journal} {Journal of High Energy Physics}\ }\textbf {\bibinfo
  {volume} {2021}},\ \bibinfo {pages} {126} (\bibinfo {year}
  {2021})}\BibitemShut {NoStop}%
\bibitem [{\citenamefont {Kang}\ and\ \citenamefont {Qiu}(2009)}]{Kang2009}%
  \BibitemOpen
  \bibfield  {author} {\bibinfo {author} {\bibfnamefont {Z.-B.}\ \bibnamefont
  {Kang}}\ and\ \bibinfo {author} {\bibfnamefont {J.-W.}\ \bibnamefont {Qiu}},\
  }\href {https://doi.org/10.1103/PhysRevLett.103.172001} {\bibfield  {journal}
  {\bibinfo  {journal} {Phys. Rev. Lett.}\ }\textbf {\bibinfo {volume} {103}},\
  \bibinfo {pages} {172001} (\bibinfo {year} {2009})}\BibitemShut {NoStop}%
\bibitem [{\citenamefont {Bury}\ \emph
  {et~al.}(2021{\natexlab{b}})\citenamefont {Bury}, \citenamefont {Prokudin},\
  and\ \citenamefont {Vladimirov}}]{Bury2021_PRL}%
  \BibitemOpen
  \bibfield  {author} {\bibinfo {author} {\bibfnamefont {M.}~\bibnamefont
  {Bury}}, \bibinfo {author} {\bibfnamefont {A.}~\bibnamefont {Prokudin}},\
  and\ \bibinfo {author} {\bibfnamefont {A.}~\bibnamefont {Vladimirov}},\
  }\href {https://doi.org/10.1103/PhysRevLett.126.112002} {\bibfield  {journal}
  {\bibinfo  {journal} {Phys. Rev. Lett.}\ }\textbf {\bibinfo {volume} {126}},\
  \bibinfo {pages} {112002} (\bibinfo {year} {2021}{\natexlab{b}})}\BibitemShut
  {NoStop}%
\bibitem [{\citenamefont {Scimemi}\ and\ \citenamefont
  {Vladimirov}(2020)}]{Scimemi_2020}%
  \BibitemOpen
  \bibfield  {author} {\bibinfo {author} {\bibfnamefont {I.}~\bibnamefont
  {Scimemi}}\ and\ \bibinfo {author} {\bibfnamefont {A.}~\bibnamefont
  {Vladimirov}},\ }\href {https://doi.org/10.1007/JHEP06(2020)137} {\bibfield
  {journal} {\bibinfo  {journal} {Journal of High Energy Physics}\ }\textbf
  {\bibinfo {volume} {2020}},\ \bibinfo {pages} {137} (\bibinfo {year}
  {2020})}\BibitemShut {NoStop}%
\bibitem [{\citenamefont {Collins}\ and\ \citenamefont {Soper}(1981)}]{JC1981}%
  \BibitemOpen
  \bibfield  {author} {\bibinfo {author} {\bibfnamefont {J.~C.}\ \bibnamefont
  {Collins}}\ and\ \bibinfo {author} {\bibfnamefont {D.~E.}\ \bibnamefont
  {Soper}},\ }\href
  {https://doi.org/https://doi.org/10.1016/0550-3213(81)90339-4} {\bibfield
  {journal} {\bibinfo  {journal} {Nuclear Physics B}\ }\textbf {\bibinfo
  {volume} {193}},\ \bibinfo {pages} {381} (\bibinfo {year}
  {1981})}\BibitemShut {NoStop}%
\bibitem [{\citenamefont {Collins}\ and\ \citenamefont {Soper}(1983)}]{JC1983}%
  \BibitemOpen
  \bibfield  {author} {\bibinfo {author} {\bibfnamefont {J.~C.}\ \bibnamefont
  {Collins}}\ and\ \bibinfo {author} {\bibfnamefont {D.~E.}\ \bibnamefont
  {Soper}},\ }\href
  {https://doi.org/https://doi.org/10.1016/0550-3213(83)90235-3} {\bibfield
  {journal} {\bibinfo  {journal} {Nuclear Physics B}\ }\textbf {\bibinfo
  {volume} {213}},\ \bibinfo {pages} {545} (\bibinfo {year}
  {1983})}\BibitemShut {NoStop}%
\bibitem [{\citenamefont {Collins}\ \emph {et~al.}(1985)\citenamefont
  {Collins}, \citenamefont {Soper},\ and\ \citenamefont {Sterman}}]{JC1985}%
  \BibitemOpen
  \bibfield  {author} {\bibinfo {author} {\bibfnamefont {J.~C.}\ \bibnamefont
  {Collins}}, \bibinfo {author} {\bibfnamefont {D.~E.}\ \bibnamefont {Soper}},\
  and\ \bibinfo {author} {\bibfnamefont {G.}~\bibnamefont {Sterman}},\ }\href
  {https://doi.org/https://doi.org/10.1016/0550-3213(85)90479-1} {\bibfield
  {journal} {\bibinfo  {journal} {Nuclear Physics B}\ }\textbf {\bibinfo
  {volume} {250}},\ \bibinfo {pages} {199} (\bibinfo {year}
  {1985})}\BibitemShut {NoStop}%
\bibitem [{\citenamefont {Collins}(2011)}]{JC2011}%
  \BibitemOpen
  \bibfield  {author} {\bibinfo {author} {\bibfnamefont {J.}~\bibnamefont
  {Collins}},\ }\href {https://doi.org/DOI: 10.1017/CBO9780511975592} {\emph
  {\bibinfo {title} {Foundations of Perturbative QCD}}}\ (\bibinfo  {publisher}
  {Cambridge University Press},\ \bibinfo {year} {2011})\BibitemShut {NoStop}%
\bibitem [{\citenamefont {Meng}\ \emph {et~al.}(1992)\citenamefont {Meng},
  \citenamefont {Olness},\ and\ \citenamefont {Soper}}]{Meng1992}%
  \BibitemOpen
  \bibfield  {author} {\bibinfo {author} {\bibfnamefont {R.}~\bibnamefont
  {Meng}}, \bibinfo {author} {\bibfnamefont {F.~I.}\ \bibnamefont {Olness}},\
  and\ \bibinfo {author} {\bibfnamefont {D.~E.}\ \bibnamefont {Soper}},\ }\href
  {https://doi.org/https://doi.org/10.1016/0550-3213(92)90230-9} {\bibfield
  {journal} {\bibinfo  {journal} {Nuclear Physics B}\ }\textbf {\bibinfo
  {volume} {371}},\ \bibinfo {pages} {79} (\bibinfo {year} {1992})}\BibitemShut
  {NoStop}%
\bibitem [{\citenamefont {Ji}\ \emph {et~al.}(2004)\citenamefont {Ji},
  \citenamefont {Ma},\ and\ \citenamefont {Yuan}}]{Ji2004}%
  \BibitemOpen
  \bibfield  {author} {\bibinfo {author} {\bibfnamefont {X.}~\bibnamefont
  {Ji}}, \bibinfo {author} {\bibfnamefont {J.-P.}\ \bibnamefont {Ma}},\ and\
  \bibinfo {author} {\bibfnamefont {F.}~\bibnamefont {Yuan}},\ }\href
  {https://doi.org/https://doi.org/10.1016/j.physletb.2004.07.026} {\bibfield
  {journal} {\bibinfo  {journal} {Physics Letters B}\ }\textbf {\bibinfo
  {volume} {597}},\ \bibinfo {pages} {299} (\bibinfo {year}
  {2004})}\BibitemShut {NoStop}%
\bibitem [{\citenamefont {Ji}\ \emph {et~al.}(2005{\natexlab{a}})\citenamefont
  {Ji}, \citenamefont {Ma},\ and\ \citenamefont {Yuan}}]{Ji2005}%
  \BibitemOpen
  \bibfield  {author} {\bibinfo {author} {\bibfnamefont {X.}~\bibnamefont
  {Ji}}, \bibinfo {author} {\bibfnamefont {J.-P.}\ \bibnamefont {Ma}},\ and\
  \bibinfo {author} {\bibfnamefont {F.}~\bibnamefont {Yuan}},\ }\href
  {https://doi.org/10.1103/PhysRevD.71.034005} {\bibfield  {journal} {\bibinfo
  {journal} {Phys. Rev. D}\ }\textbf {\bibinfo {volume} {71}},\ \bibinfo
  {pages} {034005} (\bibinfo {year} {2005}{\natexlab{a}})}\BibitemShut
  {NoStop}%
\bibitem [{\citenamefont {Ji}\ \emph {et~al.}(2005{\natexlab{b}})\citenamefont
  {Ji}, \citenamefont {Ma},\ and\ \citenamefont {Yuan}}]{XJi2005}%
  \BibitemOpen
  \bibfield  {author} {\bibinfo {author} {\bibfnamefont {X.}~\bibnamefont
  {Ji}}, \bibinfo {author} {\bibfnamefont {J.-P.}\ \bibnamefont {Ma}},\ and\
  \bibinfo {author} {\bibfnamefont {F.}~\bibnamefont {Yuan}},\ }\href
  {https://doi.org/10.1088/1126-6708/2005/07/020} {\bibfield  {journal}
  {\bibinfo  {journal} {Journal of High Energy Physics}\ }\textbf {\bibinfo
  {volume} {2005}},\ \bibinfo {pages} {020} (\bibinfo {year}
  {2005}{\natexlab{b}})}\BibitemShut {NoStop}%
\bibitem [{\citenamefont {Echevarría}\ \emph {et~al.}(2012)\citenamefont
  {Echevarría}, \citenamefont {Idilbi},\ and\ \citenamefont
  {Scimemi}}]{Eche2012}%
  \BibitemOpen
  \bibfield  {author} {\bibinfo {author} {\bibfnamefont {M.~G.}\ \bibnamefont
  {Echevarría}}, \bibinfo {author} {\bibfnamefont {A.}~\bibnamefont
  {Idilbi}},\ and\ \bibinfo {author} {\bibfnamefont {I.}~\bibnamefont
  {Scimemi}},\ }\href {https://doi.org/10.1007/JHEP07(2012)002} {\bibfield
  {journal} {\bibinfo  {journal} {Journal of High Energy Physics}\ }\textbf
  {\bibinfo {volume} {2012}},\ \bibinfo {pages} {2} (\bibinfo {year}
  {2012})}\BibitemShut {NoStop}%
\bibitem [{\citenamefont {Arnold}\ \emph {et~al.}(2009)\citenamefont {Arnold},
  \citenamefont {Metz},\ and\ \citenamefont {Schlegel}}]{Arnold_2009}%
  \BibitemOpen
  \bibfield  {author} {\bibinfo {author} {\bibfnamefont {S.}~\bibnamefont
  {Arnold}}, \bibinfo {author} {\bibfnamefont {A.}~\bibnamefont {Metz}},\ and\
  \bibinfo {author} {\bibfnamefont {M.}~\bibnamefont {Schlegel}},\ }\href
  {https://doi.org/10.1103/PhysRevD.79.034005} {\bibfield  {journal} {\bibinfo
  {journal} {Phys. Rev. D}\ }\textbf {\bibinfo {volume} {79}},\ \bibinfo
  {pages} {034005} (\bibinfo {year} {2009})}\BibitemShut {NoStop}%
\bibitem [{\citenamefont {Anselmino}\ \emph
  {et~al.}(2005{\natexlab{a}})\citenamefont {Anselmino}, \citenamefont
  {Boglione}, \citenamefont {D'Alesio}, \citenamefont {Kotzinian},
  \citenamefont {Murgia},\ and\ \citenamefont {Prokudin}}]{Anselmino2005}%
  \BibitemOpen
  \bibfield  {author} {\bibinfo {author} {\bibfnamefont {M.}~\bibnamefont
  {Anselmino}}, \bibinfo {author} {\bibfnamefont {M.}~\bibnamefont {Boglione}},
  \bibinfo {author} {\bibfnamefont {U.}~\bibnamefont {D'Alesio}}, \bibinfo
  {author} {\bibfnamefont {A.}~\bibnamefont {Kotzinian}}, \bibinfo {author}
  {\bibfnamefont {F.}~\bibnamefont {Murgia}},\ and\ \bibinfo {author}
  {\bibfnamefont {A.}~\bibnamefont {Prokudin}},\ }\href
  {https://doi.org/10.1103/PhysRevD.72.094007} {\bibfield  {journal} {\bibinfo
  {journal} {Phys. Rev. D}\ }\textbf {\bibinfo {volume} {72}},\ \bibinfo
  {pages} {094007} (\bibinfo {year} {2005}{\natexlab{a}})}\BibitemShut
  {NoStop}%
\bibitem [{\citenamefont {Collins}\ \emph
  {et~al.}(2006{\natexlab{a}})\citenamefont {Collins}, \citenamefont {Efremov},
  \citenamefont {Goeke}, \citenamefont {Menzel}, \citenamefont {Metz},\ and\
  \citenamefont {Schweitzer}}]{Collins2006}%
  \BibitemOpen
  \bibfield  {author} {\bibinfo {author} {\bibfnamefont {J.~C.}\ \bibnamefont
  {Collins}}, \bibinfo {author} {\bibfnamefont {A.~V.}\ \bibnamefont
  {Efremov}}, \bibinfo {author} {\bibfnamefont {K.}~\bibnamefont {Goeke}},
  \bibinfo {author} {\bibfnamefont {S.}~\bibnamefont {Menzel}}, \bibinfo
  {author} {\bibfnamefont {A.}~\bibnamefont {Metz}},\ and\ \bibinfo {author}
  {\bibfnamefont {P.}~\bibnamefont {Schweitzer}},\ }\href
  {https://doi.org/10.1103/PhysRevD.73.014021} {\bibfield  {journal} {\bibinfo
  {journal} {Phys. Rev. D}\ }\textbf {\bibinfo {volume} {73}},\ \bibinfo
  {pages} {014021} (\bibinfo {year} {2006}{\natexlab{a}})}\BibitemShut
  {NoStop}%
\bibitem [{\citenamefont {Vogelsang}\ and\ \citenamefont
  {Yuan}(2005)}]{Vogelsang_2005}%
  \BibitemOpen
  \bibfield  {author} {\bibinfo {author} {\bibfnamefont {W.}~\bibnamefont
  {Vogelsang}}\ and\ \bibinfo {author} {\bibfnamefont {F.}~\bibnamefont
  {Yuan}},\ }\href {https://doi.org/10.1103/PhysRevD.72.054028} {\bibfield
  {journal} {\bibinfo  {journal} {Phys. Rev. D}\ }\textbf {\bibinfo {volume}
  {72}},\ \bibinfo {pages} {054028} (\bibinfo {year} {2005})}\BibitemShut
  {NoStop}%
\bibitem [{\citenamefont {Anselmino}\ \emph
  {et~al.}(2009{\natexlab{b}})\citenamefont {Anselmino}, \citenamefont
  {Boglione}, \citenamefont {D’Alesio}, \citenamefont {Kotzinian},
  \citenamefont {Melis}, \citenamefont {Murgia}, \citenamefont {Prokudin},\
  and\ \citenamefont {Türk}}]{Anselmino2009}%
  \BibitemOpen
  \bibfield  {author} {\bibinfo {author} {\bibfnamefont {M.}~\bibnamefont
  {Anselmino}}, \bibinfo {author} {\bibfnamefont {M.}~\bibnamefont {Boglione}},
  \bibinfo {author} {\bibfnamefont {U.}~\bibnamefont {D’Alesio}}, \bibinfo
  {author} {\bibfnamefont {A.}~\bibnamefont {Kotzinian}}, \bibinfo {author}
  {\bibfnamefont {S.}~\bibnamefont {Melis}}, \bibinfo {author} {\bibfnamefont
  {F.}~\bibnamefont {Murgia}}, \bibinfo {author} {\bibfnamefont
  {A.}~\bibnamefont {Prokudin}},\ and\ \bibinfo {author} {\bibfnamefont
  {C.}~\bibnamefont {Türk}},\ }\href
  {https://doi.org/10.1140/epja/i2008-10697-y} {\bibfield  {journal} {\bibinfo
  {journal} {The European Physical Journal A}\ }\textbf {\bibinfo {volume}
  {39}},\ \bibinfo {pages} {89} (\bibinfo {year}
  {2009}{\natexlab{b}})}\BibitemShut {NoStop}%
\bibitem [{\citenamefont {Bacchetta}\ and\ \citenamefont
  {Radici}(2011)}]{bachetta2011}%
  \BibitemOpen
  \bibfield  {author} {\bibinfo {author} {\bibfnamefont {A.}~\bibnamefont
  {Bacchetta}}\ and\ \bibinfo {author} {\bibfnamefont {M.}~\bibnamefont
  {Radici}},\ }\href {https://doi.org/10.1103/PhysRevLett.107.212001}
  {\bibfield  {journal} {\bibinfo  {journal} {Phys. Rev. Lett.}\ }\textbf
  {\bibinfo {volume} {107}},\ \bibinfo {pages} {212001} (\bibinfo {year}
  {2011})}\BibitemShut {NoStop}%
\bibitem [{\citenamefont {Hornik}\ \emph {et~al.}(1989)\citenamefont {Hornik},
  \citenamefont {Stinchcombe},\ and\ \citenamefont {White}}]{UAT89}%
  \BibitemOpen
  \bibfield  {author} {\bibinfo {author} {\bibfnamefont {K.}~\bibnamefont
  {Hornik}}, \bibinfo {author} {\bibfnamefont {M.}~\bibnamefont
  {Stinchcombe}},\ and\ \bibinfo {author} {\bibfnamefont {H.}~\bibnamefont
  {White}},\ }\href
  {https://doi.org/https://doi.org/10.1016/0893-6080(89)90020-8} {\bibfield
  {journal} {\bibinfo  {journal} {Neural Networks}\ }\textbf {\bibinfo {volume}
  {2}},\ \bibinfo {pages} {359} (\bibinfo {year} {1989})}\BibitemShut {NoStop}%
\bibitem [{\citenamefont {Cs{\'a}ji}(2001)}]{UAT01}%
  \BibitemOpen
  \bibfield  {author} {\bibinfo {author} {\bibfnamefont {B.~C.}\ \bibnamefont
  {Cs{\'a}ji}},\ }\bibfield  {title} {\bibinfo {title} {Approximation with
  artificial neural networks},\ }\href@noop {} {\bibfield  {journal} {\bibinfo
  {journal} {MS'Thesis, Dept. Science, Eotvos Lorand Univ., Budapest, Hungary}\
  } (\bibinfo {year} {2001})}\BibitemShut {NoStop}%
\bibitem [{\citenamefont {Collins}(2002)}]{JC2002}%
  \BibitemOpen
  \bibfield  {author} {\bibinfo {author} {\bibfnamefont {J.~C.}\ \bibnamefont
  {Collins}},\ }\href
  {https://doi.org/https://doi.org/10.1016/S0370-2693(02)01819-1} {\bibfield
  {journal} {\bibinfo  {journal} {Physics Letters B}\ }\textbf {\bibinfo
  {volume} {536}},\ \bibinfo {pages} {43} (\bibinfo {year} {2002})}\BibitemShut
  {NoStop}%
\bibitem [{\citenamefont {Kang}\ and\ \citenamefont {Qiu}(2010)}]{Kang2010}%
  \BibitemOpen
  \bibfield  {author} {\bibinfo {author} {\bibfnamefont {Z.-B.}\ \bibnamefont
  {Kang}}\ and\ \bibinfo {author} {\bibfnamefont {J.-W.}\ \bibnamefont {Qiu}},\
  }\href {https://doi.org/10.1103/PhysRevD.81.054020} {\bibfield  {journal}
  {\bibinfo  {journal} {Phys. Rev. D}\ }\textbf {\bibinfo {volume} {81}},\
  \bibinfo {pages} {054020} (\bibinfo {year} {2010})}\BibitemShut {NoStop}%
\bibitem [{\citenamefont {Bacchetta}\ \emph {et~al.}(2004)\citenamefont
  {Bacchetta}, \citenamefont {D'Alesio}, \citenamefont {Diehl},\ and\
  \citenamefont {Miller}}]{bachetta2004}%
  \BibitemOpen
  \bibfield  {author} {\bibinfo {author} {\bibfnamefont {A.}~\bibnamefont
  {Bacchetta}}, \bibinfo {author} {\bibfnamefont {U.}~\bibnamefont {D'Alesio}},
  \bibinfo {author} {\bibfnamefont {M.}~\bibnamefont {Diehl}},\ and\ \bibinfo
  {author} {\bibfnamefont {C.~A.}\ \bibnamefont {Miller}},\ }\bibfield  {title}
  {\bibinfo {title} {Single-spin asymmetries: The trento conventions},\ }\href
  {https://doi.org/10.1103/PhysRevD.70.117504} {\bibfield  {journal} {\bibinfo
  {journal} {Phys. Rev. D}\ }\textbf {\bibinfo {volume} {70}},\ \bibinfo
  {pages} {117504} (\bibinfo {year} {2004})}\BibitemShut {NoStop}%
\bibitem [{\citenamefont {Anselmino}\ \emph {et~al.}(2003)\citenamefont
  {Anselmino}, \citenamefont {D'Alesio},\ and\ \citenamefont
  {Murgia}}]{Anselmino_DY_2003}%
  \BibitemOpen
  \bibfield  {author} {\bibinfo {author} {\bibfnamefont {M.}~\bibnamefont
  {Anselmino}}, \bibinfo {author} {\bibfnamefont {U.}~\bibnamefont
  {D'Alesio}},\ and\ \bibinfo {author} {\bibfnamefont {F.}~\bibnamefont
  {Murgia}},\ }\href {https://doi.org/10.1103/PhysRevD.67.074010} {\bibfield
  {journal} {\bibinfo  {journal} {Phys. Rev. D}\ }\textbf {\bibinfo {volume}
  {67}},\ \bibinfo {pages} {074010} (\bibinfo {year} {2003})}\BibitemShut
  {NoStop}%
\bibitem [{\citenamefont {Anselmino}\ \emph {et~al.}(2015)\citenamefont
  {Anselmino}, \citenamefont {Boglione}, \citenamefont {D'Alesio},
  \citenamefont {Hernandez}, \citenamefont {Melis}, \citenamefont {Murgia},\
  and\ \citenamefont {Prokudin}}]{Anselmino_2015}%
  \BibitemOpen
  \bibfield  {author} {\bibinfo {author} {\bibfnamefont {M.}~\bibnamefont
  {Anselmino}}, \bibinfo {author} {\bibfnamefont {M.}~\bibnamefont {Boglione}},
  \bibinfo {author} {\bibfnamefont {U.}~\bibnamefont {D'Alesio}}, \bibinfo
  {author} {\bibfnamefont {J.~O.~G.}\ \bibnamefont {Hernandez}}, \bibinfo
  {author} {\bibfnamefont {S.}~\bibnamefont {Melis}}, \bibinfo {author}
  {\bibfnamefont {F.}~\bibnamefont {Murgia}},\ and\ \bibinfo {author}
  {\bibfnamefont {A.}~\bibnamefont {Prokudin}},\ }\href
  {https://doi.org/10.1103/PhysRevD.92.114023} {\bibfield  {journal} {\bibinfo
  {journal} {Phys. Rev. D}\ }\textbf {\bibinfo {volume} {92}},\ \bibinfo
  {pages} {114023} (\bibinfo {year} {2015})}\BibitemShut {NoStop}%
\bibitem [{\citenamefont {Anselmino}\ \emph
  {et~al.}(2005{\natexlab{b}})\citenamefont {Anselmino}, \citenamefont
  {Boglione}, \citenamefont {D'Alesio}, \citenamefont {Kotzinian},
  \citenamefont {Murgia},\ and\ \citenamefont
  {Prokudin}}]{Anselmino_2005_April}%
  \BibitemOpen
  \bibfield  {author} {\bibinfo {author} {\bibfnamefont {M.}~\bibnamefont
  {Anselmino}}, \bibinfo {author} {\bibfnamefont {M.}~\bibnamefont {Boglione}},
  \bibinfo {author} {\bibfnamefont {U.}~\bibnamefont {D'Alesio}}, \bibinfo
  {author} {\bibfnamefont {A.}~\bibnamefont {Kotzinian}}, \bibinfo {author}
  {\bibfnamefont {F.}~\bibnamefont {Murgia}},\ and\ \bibinfo {author}
  {\bibfnamefont {A.}~\bibnamefont {Prokudin}},\ }\href
  {https://doi.org/10.1103/PhysRevD.71.074006} {\bibfield  {journal} {\bibinfo
  {journal} {Phys. Rev. D}\ }\textbf {\bibinfo {volume} {71}},\ \bibinfo
  {pages} {074006} (\bibinfo {year} {2005}{\natexlab{b}})}\BibitemShut
  {NoStop}%
\bibitem [{\citenamefont {Stump}\ \emph {et~al.}(2003)\citenamefont {Stump},
  \citenamefont {Huston}, \citenamefont {Pumplin}, \citenamefont {Tung},
  \citenamefont {Lai}, \citenamefont {Kuhlmann},\ and\ \citenamefont
  {Owens}}]{CTEQ6l_2003}%
  \BibitemOpen
  \bibfield  {author} {\bibinfo {author} {\bibfnamefont {D.}~\bibnamefont
  {Stump}}, \bibinfo {author} {\bibfnamefont {J.}~\bibnamefont {Huston}},
  \bibinfo {author} {\bibfnamefont {J.}~\bibnamefont {Pumplin}}, \bibinfo
  {author} {\bibfnamefont {W.-K.}\ \bibnamefont {Tung}}, \bibinfo {author}
  {\bibfnamefont {H.~L.}\ \bibnamefont {Lai}}, \bibinfo {author} {\bibfnamefont
  {S.}~\bibnamefont {Kuhlmann}},\ and\ \bibinfo {author} {\bibfnamefont
  {J.~F.}\ \bibnamefont {Owens}},\ }\href
  {https://doi.org/10.1088/1126-6708/2003/10/046} {\bibfield  {journal}
  {\bibinfo  {journal} {Journal of High Energy Physics}\ }\textbf {\bibinfo
  {volume} {2003}},\ \bibinfo {pages} {046} (\bibinfo {year}
  {2003})}\BibitemShut {NoStop}%
\bibitem [{\citenamefont {Buckley}\ \emph {et~al.}(2015)\citenamefont
  {Buckley}, \citenamefont {Ferrando}, \citenamefont {Lloyd}, \citenamefont
  {Nordström}, \citenamefont {Page}, \citenamefont {Rüfenacht}, \citenamefont
  {Schönherr},\ and\ \citenamefont {Watt}}]{LHAPDF6}%
  \BibitemOpen
  \bibfield  {author} {\bibinfo {author} {\bibfnamefont {A.}~\bibnamefont
  {Buckley}}, \bibinfo {author} {\bibfnamefont {J.}~\bibnamefont {Ferrando}},
  \bibinfo {author} {\bibfnamefont {S.}~\bibnamefont {Lloyd}}, \bibinfo
  {author} {\bibfnamefont {K.}~\bibnamefont {Nordström}}, \bibinfo {author}
  {\bibfnamefont {B.}~\bibnamefont {Page}}, \bibinfo {author} {\bibfnamefont
  {M.}~\bibnamefont {Rüfenacht}}, \bibinfo {author} {\bibfnamefont
  {M.}~\bibnamefont {Schönherr}},\ and\ \bibinfo {author} {\bibfnamefont
  {G.}~\bibnamefont {Watt}},\ }\bibfield  {journal} {\bibinfo  {journal} {The
  European Physical Journal C}\ }\textbf {\bibinfo {volume} {75}},\ \href
  {https://doi.org/10.1140/epjc/s10052-015-3318-8}
  {10.1140/epjc/s10052-015-3318-8} (\bibinfo {year} {2015})\BibitemShut
  {NoStop}%
\bibitem [{\citenamefont {Ball}\ and\ \citenamefont
  {et~al}(2022)}]{NNPDF40_2022}%
  \BibitemOpen
  \bibfield  {author} {\bibinfo {author} {\bibfnamefont {R.~D.}\ \bibnamefont
  {Ball}}\ and\ \bibinfo {author} {\bibnamefont {et~al}},\ }\href
  {https://doi.org/10.1140/epjc/s10052-022-10328-7} {\bibfield  {journal}
  {\bibinfo  {journal} {The European Physical Journal C}\ }\textbf {\bibinfo
  {volume} {82}},\ \bibinfo {pages} {428} (\bibinfo {year} {2022})}\BibitemShut
  {NoStop}%
\bibitem [{\citenamefont {de~Florian}\ \emph {et~al.}(2015)\citenamefont
  {de~Florian}, \citenamefont {Sassot}, \citenamefont {Epele}, \citenamefont
  {Hern\'andez-Pinto},\ and\ \citenamefont {Stratmann}}]{DSS14}%
  \BibitemOpen
  \bibfield  {author} {\bibinfo {author} {\bibfnamefont {D.}~\bibnamefont
  {de~Florian}}, \bibinfo {author} {\bibfnamefont {R.}~\bibnamefont {Sassot}},
  \bibinfo {author} {\bibfnamefont {M.}~\bibnamefont {Epele}}, \bibinfo
  {author} {\bibfnamefont {R.~J.}\ \bibnamefont {Hern\'andez-Pinto}},\ and\
  \bibinfo {author} {\bibfnamefont {M.}~\bibnamefont {Stratmann}},\ }\href
  {https://doi.org/10.1103/PhysRevD.91.014035} {\bibfield  {journal} {\bibinfo
  {journal} {Phys. Rev. D}\ }\textbf {\bibinfo {volume} {91}},\ \bibinfo
  {pages} {014035} (\bibinfo {year} {2015})}\BibitemShut {NoStop}%
\bibitem [{\citenamefont {de~Florian}\ \emph {et~al.}(2017)\citenamefont
  {de~Florian}, \citenamefont {Epele}, \citenamefont {Hern\'andez-Pinto},
  \citenamefont {Sassot},\ and\ \citenamefont {Stratmann}}]{DSS17}%
  \BibitemOpen
  \bibfield  {author} {\bibinfo {author} {\bibfnamefont {D.}~\bibnamefont
  {de~Florian}}, \bibinfo {author} {\bibfnamefont {M.}~\bibnamefont {Epele}},
  \bibinfo {author} {\bibfnamefont {R.~J.}\ \bibnamefont {Hern\'andez-Pinto}},
  \bibinfo {author} {\bibfnamefont {R.}~\bibnamefont {Sassot}},\ and\ \bibinfo
  {author} {\bibfnamefont {M.}~\bibnamefont {Stratmann}},\ }\href
  {https://doi.org/10.1103/PhysRevD.95.094019} {\bibfield  {journal} {\bibinfo
  {journal} {Phys. Rev. D}\ }\textbf {\bibinfo {volume} {95}},\ \bibinfo
  {pages} {094019} (\bibinfo {year} {2017})}\BibitemShut {NoStop}%
\bibitem [{\citenamefont {Signori}\ \emph {et~al.}(2013)\citenamefont
  {Signori}, \citenamefont {Bacchetta}, \citenamefont {Radici},\ and\
  \citenamefont {Schnell}}]{Signori_2013}%
  \BibitemOpen
  \bibfield  {author} {\bibinfo {author} {\bibfnamefont {A.}~\bibnamefont
  {Signori}}, \bibinfo {author} {\bibfnamefont {A.}~\bibnamefont {Bacchetta}},
  \bibinfo {author} {\bibfnamefont {M.}~\bibnamefont {Radici}},\ and\ \bibinfo
  {author} {\bibfnamefont {G.}~\bibnamefont {Schnell}},\ }\href
  {https://doi.org/10.1007/JHEP11(2013)194} {\bibfield  {journal} {\bibinfo
  {journal} {Journal of High Energy Physics}\ }\textbf {\bibinfo {volume}
  {2013}},\ \bibinfo {pages} {194} (\bibinfo {year} {2013})}\BibitemShut
  {NoStop}%
\bibitem [{\citenamefont {Anselmino}\ \emph {et~al.}(2014)\citenamefont
  {Anselmino}, \citenamefont {Boglione}, \citenamefont {H.}, \citenamefont
  {Melis},\ and\ \citenamefont {Prokudin}}]{Anselmino_2014}%
  \BibitemOpen
  \bibfield  {author} {\bibinfo {author} {\bibfnamefont {M.}~\bibnamefont
  {Anselmino}}, \bibinfo {author} {\bibfnamefont {M.}~\bibnamefont {Boglione}},
  \bibinfo {author} {\bibfnamefont {J.~O.~G.}\ \bibnamefont {H.}}, \bibinfo
  {author} {\bibfnamefont {S.}~\bibnamefont {Melis}},\ and\ \bibinfo {author}
  {\bibfnamefont {A.}~\bibnamefont {Prokudin}},\ }\href
  {https://doi.org/10.1007/JHEP04(2014)005} {\bibfield  {journal} {\bibinfo
  {journal} {Journal of High Energy Physics}\ }\textbf {\bibinfo {volume}
  {2014}},\ \bibinfo {pages} {5} (\bibinfo {year} {2014})}\BibitemShut
  {NoStop}%
\bibitem [{\citenamefont {Airapetian}\ and\ \citenamefont
  {et~al}(2013)}]{HERMES_2013}%
  \BibitemOpen
  \bibfield  {author} {\bibinfo {author} {\bibfnamefont {A.}~\bibnamefont
  {Airapetian}}\ and\ \bibinfo {author} {\bibnamefont {et~al}} (\bibinfo
  {collaboration} {HERMES Collaboration}),\ }\href
  {https://doi.org/10.1103/PhysRevD.87.074029} {\bibfield  {journal} {\bibinfo
  {journal} {Phys. Rev. D}\ }\textbf {\bibinfo {volume} {87}},\ \bibinfo
  {pages} {074029} (\bibinfo {year} {2013})}\BibitemShut {NoStop}%
\bibitem [{\citenamefont {Collins}\ \emph
  {et~al.}(2006{\natexlab{b}})\citenamefont {Collins}, \citenamefont {Efremov},
  \citenamefont {Goeke}, \citenamefont {Perdekamp}, \citenamefont {Menzel},
  \citenamefont {Meredith}, \citenamefont {Metz},\ and\ \citenamefont
  {Schweitzer}}]{JC_2006_PRD73-094023}%
  \BibitemOpen
  \bibfield  {author} {\bibinfo {author} {\bibfnamefont {J.~C.}\ \bibnamefont
  {Collins}}, \bibinfo {author} {\bibfnamefont {A.~V.}\ \bibnamefont
  {Efremov}}, \bibinfo {author} {\bibfnamefont {K.}~\bibnamefont {Goeke}},
  \bibinfo {author} {\bibfnamefont {M.~G.}\ \bibnamefont {Perdekamp}}, \bibinfo
  {author} {\bibfnamefont {S.}~\bibnamefont {Menzel}}, \bibinfo {author}
  {\bibfnamefont {B.}~\bibnamefont {Meredith}}, \bibinfo {author}
  {\bibfnamefont {A.}~\bibnamefont {Metz}},\ and\ \bibinfo {author}
  {\bibfnamefont {P.}~\bibnamefont {Schweitzer}},\ }\href
  {https://doi.org/10.1103/PhysRevD.73.094023} {\bibfield  {journal} {\bibinfo
  {journal} {Phys. Rev. D}\ }\textbf {\bibinfo {volume} {73}},\ \bibinfo
  {pages} {094023} (\bibinfo {year} {2006}{\natexlab{b}})}\BibitemShut
  {NoStop}%
\bibitem [{\citenamefont {Efremov}\ \emph {et~al.}(2005)\citenamefont
  {Efremov}, \citenamefont {Goeke}, \citenamefont {Menzel}, \citenamefont
  {Metz},\ and\ \citenamefont {Schweitzer}}]{EFREMOV2005233}%
  \BibitemOpen
  \bibfield  {author} {\bibinfo {author} {\bibfnamefont {A.~V.}\ \bibnamefont
  {Efremov}}, \bibinfo {author} {\bibfnamefont {K.}~\bibnamefont {Goeke}},
  \bibinfo {author} {\bibfnamefont {S.}~\bibnamefont {Menzel}}, \bibinfo
  {author} {\bibfnamefont {A.}~\bibnamefont {Metz}},\ and\ \bibinfo {author}
  {\bibfnamefont {P.}~\bibnamefont {Schweitzer}},\ }\href
  {https://doi.org/https://doi.org/10.1016/j.physletb.2005.03.010} {\bibfield
  {journal} {\bibinfo  {journal} {Physics Letters B}\ }\textbf {\bibinfo
  {volume} {612}},\ \bibinfo {pages} {233} (\bibinfo {year}
  {2005})}\BibitemShut {NoStop}%
\bibitem [{\citenamefont {Dembinski}\ \emph {et~al.}(2020)\citenamefont
  {Dembinski}, \citenamefont {Ongmongkolkul},\ and\ \citenamefont
  {et~al}}]{iminuit_ref}%
  \BibitemOpen
  \bibfield  {author} {\bibinfo {author} {\bibfnamefont {H.}~\bibnamefont
  {Dembinski}}, \bibinfo {author} {\bibfnamefont {P.}~\bibnamefont
  {Ongmongkolkul}},\ and\ \bibinfo {author} {\bibnamefont {et~al}},\ }\bibfield
   {journal} {\bibinfo  {journal} {scikit-hep/iminuit}\ }\href
  {https://doi.org/10.5281/zenodo.3949207} {10.5281/zenodo.3949207} (\bibinfo
  {year} {2020})\BibitemShut {NoStop}%
\bibitem [{\citenamefont {Collins}\ \emph
  {et~al.}(2006{\natexlab{c}})\citenamefont {Collins}, \citenamefont {Efremov},
  \citenamefont {Goeke}, \citenamefont {Menzel}, \citenamefont {Metz},\ and\
  \citenamefont {Schweitzer}}]{Collins_2006}%
  \BibitemOpen
  \bibfield  {author} {\bibinfo {author} {\bibfnamefont {J.~C.}\ \bibnamefont
  {Collins}}, \bibinfo {author} {\bibfnamefont {A.~V.}\ \bibnamefont
  {Efremov}}, \bibinfo {author} {\bibfnamefont {K.}~\bibnamefont {Goeke}},
  \bibinfo {author} {\bibfnamefont {S.}~\bibnamefont {Menzel}}, \bibinfo
  {author} {\bibfnamefont {A.}~\bibnamefont {Metz}},\ and\ \bibinfo {author}
  {\bibfnamefont {P.}~\bibnamefont {Schweitzer}},\ }\href
  {https://doi.org/10.1103/PhysRevD.73.014021} {\bibfield  {journal} {\bibinfo
  {journal} {Phys. Rev. D}\ }\textbf {\bibinfo {volume} {73}},\ \bibinfo
  {pages} {014021} (\bibinfo {year} {2006}{\natexlab{c}})}\BibitemShut
  {NoStop}%
\bibitem [{\citenamefont {Anselmino}\ \emph {et~al.}(2012)\citenamefont
  {Anselmino}, \citenamefont {Boglione},\ and\ \citenamefont
  {Melis}}]{Anselmino2012}%
  \BibitemOpen
  \bibfield  {author} {\bibinfo {author} {\bibfnamefont {M.}~\bibnamefont
  {Anselmino}}, \bibinfo {author} {\bibfnamefont {M.}~\bibnamefont
  {Boglione}},\ and\ \bibinfo {author} {\bibfnamefont {S.}~\bibnamefont
  {Melis}},\ }\href {https://doi.org/10.1103/PhysRevD.86.014028} {\bibfield
  {journal} {\bibinfo  {journal} {Phys. Rev. D}\ }\textbf {\bibinfo {volume}
  {86}},\ \bibinfo {pages} {014028} (\bibinfo {year} {2012})}\BibitemShut
  {NoStop}%
\bibitem [{\citenamefont {Sun}\ and\ \citenamefont {Yuan}(2013)}]{Sun_2013}%
  \BibitemOpen
  \bibfield  {author} {\bibinfo {author} {\bibfnamefont {P.}~\bibnamefont
  {Sun}}\ and\ \bibinfo {author} {\bibfnamefont {F.}~\bibnamefont {Yuan}},\
  }\href {https://doi.org/10.1103/PhysRevD.88.114012} {\bibfield  {journal}
  {\bibinfo  {journal} {Phys. Rev. D}\ }\textbf {\bibinfo {volume} {88}},\
  \bibinfo {pages} {114012} (\bibinfo {year} {2013})}\BibitemShut {NoStop}%
\bibitem [{\citenamefont {Luo}\ and\ \citenamefont {Sun}(2020)}]{Luo_2020}%
  \BibitemOpen
  \bibfield  {author} {\bibinfo {author} {\bibfnamefont {X.}~\bibnamefont
  {Luo}}\ and\ \bibinfo {author} {\bibfnamefont {H.}~\bibnamefont {Sun}},\
  }\href {https://doi.org/10.1103/PhysRevD.101.074016} {\bibfield  {journal}
  {\bibinfo  {journal} {Phys. Rev. D}\ }\textbf {\bibinfo {volume} {101}},\
  \bibinfo {pages} {074016} (\bibinfo {year} {2020})}\BibitemShut {NoStop}%
\bibitem [{\citenamefont {Bacchetta}\ \emph
  {et~al.}(2022{\natexlab{a}})\citenamefont {Bacchetta}, \citenamefont
  {Delcarro}, \citenamefont {Pisano},\ and\ \citenamefont
  {Radici}}]{bachetta2022}%
  \BibitemOpen
  \bibfield  {author} {\bibinfo {author} {\bibfnamefont {A.}~\bibnamefont
  {Bacchetta}}, \bibinfo {author} {\bibfnamefont {F.}~\bibnamefont {Delcarro}},
  \bibinfo {author} {\bibfnamefont {C.}~\bibnamefont {Pisano}},\ and\ \bibinfo
  {author} {\bibfnamefont {M.}~\bibnamefont {Radici}},\ }\href
  {https://doi.org/https://doi.org/10.1016/j.physletb.2022.136961} {\bibfield
  {journal} {\bibinfo  {journal} {Physics Letters B}\ }\textbf {\bibinfo
  {volume} {827}},\ \bibinfo {pages} {136961} (\bibinfo {year}
  {2022}{\natexlab{a}})}\BibitemShut {NoStop}%
\bibitem [{\citenamefont {Alekseev}\ and\ \citenamefont
  {et~al.}(2009)}]{COMPAS_Data_Alekseev_2009}%
  \BibitemOpen
  \bibfield  {author} {\bibinfo {author} {\bibfnamefont {M.}~\bibnamefont
  {Alekseev}}\ and\ \bibinfo {author} {\bibnamefont {et~al.}},\ }\href
  {https://doi.org/10.1016/j.physletb.2009.01.060} {\bibfield  {journal}
  {\bibinfo  {journal} {Physics Letters B}\ }\textbf {\bibinfo {volume}
  {673}},\ \bibinfo {pages} {127–135} (\bibinfo {year} {2009})}\BibitemShut
  {NoStop}%
\bibitem [{\citenamefont {Aghasyan}\ and\ \citenamefont
  {et~al.}(2017)}]{COMPASS_DY_2017}%
  \BibitemOpen
  \bibfield  {author} {\bibinfo {author} {\bibfnamefont {M.}~\bibnamefont
  {Aghasyan}}\ and\ \bibinfo {author} {\bibnamefont {et~al.}} (\bibinfo
  {collaboration} {COMPASS Collaboration}),\ }\href
  {https://doi.org/10.1103/PhysRevLett.119.112002} {\bibfield  {journal}
  {\bibinfo  {journal} {Phys. Rev. Lett.}\ }\textbf {\bibinfo {volume} {119}},\
  \bibinfo {pages} {112002} (\bibinfo {year} {2017})}\BibitemShut {NoStop}%
\bibitem [{\citenamefont {Abadi}\ and\ \citenamefont
  {et~al}(2015)}]{tensorflow2015-whitepaper}%
  \BibitemOpen
  \bibfield  {author} {\bibinfo {author} {\bibfnamefont {M.}~\bibnamefont
  {Abadi}}\ and\ \bibinfo {author} {\bibnamefont {et~al}},\ }\bibfield  {title}
  {\bibinfo {title} {{TensorFlow}: Large-scale machine learning on
  heterogeneous systems},\ }\href {https://www.tensorflow.org/} {\bibfield
  {journal} {\bibinfo  {journal} {Software available from tensorflow.org}\ }
  (\bibinfo {year} {2015})}\BibitemShut {NoStop}%
\bibitem [{\citenamefont {Geron}(2019)}]{Geron_2019}%
  \BibitemOpen
  \bibfield  {author} {\bibinfo {author} {\bibfnamefont {A.}~\bibnamefont
  {Geron}},\ }\href@noop {} {\emph {\bibinfo {title} {Hands-on machine learning
  with scikit-learn, keras, and TensorFlow: Concepts, tools, and techniques to
  build intelligent systems}}},\ \bibinfo {edition} {2nd}\ ed.\ (\bibinfo
  {publisher} {O'Reilly Media},\ \bibinfo {year} {2019})\BibitemShut {NoStop}%
\bibitem [{\citenamefont {Airapetian}\ \emph {et~al.}(2020)\citenamefont
  {Airapetian}, \citenamefont {et~al.},\ and\ \citenamefont
  {Collaboration}}]{HERMES_Data_Airapetian_2020}%
  \BibitemOpen
  \bibfield  {author} {\bibinfo {author} {\bibfnamefont {A.}~\bibnamefont
  {Airapetian}}, \bibinfo {author} {\bibnamefont {et~al.}},\ and\ \bibinfo
  {author} {\bibfnamefont {T.~H.}\ \bibnamefont {Collaboration}},\ }\href
  {https://doi.org/10.1007/JHEP12(2020)010} {\bibfield  {journal} {\bibinfo
  {journal} {Journal of High Energy Physics}\ }\textbf {\bibinfo {volume}
  {2020}},\ \bibinfo {pages} {10} (\bibinfo {year} {2020})}\BibitemShut
  {NoStop}%
\bibitem [{\citenamefont {Qian}\ and\ \citenamefont {et~al.}(2011)}]{JLab2011}%
  \BibitemOpen
  \bibfield  {author} {\bibinfo {author} {\bibfnamefont {X.}~\bibnamefont
  {Qian}}\ and\ \bibinfo {author} {\bibnamefont {et~al.}} (\bibinfo
  {collaboration} {Jefferson Lab Hall A Collaboration}),\ }\href
  {https://doi.org/10.1103/PhysRevLett.107.072003} {\bibfield  {journal}
  {\bibinfo  {journal} {Phys. Rev. Lett.}\ }\textbf {\bibinfo {volume} {107}},\
  \bibinfo {pages} {072003} (\bibinfo {year} {2011})}\BibitemShut {NoStop}%
\bibitem [{\citenamefont {Zhao}\ and\ \citenamefont {et~al.}(2014)}]{JLab2014}%
  \BibitemOpen
  \bibfield  {author} {\bibinfo {author} {\bibfnamefont {Y.~X.}\ \bibnamefont
  {Zhao}}\ and\ \bibinfo {author} {\bibnamefont {et~al.}},\ }\href
  {https://doi.org/10.1103/PhysRevC.90.055201} {\bibfield  {journal} {\bibinfo
  {journal} {Phys. Rev. C}\ }\textbf {\bibinfo {volume} {90}},\ \bibinfo
  {pages} {055201} (\bibinfo {year} {2014})}\BibitemShut {NoStop}%
\bibitem [{\citenamefont {Airapetian}\ and\ \citenamefont
  {et~al.}(2009)}]{HERMES_Data_Airapetian_2009}%
  \BibitemOpen
  \bibfield  {author} {\bibinfo {author} {\bibfnamefont {A.}~\bibnamefont
  {Airapetian}}\ and\ \bibinfo {author} {\bibnamefont {et~al.}},\ }\href
  {http://dx.doi.org/10.1103/PhysRevLett.103.152002} {\bibfield  {journal}
  {\bibinfo  {journal} {Physical Review Letters}\ }\textbf {\bibinfo {volume}
  {103}} (\bibinfo {year} {2009})}\BibitemShut {NoStop}%
\bibitem [{\citenamefont {Adolph}\ and\ \citenamefont
  {et~al.}(2015)}]{COMPAS_Data_Adolph_2015}%
  \BibitemOpen
  \bibfield  {author} {\bibinfo {author} {\bibfnamefont {C.}~\bibnamefont
  {Adolph}}\ and\ \bibinfo {author} {\bibnamefont {et~al.}},\ }\href
  {https://doi.org/https://doi.org/10.1016/j.physletb.2015.03.056} {\bibfield
  {journal} {\bibinfo  {journal} {Physics Letters B}\ }\textbf {\bibinfo
  {volume} {744}},\ \bibinfo {pages} {250} (\bibinfo {year}
  {2015})}\BibitemShut {NoStop}%
\bibitem [{\citenamefont {Boglione}\ \emph {et~al.}(2018)\citenamefont
  {Boglione}, \citenamefont {D’Alesio}, \citenamefont {Flore},\ and\
  \citenamefont {Gonzalez-Hernandez}}]{Boglione_2018}%
  \BibitemOpen
  \bibfield  {author} {\bibinfo {author} {\bibfnamefont {M.}~\bibnamefont
  {Boglione}}, \bibinfo {author} {\bibfnamefont {U.}~\bibnamefont
  {D’Alesio}}, \bibinfo {author} {\bibfnamefont {C.}~\bibnamefont {Flore}},\
  and\ \bibinfo {author} {\bibfnamefont {J.~O.}\ \bibnamefont
  {Gonzalez-Hernandez}},\ }\href {https://doi.org/10.1007/JHEP07(2018)148}
  {\bibfield  {journal} {\bibinfo  {journal} {Journal of High Energy Physics}\
  }\textbf {\bibinfo {volume} {2018}},\ \bibinfo {pages} {148} (\bibinfo {year}
  {2018})}\BibitemShut {NoStop}%
\bibitem [{\citenamefont {Pobylitsa}(2003)}]{Pobylitsa_2003}%
  \BibitemOpen
  \bibfield  {author} {\bibinfo {author} {\bibfnamefont {P.~V.}\ \bibnamefont
  {Pobylitsa}},\ }\bibfield  {journal} {\bibinfo  {journal} {ArXiv}\ }\href
  {https://doi.org/arXiv:hep-ph/0301236} {arXiv:hep-ph/0301236} (\bibinfo
  {year} {2003})\BibitemShut {NoStop}%
\bibitem [{\citenamefont {He}\ and\ \citenamefont {Wang}(2019)}]{He_2019}%
  \BibitemOpen
  \bibfield  {author} {\bibinfo {author} {\bibfnamefont {F.}~\bibnamefont
  {He}}\ and\ \bibinfo {author} {\bibfnamefont {P.}~\bibnamefont {Wang}},\
  }\href {https://doi.org/10.1103/PhysRevD.100.074032} {\bibfield  {journal}
  {\bibinfo  {journal} {Phys. Rev. D}\ }\textbf {\bibinfo {volume} {100}},\
  \bibinfo {pages} {074032} (\bibinfo {year} {2019})}\BibitemShut {NoStop}%
\bibitem [{\citenamefont {Klein}\ \emph {et~al.}(2016)\citenamefont {Klein},
  \citenamefont {Keller}, \citenamefont {Liu},\ and\ \citenamefont
  {et~al}}]{SpinQuest_Proposal}%
  \BibitemOpen
  \bibfield  {author} {\bibinfo {author} {\bibfnamefont {A.}~\bibnamefont
  {Klein}}, \bibinfo {author} {\bibfnamefont {D.}~\bibnamefont {Keller}},
  \bibinfo {author} {\bibfnamefont {K.}~\bibnamefont {Liu}},\ and\ \bibinfo
  {author} {\bibnamefont {et~al}},\ }\href@noop {} {\bibfield  {journal}
  {\bibinfo  {journal} {E1039 FNAL proposal SEAQUEST Document 1720-v3}\ }
  (\bibinfo {year} {2016})}\BibitemShut {NoStop}%
\bibitem [{\citenamefont {Barry}\ \emph {et~al.}(2021)\citenamefont {Barry},
  \citenamefont {Ji}, \citenamefont {Sato},\ and\ \citenamefont
  {Melnitchouk}}]{JAM21}%
  \BibitemOpen
  \bibfield  {author} {\bibinfo {author} {\bibfnamefont {P.~C.}\ \bibnamefont
  {Barry}}, \bibinfo {author} {\bibfnamefont {C.-R.}\ \bibnamefont {Ji}},
  \bibinfo {author} {\bibfnamefont {N.}~\bibnamefont {Sato}},\ and\ \bibinfo
  {author} {\bibfnamefont {W.}~\bibnamefont {Melnitchouk}} (\bibinfo
  {collaboration} {JAM Collaboration}),\ }\href
  {https://doi.org/10.1103/PhysRevLett.127.232001} {\bibfield  {journal}
  {\bibinfo  {journal} {Phys. Rev. Lett.}\ }\textbf {\bibinfo {volume} {127}},\
  \bibinfo {pages} {232001} (\bibinfo {year} {2021})}\BibitemShut {NoStop}%
\bibitem [{\citenamefont {Dove}\ \emph {et~al.}(2022)\citenamefont {Dove},
  \citenamefont {et~al.},\ and\ \citenamefont {Collaboration}}]{SeaQuest_2022}%
  \BibitemOpen
  \bibfield  {author} {\bibinfo {author} {\bibfnamefont {J.}~\bibnamefont
  {Dove}}, \bibinfo {author} {\bibnamefont {et~al.}},\ and\ \bibinfo {author}
  {\bibfnamefont {T.~S.}\ \bibnamefont {Collaboration}},\ }\bibfield  {journal}
  {\bibinfo  {journal} {arXiv}\ }\href
  {https://doi.org/10.48550/ARXIV.2212.12160} {10.48550/ARXIV.2212.12160}
  (\bibinfo {year} {2022})\BibitemShut {NoStop}%
\bibitem [{\citenamefont {Chen}\ and\ \citenamefont
  {et~al.}(2019)}]{SpinQuest}%
  \BibitemOpen
  \bibfield  {author} {\bibinfo {author} {\bibfnamefont {A.}~\bibnamefont
  {Chen}}\ and\ \bibinfo {author} {\bibnamefont {et~al.}},\ }\bibfield
  {journal} {\bibinfo  {journal} {arXiv}\ }\href
  {https://doi.org/10.48550/ARXIV.1901.09994} {10.48550/ARXIV.1901.09994}
  (\bibinfo {year} {2019})\BibitemShut {NoStop}%
\bibitem [{\citenamefont {Keller}(2022)}]{Dustin_2022}%
  \BibitemOpen
  \bibfield  {author} {\bibinfo {author} {\bibfnamefont {D.}~\bibnamefont
  {Keller}},\ }\href {https://doi.org/10.48550/arXiv.2205.01249} {\bibfield
  {journal} {\bibinfo  {journal} {arXiv:2205.01249}\ } (\bibinfo {year}
  {2022})}\BibitemShut {NoStop}%
\bibitem [{\citenamefont {Meng}\ \emph {et~al.}(1996)\citenamefont {Meng},
  \citenamefont {Olness},\ and\ \citenamefont {Soper}}]{Meng1996}%
  \BibitemOpen
  \bibfield  {author} {\bibinfo {author} {\bibfnamefont {R.}~\bibnamefont
  {Meng}}, \bibinfo {author} {\bibfnamefont {F.~I.}\ \bibnamefont {Olness}},\
  and\ \bibinfo {author} {\bibfnamefont {D.~E.}\ \bibnamefont {Soper}},\
  }\bibfield  {title} {\bibinfo {title} {Semi-inclusive deeply inelastic
  scattering at small ${q}_{T}$},\ }\href
  {https://doi.org/10.1103/PhysRevD.54.1919} {\bibfield  {journal} {\bibinfo
  {journal} {Phys. Rev. D}\ }\textbf {\bibinfo {volume} {54}},\ \bibinfo
  {pages} {1919} (\bibinfo {year} {1996})}\BibitemShut {NoStop}%
\bibitem [{\citenamefont {Scimemi}\ and\ \citenamefont
  {Vladimirov}(2018)}]{Scimemi_2018}%
  \BibitemOpen
  \bibfield  {author} {\bibinfo {author} {\bibfnamefont {I.}~\bibnamefont
  {Scimemi}}\ and\ \bibinfo {author} {\bibfnamefont {A.}~\bibnamefont
  {Vladimirov}},\ }\bibfield  {title} {\bibinfo {title} {Analysis of vector
  boson production within tmd factorization},\ }\href
  {https://doi.org/10.1140/epjc/s10052-018-5557-y} {\bibfield  {journal}
  {\bibinfo  {journal} {The European Physical Journal C}\ }\textbf {\bibinfo
  {volume} {78}},\ \bibinfo {pages} {89} (\bibinfo {year} {2018})}\BibitemShut
  {NoStop}%
\bibitem [{\citenamefont {Bacchetta}\ \emph {et~al.}(2020)\citenamefont
  {Bacchetta}, \citenamefont {Bertone}, \citenamefont {Bissolotti},
  \citenamefont {Bozzi}, \citenamefont {Delcarro}, \citenamefont {Piacenza},\
  and\ \citenamefont {Radici}}]{Bacchetta_2020}%
  \BibitemOpen
  \bibfield  {author} {\bibinfo {author} {\bibfnamefont {A.}~\bibnamefont
  {Bacchetta}}, \bibinfo {author} {\bibfnamefont {V.}~\bibnamefont {Bertone}},
  \bibinfo {author} {\bibfnamefont {C.}~\bibnamefont {Bissolotti}}, \bibinfo
  {author} {\bibfnamefont {G.}~\bibnamefont {Bozzi}}, \bibinfo {author}
  {\bibfnamefont {F.}~\bibnamefont {Delcarro}}, \bibinfo {author}
  {\bibfnamefont {F.}~\bibnamefont {Piacenza}},\ and\ \bibinfo {author}
  {\bibfnamefont {M.}~\bibnamefont {Radici}},\ }\bibfield  {title} {\bibinfo
  {title} {Transverse-momentum-dependent parton distributions up to n3ll from
  drell-yan data},\ }\href {https://doi.org/10.1007/JHEP07(2020)117} {\bibfield
   {journal} {\bibinfo  {journal} {Journal of High Energy Physics}\ }\textbf
  {\bibinfo {volume} {2020}},\ \bibinfo {pages} {117} (\bibinfo {year}
  {2020})}\BibitemShut {NoStop}%
\bibitem [{\citenamefont {Bacchetta}\ \emph
  {et~al.}(2022{\natexlab{b}})\citenamefont {Bacchetta}, \citenamefont
  {Bertone}, \citenamefont {Bissolotti}, \citenamefont {Bozzi}, \citenamefont
  {Cerutti}, \citenamefont {Piacenza}, \citenamefont {Radici}, \citenamefont
  {Signori},\ and\ \citenamefont {Collaboration}}]{Bacchetta2022_unpolTMD}%
  \BibitemOpen
  \bibfield  {author} {\bibinfo {author} {\bibfnamefont {A.}~\bibnamefont
  {Bacchetta}}, \bibinfo {author} {\bibfnamefont {V.}~\bibnamefont {Bertone}},
  \bibinfo {author} {\bibfnamefont {C.}~\bibnamefont {Bissolotti}}, \bibinfo
  {author} {\bibfnamefont {G.}~\bibnamefont {Bozzi}}, \bibinfo {author}
  {\bibfnamefont {M.}~\bibnamefont {Cerutti}}, \bibinfo {author} {\bibfnamefont
  {F.}~\bibnamefont {Piacenza}}, \bibinfo {author} {\bibfnamefont
  {M.}~\bibnamefont {Radici}}, \bibinfo {author} {\bibfnamefont
  {A.}~\bibnamefont {Signori}},\ and\ \bibinfo {author} {\bibfnamefont {T.~M.
  A.~P.}\ \bibnamefont {Collaboration}},\ }\bibfield  {title} {\bibinfo {title}
  {Unpolarized transverse momentum distributions from a global fit of drell-yan
  and semi-inclusive deep-inelastic scattering data},\ }\href
  {https://doi.org/10.1007/JHEP10(2022)127} {\bibfield  {journal} {\bibinfo
  {journal} {Journal of High Energy Physics}\ }\textbf {\bibinfo {volume}
  {2022}},\ \bibinfo {pages} {127} (\bibinfo {year}
  {2022}{\natexlab{b}})}\BibitemShut {NoStop}%
\bibitem [{\citenamefont {Boussarie}\ and\ \citenamefont
  {et~al}(2023)}]{Boussarie_2023}%
  \BibitemOpen
  \bibfield  {author} {\bibinfo {author} {\bibfnamefont {R.}~\bibnamefont
  {Boussarie}}\ and\ \bibinfo {author} {\bibnamefont {et~al}},\ }\href
  {https://doi.org/10.48550/arXiv.2304.03302} {\bibfield  {journal} {\bibinfo
  {journal} {arXiv:2304.03302}\ } (\bibinfo {year} {2023})}\BibitemShut
  {NoStop}%
\bibitem [{\citenamefont {Bhattacharya}\ \emph {et~al.}(2022)\citenamefont
  {Bhattacharya}, \citenamefont {Kang}, \citenamefont {Metz}, \citenamefont
  {Penn},\ and\ \citenamefont {Pitonyak}}]{Bhattacharya_2022}%
  \BibitemOpen
  \bibfield  {author} {\bibinfo {author} {\bibfnamefont {S.}~\bibnamefont
  {Bhattacharya}}, \bibinfo {author} {\bibfnamefont {Z.-B.}\ \bibnamefont
  {Kang}}, \bibinfo {author} {\bibfnamefont {A.}~\bibnamefont {Metz}}, \bibinfo
  {author} {\bibfnamefont {G.}~\bibnamefont {Penn}},\ and\ \bibinfo {author}
  {\bibfnamefont {D.}~\bibnamefont {Pitonyak}},\ }\href
  {https://doi.org/10.1103/PhysRevD.105.034007} {\bibfield  {journal} {\bibinfo
   {journal} {Phys. Rev. D}\ }\textbf {\bibinfo {volume} {105}},\ \bibinfo
  {pages} {034007} (\bibinfo {year} {2022})}\BibitemShut {NoStop}%
\bibitem [{\citenamefont {Boer}\ and\ \citenamefont
  {et~al}(2010)}]{EIC_Boer_2011}%
  \BibitemOpen
  \bibfield  {author} {\bibinfo {author} {\bibfnamefont {D.}~\bibnamefont
  {Boer}}\ and\ \bibinfo {author} {\bibnamefont {et~al}},\ }\href@noop {}
  {\bibfield  {journal} {\bibinfo  {journal} {A report on the joint
  BNL/INT/Jlab program on the science case for an Electron-Ion Collider.}\ }
  (\bibinfo {year} {2010})}\BibitemShut {NoStop}%
\bibitem [{\citenamefont {Accardi}\ and\ \citenamefont
  {et~al}(2016)}]{EIC_Accardi_2016}%
  \BibitemOpen
  \bibfield  {author} {\bibinfo {author} {\bibfnamefont {A.}~\bibnamefont
  {Accardi}}\ and\ \bibinfo {author} {\bibnamefont {et~al}},\ }\href
  {https://doi.org/10.1140/epja/i2016-16268-9} {\bibfield  {journal} {\bibinfo
  {journal} {The European Physical Journal A}\ }\textbf {\bibinfo {volume}
  {52}},\ \bibinfo {pages} {268} (\bibinfo {year} {2016})}\BibitemShut
  {NoStop}%
\bibitem [{\citenamefont {Khalek}\ and\ \citenamefont
  {et~al}(2022)}]{EIC_Abdul_2022}%
  \BibitemOpen
  \bibfield  {author} {\bibinfo {author} {\bibfnamefont {R.~A.}\ \bibnamefont
  {Khalek}}\ and\ \bibinfo {author} {\bibnamefont {et~al}},\ }\href
  {https://doi.org/https://doi.org/10.1016/j.nuclphysa.2022.122447} {\bibfield
  {journal} {\bibinfo  {journal} {Nuclear Physics A}\ }\textbf {\bibinfo
  {volume} {1026}},\ \bibinfo {pages} {122447} (\bibinfo {year}
  {2022})}\BibitemShut {NoStop}%
\bibitem [{\citenamefont {Ji}(1997)}]{Ji_1997}%
  \BibitemOpen
  \bibfield  {author} {\bibinfo {author} {\bibfnamefont {X.}~\bibnamefont
  {Ji}},\ }\href {https://doi.org/10.1103/PhysRevLett.78.610} {\bibfield
  {journal} {\bibinfo  {journal} {Phys. Rev. Lett.}\ }\textbf {\bibinfo
  {volume} {78}},\ \bibinfo {pages} {610} (\bibinfo {year} {1997})}\BibitemShut
  {NoStop}%
\bibitem [{\citenamefont {Lorc\'e}\ \emph {et~al.}(2012)\citenamefont
  {Lorc\'e}, \citenamefont {Pasquini}, \citenamefont {Xiong},\ and\
  \citenamefont {Yuan}}]{Lorce_2012}%
  \BibitemOpen
  \bibfield  {author} {\bibinfo {author} {\bibfnamefont {C.}~\bibnamefont
  {Lorc\'e}}, \bibinfo {author} {\bibfnamefont {B.}~\bibnamefont {Pasquini}},
  \bibinfo {author} {\bibfnamefont {X.}~\bibnamefont {Xiong}},\ and\ \bibinfo
  {author} {\bibfnamefont {F.}~\bibnamefont {Yuan}},\ }\href
  {https://doi.org/10.1103/PhysRevD.85.114006} {\bibfield  {journal} {\bibinfo
  {journal} {Phys. Rev. D}\ }\textbf {\bibinfo {volume} {85}},\ \bibinfo
  {pages} {114006} (\bibinfo {year} {2012})}\BibitemShut {NoStop}%
\bibitem [{\citenamefont {Radici}(2014)}]{Radici_2014}%
  \BibitemOpen
  \bibfield  {author} {\bibinfo {author} {\bibfnamefont {M.}~\bibnamefont
  {Radici}},\ }\href {https://doi.org/10.1088/1742-6596/527/1/012025}
  {\bibfield  {journal} {\bibinfo  {journal} {Journal of Physics: Conference
  Series}\ }\textbf {\bibinfo {volume} {527}},\ \bibinfo {pages} {012025}
  (\bibinfo {year} {2014})}\BibitemShut {NoStop}%
\bibitem [{\citenamefont {Stefanis}\ \emph {et~al.}(2017)\citenamefont
  {Stefanis}, \citenamefont {Alexandrou}, \citenamefont {Horn}, \citenamefont
  {Moutarde},\ and\ \citenamefont {Scimemi}}]{Stefanis_2017}%
  \BibitemOpen
  \bibfield  {author} {\bibinfo {author} {\bibfnamefont {N.~G.}\ \bibnamefont
  {Stefanis}}, \bibinfo {author} {\bibfnamefont {C.}~\bibnamefont
  {Alexandrou}}, \bibinfo {author} {\bibfnamefont {T.}~\bibnamefont {Horn}},
  \bibinfo {author} {\bibfnamefont {H.}~\bibnamefont {Moutarde}},\ and\
  \bibinfo {author} {\bibfnamefont {I.}~\bibnamefont {Scimemi}},\ }\bibfield
  {journal} {\bibinfo  {journal} {EPJ Web Conf.}\ }\textbf {\bibinfo {volume}
  {137}},\ \href {https://doi.org/10.1051/epjconf/201713701003}
  {10.1051/epjconf/201713701003} (\bibinfo {year} {2017})\BibitemShut {NoStop}%
\bibitem [{\citenamefont {Yoon}\ and\ \citenamefont {et~al}(2017)}]{Yoon_2017}%
  \BibitemOpen
  \bibfield  {author} {\bibinfo {author} {\bibfnamefont {B.}~\bibnamefont
  {Yoon}}\ and\ \bibinfo {author} {\bibnamefont {et~al}},\ }\href
  {https://doi.org/10.1103/PhysRevD.96.094508} {\bibfield  {journal} {\bibinfo
  {journal} {Phys. Rev. D}\ }\textbf {\bibinfo {volume} {96}},\ \bibinfo
  {pages} {094508} (\bibinfo {year} {2017})}\BibitemShut {NoStop}%
\bibitem [{\citenamefont {Lin}(2021)}]{Lin_2021}%
  \BibitemOpen
  \bibfield  {author} {\bibinfo {author} {\bibfnamefont {H.-W.}\ \bibnamefont
  {Lin}},\ }\href {https://doi.org/10.1103/PhysRevLett.127.182001} {\bibfield
  {journal} {\bibinfo  {journal} {Phys. Rev. Lett.}\ }\textbf {\bibinfo
  {volume} {127}},\ \bibinfo {pages} {182001} (\bibinfo {year}
  {2021})}\BibitemShut {NoStop}%
\bibitem [{\citenamefont {Shanahan}\ \emph {et~al.}(2020)\citenamefont
  {Shanahan}, \citenamefont {Wagman},\ and\ \citenamefont
  {Zhao}}]{Shanahan_2020}%
  \BibitemOpen
  \bibfield  {author} {\bibinfo {author} {\bibfnamefont {P.}~\bibnamefont
  {Shanahan}}, \bibinfo {author} {\bibfnamefont {M.}~\bibnamefont {Wagman}},\
  and\ \bibinfo {author} {\bibfnamefont {Y.}~\bibnamefont {Zhao}},\ }\bibfield
  {title} {\bibinfo {title} {Collins-soper kernel for tmd evolution from
  lattice qcd},\ }\href {https://doi.org/10.1103/PhysRevD.102.014511}
  {\bibfield  {journal} {\bibinfo  {journal} {Phys. Rev. D}\ }\textbf {\bibinfo
  {volume} {102}},\ \bibinfo {pages} {014511} (\bibinfo {year}
  {2020})}\BibitemShut {NoStop}%
\bibitem [{\citenamefont {Aybat}\ \emph
  {et~al.}(2012{\natexlab{b}})\citenamefont {Aybat}, \citenamefont {Collins},
  \citenamefont {Qiu},\ and\ \citenamefont {Rogers}}]{Aybat_2012Feb}%
  \BibitemOpen
  \bibfield  {author} {\bibinfo {author} {\bibfnamefont {S.~M.}\ \bibnamefont
  {Aybat}}, \bibinfo {author} {\bibfnamefont {J.~C.}\ \bibnamefont {Collins}},
  \bibinfo {author} {\bibfnamefont {J.~W.}\ \bibnamefont {Qiu}},\ and\ \bibinfo
  {author} {\bibfnamefont {T.~C.}\ \bibnamefont {Rogers}},\ }\href
  {https://doi.org/10.1103/PhysRevD.85.034043} {\bibfield  {journal} {\bibinfo
  {journal} {Phys. Rev. D}\ }\textbf {\bibinfo {volume} {85}},\ \bibinfo
  {pages} {034043} (\bibinfo {year} {2012}{\natexlab{b}})}\BibitemShut
  {NoStop}%
\end{thebibliography}%

\end{document}